\documentclass[acmsmall]{acmart}

\AtBeginDocument{%
  }
\usepackage{threeparttable}
\usepackage{multirow}
\usepackage{longtable}
\usepackage{ragged2e}
\usepackage{array}

\setcopyright{none}
\begin{document}

\title{Before You Scroll Again: Predicting Regretful Social Media Sessions
       from In-the-Wild Contextual and Wearable Sensing}

\authornote{Authors contributed equally to this work.}

\author{Sally Ahmed}
\authornotemark[1]
\email{sallyelf@media.mit.edu}
\orcid{0009-0005-3399-0092}
\affiliation{%
  \institution{MIT Media Lab}
  \city{Cambridge}
  \country{USA}}

\author{Jan Enkmann}
\authornotemark[1]
\email{jan.enkmann@hpi.de}
\affiliation{%
  \institution{Hasso Plattner Institute}
  \city{Potsdam}
  \country{Germany}}

\author{Kye Shimizu}
\authornotemark[1]
\email{kyeshmz@media.mit.edu}
\orcid{0000-0002-5017-6966}
\affiliation{%
  \institution{MIT Media Lab}
  \city{Cambridge}
  \country{USA}}

\author{Ivy Yip}
\email{ivysyip@mit.edu}
\orcid{0009-0006-7958-7730}
\affiliation{%
  \institution{MIT Media Lab}
  \city{Cambridge}
  \country{USA}}

\author{Vincent Beermann}
\email{vincent.beermann@hpi.de}
\affiliation{%
  \institution{Hasso Plattner Institute}
  \city{Potsdam}
  \country{Germany}}

\author{Ayse Alomar}
\email{ayse.alomar@student.hpi.uni-potsdam.de}
\affiliation{%
  \institution{Hasso Plattner Institute}
  \city{Potsdam}
  \country{Germany}}

\author{Falk Uebernickel}
\email{falk.uebernickel@hpi.de}
\affiliation{%
  \institution{Hasso Plattner Institute}
  \city{Potsdam}
  \country{Germany}}

\author{Pattie Maes}
\email{pattie@media.mit.edu}
\orcid{0000-0002-7722-6038}
\affiliation{%
  \institution{MIT Media Lab}
  \city{Cambridge}
  \country{USA}}

\renewcommand{\shortauthors}{Ahmed, Enkmann, Shimizu, et. al}

\begin{abstract}
Users often feel regret after using social media, making regret a more ecologically valid target than screen time for understanding when phone use becomes problematic. Existing self-monitoring tools cannot anticipate regret before it occurs, and prior physiological work on social media use has been confined to the lab with research-grade sensors and curated content, leaving the question of in-the-wild prediction open. We deployed a 7-day in-the-wild experience sampling study with 21 participants, combining passive smartphone logging, a low-cost consumer smartwatch (Bangle.js 2, \$80), session-level surveys (1,445 sessions), and exit interviews to investigate when and why social media sessions become regretful, and whether regret can be anticipated before a session begins. Three findings stand out: (i) the gap between intended and actual use predicts regret far more strongly than session duration, with duration's apparent effect collapsing once intention is modeled; (ii) regret is amplified when sessions displace a valued alternative, particularly at night and following productivity-app use; and (iii) pre-session contextual features generalize across participants while physiological signals add person-specific lift, pointing toward a two-layer architecture for just-in-time adaptive interventions. Interview themes of scrolling-as-avoidance and time blindness contextualize these patterns and surface design opportunities beyond timer-based interventions.
\end{abstract}



\received{20 February 2007}
\received[revised]{12 March 2009}
\received[accepted]{5 June 2009}

\maketitle

\section{Introduction}

Social media use has become deeply embedded in daily life, yet not all engagement is experienced equally. While platforms were originally designed to foster connection and communication, growing evidence suggests that a significant portion of use has become compulsive and regret-inducing, driven less by user intention than by platform design features such as infinite scroll, autoplay, and algorithmically personalized feeds \cite{lukoff_what_2018, roffarello_achieving_2023,rixen2023loop}. Existing interventions, most commonly screen time dashboards and app timers, operate on a blunt assumption: that reducing total usage time is equivalent to reducing harm. This conflates quantity with quality, treating thirty minutes of an intentional catch-up with a friend the same as thirty minutes of late-night doomscrolling \cite{lukoff_what_2018}.

A more precise outcome measure is needed. We focus on regret: a counterfactual cognitive-affective state in which a person evaluates their past behavior against an imagined alternative and concludes that a different action would have been preferable \citep{zeelenberg_anticipated_1999, gilovich_experience_1995}. Unlike screen time, regret directly captures subjective harm: a session can be long without being regretted, and short sessions can still leave users feeling their time was wasted. Measuring and predicting regret therefore offers a more ecologically valid signal for just-in-time adaptive interventions (JITAIs), which must act before problematic behavior occurs, not after it \cite{nahum-shani_just--time_2025}.

Prior work has begun to characterize regret in social media contexts. \citet{cho2021reflect} used experience sampling to identify correlates of post-session regret, and \citet{guo_what_2025} examined how regret varies across app types and contexts. Both, however, treat regret as an outcome to be explained after the fact rather than a state to be predicted before a session begins, and neither incorporates passive physiological sensing — a critical gap, since by the time a session is underway the JITAI intervention window has already narrowed \cite{orzikulova_time2stop_2024}.

Passive smartphone logs offer scalable behavioral context — prior app usage, time of day, session frequency, app-switching — but they cannot observe the internal state a person brings into a session. Consumer-grade smartwatches can complement this with continuous heart rate, accelerometry, and skin temperature data that may reflect stress, fatigue, and arousal in the moments before a session begins \cite{sarsenbayeva2020smartphone}. Recent work shows that even without specialized processing such as HRV, consumer wearables can predict affective and behavioral states in naturalistic settings \cite{pakhomov2020consumer, quiroz2018emotion, tutunji2023detecting}. Whether these signals, combined with behavioral context, can predict regret in naturalistic social media use remains open.

We address this question through a 7-day in-the-wild experience sampling study with $N = 21$ participants, combining passive smartphone logging, consumer-grade smartwatch sensing (Bangle.js 2), post-session and end-of-day surveys, and exit interviews, with all engagement occurring on participants' own social media accounts.

Building on prior work~\cite{cho2021reflect, guo_what_2025, lukoff_what_2018, orzikulova_time2stop_2024}, we investigate the following research questions:

\textbf{RQ1:} What contextual and behavioral factors are associated with session-level regret?

\textbf{RQ2:} How accurately can pre-session variables predict upcoming regret, and which variable types contribute most to prediction?

Three findings emerge with implications for sensing-informed interventions:
\begin{enumerate}
    \item \textbf{Regret tracks the intention-usage gap, not session duration.} The perceived gap between intended and actual time predicted regret roughly twice as strongly as duration ($\beta_{\text{IUG}} = 0.76$ vs.\ $\beta_{\text{DUR}} = 0.36$); entered jointly, duration's effect shrank to $\beta = 0.18$ while the intention-usage gap remained dominant. The qualitative theme of \textit{time blindness} converges on the same reading.

    \item \textbf{Regret is amplified when opportunity costs are salient.} Sessions at night and sessions following productivity-app use carried elevated regret, echoing the qualitative \textit{escape-hatch} theme of turning to social media to defer an unwanted task. The felt cost of a session tracks the value of what it competes with.

    \item \textbf{Pre-session passive signals modestly predict above-median regret, but deployability is heterogeneous.} Combined phone-context and smartwatch features reach a within-person mean AUC of $.740$ vs.\ $.732$ for phone context alone. On the smartwatch-eligible subset, both variants qualify $5/19$ participants individually (AUC\,$\ge\,.70$), but the deployable sets differ by two participants: the smartwatch channel shifts coverage rather than expanding it. A sensing-based JITAI needs per-user qualification, not a blanket rollout.
\end{enumerate}

\section{Related Work}

\subsection{Problematic Social Media Use from a Self-Control Theory Perspective}

Excessive social-media use is typically studied in HCI and psychology under the label of \textit{problematic social media use} (PSMU), defined as ``compulsive and addictive patterns of use that lead to distress or impairment'' \citep[p.~14]{Shannon2025}. PSMU has been linked to elevated rates of depression, anxiety, and stress \citep{Shannon2022} and to reduced life satisfaction and self-esteem \citep{Huang2022}, and the rise of always-available smartphones has amplified the problem \citep{Marino2021, Montag2021}.

Two complementary mechanistic accounts of PSMU dominate the literature. The first treats PSMU as a behavioral addiction sustained by reward learning in mesocorticolimbic systems and the platforms' near-infinite stream of low-effort potential rewards \citep{Andreassen2016a, Robinson2025, Shannon2025, Flayelle2023}. The second treats PSMU as habit, with notifications and app icons acting as contextual cues that reliably predict reward and so trigger automatic engagement \citep{Bayer2022, Lindstrom2021, Oulasvirta2012, Wood2016}. Both accounts converge on the same point: repeated engagement becomes \textit{problematic} not in itself, but when ``specific habit sequences consistently undercut users' goals'' \citep[p.~3]{Bayer2022}. Self-control theory provides the language to describe that conflict and the leverage points for intervening on it.

Following \citet{Inzlicht2021}, we distinguish \textit{self-regulation}, the broad process of steering behavior toward valued goals, from \textit{self-control}, which refers specifically to the resolution of conflict between two mutually exclusive options where one delivers an immediately available, lower-priority reward and the other serves a more valued but more distal goal (see also \citealp{Duckworth2016}). Among competing accounts of how this conflict is resolved, the \textit{value-based choice} model \citep{Berkman2017} has received the strongest recent neuroscientific support \citep{Cosme2019, Scholz2022}. On this account, self-control is not a contest between two systems \citep{Hofmann2009} or the depletion of a willpower resource \citep{Baumeister2007} but the behavioral consequence of a momentary value computation that integrates expected reward, required effort, and the opportunity cost of forgoing alternatives. Social media is engineered to win this computation: reward is high, effort is near-zero, and at many decision points (waiting for a bus, an ad break, a quiet pause) the salient opportunity cost is itself low. Use becomes subjectively problematic when it competes with a high-value long-term goal whose component actions are effortful and whose payoffs are uncertain \citep{Reinecke2016}; in that competition long-term goals often lose, and the loss is what we recognize as a failure of self-control \citep{Kotabe2015}.

The empirical relevance of this framework is well established. \citet{Hofmann2012} sampled 7{,}827 desire reports from 205 adults and found that media-related desires were among the most frequent, conflicted with other goals more often than most categories, and were resisted less successfully than urges in many other domains. \citet{Reinecke2016} showed that more than half of all social media sessions conflict with other important goals. Trait-level evidence adds a useful qualification: across three studies of university students ($N = 973$), \citet{Troll2021} found that trait self-control predicted academic performance, but the effect was \textit{not} mediated by aggregate screen time or pickups; what mediated it was \textit{how} the phone was used and whether students adopted placement habits that pre-empted conflict. As the authors put it, ``it is less a matter of how much but rather a matter of how students are using their smartphones'' \citep[p.~7]{Troll2021}. Together, these findings motivate the rest of this paper: whether an episode is problematic depends not on its duration but on whether it conflicts with more important goals at that moment, a property aggregate screen-time metrics cannot capture and that motivates measurement at the level of the individual session.

\subsection{Regret as Cognitive Affect}

Self-control theory tells us that media use becomes problematic when a momentary value computation favours an immediately available, low-effort reward over a more valued long-term goal. Regret is the user's own readout of that computation in retrospect: a quantification of the subjective experience that follows a failure of self-control. In the psychological and economic tradition, regret is a cognitively based emotion that arises when a person realizes or imagines that the present situation would have been better had a different choice been made \citep{zeelenberg_anticipated_1999, connolly_regret_2002}. The construct is structurally an opportunity-cost evaluation, the same kind of cross-option comparison that drives the value-based-choice account of self-control, only carried out after the fact: what was obtained is weighed against what could have been obtained, and the gap between them is the regret signal.

This anchors regret cleanly against the dimensional affect captured by valence and arousal. Where valence and arousal index a momentary feeling state, regret implicates a comparison between what happened and what could have happened, and that comparison is what gives the affect its cognitive content. Anticipated regret, the prospective form of the same comparison, has been shown to shape choices under uncertainty across domains including consumer and interpersonal decisions \citep{zeelenberg_anticipated_1999}, while the experienced form follows a characteristic temporal pattern in which actions generate more short-term regret and inactions accumulate regret over the long run \citep{gilovich_experience_1995}. For digital wellbeing, the experienced form is the more relevant signal: it indexes the user's own assessment that a session conflicted with what they otherwise valued.

Recent HCI work has operationalized this construct in the wild. \citet{lukoff_what_2018} sampled 86{,}402 sessions of app use and found that a session's duration predicted neither how meaningful nor how problematic it felt; what predicted those judgments was whether the user had opened the app instrumentally or out of habit, with passive social-media browsing producing the lowest meaningfulness ratings and participants describing the experience as a loss of autonomy rather than a problem of duration. \citet{cho2021reflect} extended the picture to the feature level, showing that users distinguish active features (messaging, posting), rarely regretted, from passive features (feed scrolling, recommendations), frequently regretted, and that most regretted sessions are partial collapses of a meaningful intent into mindless consumption mid-session. A multimodal-LLM analysis of more than 34{,}000 smartphone screenshots reaches the same conclusion at scale and ties regret directly to intention: regret is highest when behavior deviates from the goal that motivated the session, when use is non-intentional, and when algorithmic recommendations pull attention from the original purpose, with about 60\% of sessions showing such drift \citep{guo_what_2025}. Regret in this literature thus serves as the user's own readout of an intention--behavior mismatch, and the construct lends itself to event-level measurement that aggregate-level metrics cannot provide.

\subsection{In-the-Wild Studies of Phone Use and Affect}

Beyond the regret-specific work just discussed, a parallel body of in-the-wild research has used passive sensing and experience sampling to characterize how phone and social-media use unfold in everyday life. \citet{ruensuk_how_2020} demonstrated that built-in smartphone sensors can predict transitory affective states during social-media browsing with high accuracy, and \citet{gebhardt_detecting_2024} extended this to passive consumption, classifying up to eight emotional states from behavioral and physiological signals during scrolling, with behavioral features alone proving sufficient for robust detection. These studies establish that affective signals are recoverable in naturalistic feed-browsing settings, but they target emotion \textit{during} use rather than the conditions preceding a session.

Other work has shifted attention to user-level structure rather than within-session affect. \citet{sramek_beyond_2025} clustered Instagram users by feature-level usage patterns over two weeks ($N = 108$) and found previously unidentified subgroups whose maladaptive use was not captured by aggregate duration: feature composition, not amount of use, defined the high-risk group. \citet{kim_beneficial_2021} similarly showed in a fourteen-week study of college students that what they termed \textit{willful neglect} of incoming notifications was positively related to academic performance, independent of overall screen time. \citet{lee_leveraging_2025} extended this stream by extracting fine-grained interaction routines from app sequences to predict daily stress in a one-month deployment, achieving 75\% accuracy with a small set of routine-based features.

Three observations follow from this literature. First, behavioral logs are most informative when analyzed at finer-than-app granularity, whether at the feature level \citep{cho2021reflect, sramek_beyond_2025} or at the level of within-app interaction routines \citep{lee_leveraging_2025}. Second, in-the-wild affect detection during phone use is feasible with consumer-accessible sensors \citep{ruensuk_how_2020, gebhardt_detecting_2024}. Third, no in-the-wild study has yet asked whether the user's physiological state \textit{before} a session, measured passively through a wearable, predicts whether that session will be regretted. Bridging this gap is the central empirical question of the present paper.

\subsection{Wearable Physiological Sensing for Inferring User States}
Affective computing \citep{picard1997affective} laid the foundation for systems that recognize human emotion, and physiological signals (heart rate, skin temperature, accelerometry) emerged early as a promising path because they support passive, continuous sensing \citep{picard2001toward, hickey2021ppgstress}. Much of this work has relied on research-grade devices like the Empatica E4 \citep{saganowski2022emotion}, but consumer smartwatches with limited sensor sets have shown comparable promise in everyday settings \citep{quiroz2018emotion, kapogianni2025using}, and multi-modal pipelines fusing wearable signals with smartphone logs have been used as digital biomarkers of stress, anxiety, and affect \citep{saylam2023quantifying, saganowski2020consumer}.

Three challenges shape the design space for the present study. First, performance degrades from lab to field: wrist-worn sensors achieve strong affect prediction in controlled settings \citep{siirtola2023wesad, schmidt2018wesad} but suffer in deployment from motion artifacts, non-wear, and contextual confounds \citep{can2019continuous, saganowski2022emotion, hovsepian2015cstress, toshnazarov2024sosw}, with field performance varying widely across studies \citep{harper2019endtoend, dai2021comparing, park2020wellbeat, yu2023semisupervised, rashid2024reproducible}. Second, fusing physiological and behavioral data helps: \citet{wang2014studentlife} established the gold standard for combining passive sensing with survey data, \citet{morshed2019mood} extended this to predict mood \textit{instability} (a framing that parallels our interest in within-person variability rather than stable traits), and personalized multi-modal models on consumer wearables show meaningful gains over single-source baselines \citep{taylor2017personalized, jaques2015predicting, sano2018identifying, umematsu2020emotional, yan2020affect, shah2021personalized, schmidt2024personalized, bari2020stressful, garciaceja2018engagement, sultana2020using}. Third, generalization across individuals remains the hardest challenge: \citet{meegahapola2023generalization} (329K self-reports, 678 participants) found partially personalized models substantially outperformed global ones, and long-running deployments show patterns shift over time \citep{xu2023globem, wang2024college}. These findings motivate our dual evaluation: within-person temporal splits that mirror deployment, and Leave-One-Participant-Out cross-validation that tests transfer to new individuals.

A smaller body of work asks whether physiology measured \textit{before} an event predicts its outcome. \citet{taylor2017personalized} predicted next-day mood and stress from prior-day data, and sleep and circadian features have emerged as reliable next-day affect predictors \citep{jaques2015predicting, sano2018identifying}, but these target broad daily affect rather than specific behavioral episodes. \citet{lin2025beyond} argue the field must move beyond detection toward actionable sensing. Sensor-derived context has been used to time intervention delivery: \citet{mishra2021receptivity} showed contextual features improve mHealth receptivity by up to 40\% over random delivery, and Sense2Stop \citep{battalio2021sense2stop} demonstrated wearable-triggered just-in-time stress management. No prior work, however, has tested whether physiological state immediately before a social-media session predicts the quality of that session.

\subsection{Interventions for Smartphone Overuse}

Existing smartphone interventions range from time-based lockouts to friction-based tools, but they share a common limitation: they trigger on behavioral signals (duration, app-usage patterns) rather than on the user's internal state. A systematic review of digital self-control tools found that strong lockout mechanisms outperform notifications, though users frequently report frustration with rigid restrictions \citep{roffarello_achieving_2023}. A complementary limitation is sustained engagement: Epstein et al. \cite{epstein2016reconsidering} identified lapsing as a recurring stage in personal informatics tool use, with users abandoning self-tracking systems when the tools fail to adapt to changing contexts or goals. The same pattern recurs in screen-time interventions, which our qualitative findings (Section 5.4) corroborate from the user's perspective. Even simple pre-use friction can substantially reduce habitual app opening: the One Sec study, which intercepts an app launch with a short waiting time and an option to dismiss the attempt, decreased actual app openings by 57\% over six weeks in a field experiment with 280 participants \citep{gruning_directing_2023}. Just-in-time adaptive interventions (JITAIs) extend this trajectory \citep{nahum-shani_just--time_2025}, building on earlier evidence that context-aware prompting raises user satisfaction and lowers response times relative to context-unaware delivery \citep{pejovic2014interruptme}, that pre-delivery context shapes whether users respond to or ignore notifications in the first place \cite{mehrotra2016myphone}. Adaptive systems such as Time2Stop have shown that dynamic triggering outperforms static rules \citep{orzikulova_time2stop_2024}, while user-defined context-action rules face challenges when system responses do not align with user goals \citep{kim_navigating_2024}. SwitchTube illustrates a complementary direction: by letting users toggle in real time between a Focus Mode that hides YouTube's recommendations and an Explore Mode that surfaces them, it shifts the design problem from blanket blocking to giving users moment-to-moment control over engagement with engagement-maximizing features \citep{lukoff_switchtube_2023}. \citet{shahu2024beyond} explored smartwatch-delivered interventions for digital wellbeing and found initial reductions in screen time that diminished over two weeks, highlighting the limits of fixed prompts and the case for adaptive, state-aware triggering. What remains untested is whether physiological signals could serve as triggers for proactive intervention before a regretful session begins.

\section{Method}
\label{sec:method}

%

We conducted a one-week, in-the-wild study combining baseline psychometrics, continuous smartwatch sensing, smartphone app-usage logging, experience sampling, daily surveys, and a follow-up interview. The protocol was piloted with four participants and refined for reliability of data capture and survey clarity. The study was approved by the authors' Institutional Review Board (Protocol \#XXXXXXXXXXXXXX); all participants gave written informed consent.

\subsection{Participants}

We recruited Android users via university and community mailing lists, screening for daily social-media use of more than 1--2 hours, willingness to wear a smartwatch continuously for one week, and willingness to complete end-of-session and end-of-day surveys. Eligible respondents attended an in-person onboarding session and received a \$60 gift card on completion. The final sample comprised 21 participants (11~female, 10~male; age 18--33, $M = 21.8$, $SD = 4.5$), predominantly undergraduate students at a local university; baseline psychometric scores spanned subclinical to moderate symptom levels (Table~\ref{tab:demographics}; Appendix~\ref{sec:exclusion-flow}).

\subsection{Procedure}

At onboarding, participants gave informed consent, completed the baseline survey battery (Section~\ref{sec:baseline}), and were fitted with a Bangle.js~2 smartwatch~\cite{banglejs2}; the study app was installed on their Android phone and permissions for background usage logging and notifications were configured. We chose the Bangle.js~2 for its low cost, open-source firmware, and on-device local storage, enabling deployment without continuous network connectivity. Participants were instructed to use their phone normally for seven days, to wear the watch during waking hours, and to respond to post-session and end-of-day surveys. At offboarding, the app was de-installed, the watch was returned, and a sub-sample completed a semi-structured exit interview reflecting on their study experience, regretful and meaningful sessions, and suggestions for healthier use.
 
\subsection{Implementation}
\label{implementation}

\subsubsection{Bangle.js~2 Smartwatch}
Throughout the 7-day period, the Bangle.js~2 smartwatch~\cite{banglejs2} continuously collected physiological and behavioral signals, including heart rate, activity levels, and skin temperature. All variables (for details see Appendix~\ref{sec:smartwatch-data}) were collected for one-minute intervals followed by a 4-minute recording break, resulting in twelve one-minute measurement periods per hour. These data streams were used to characterize physiological arousal and behavioral patterns across daily life. Data were stored locally on the watch and retrieved after returning it during the offboarding session. Raw data were subsequently preprocessed to remove noise, identify outliers and hardware artifacts, and ensure analytical validity. The watch has no independent Wifi, so its clock was paired with the participant's phone over Bluetooth using Gadgetbridge~\citep{gadgetbridge}, an open-source companion app that periodically pushes the phone's time to the watch. This ensured that smartwatch samples and phone app-usage events shared a common timeline for downstream synchronization.

\subsubsection{Android Observation Application}
\begin{figure}[t]
  \centering
  \begin{minipage}[t]{0.23\columnwidth}
    \centering
    \includegraphics[width=\linewidth]{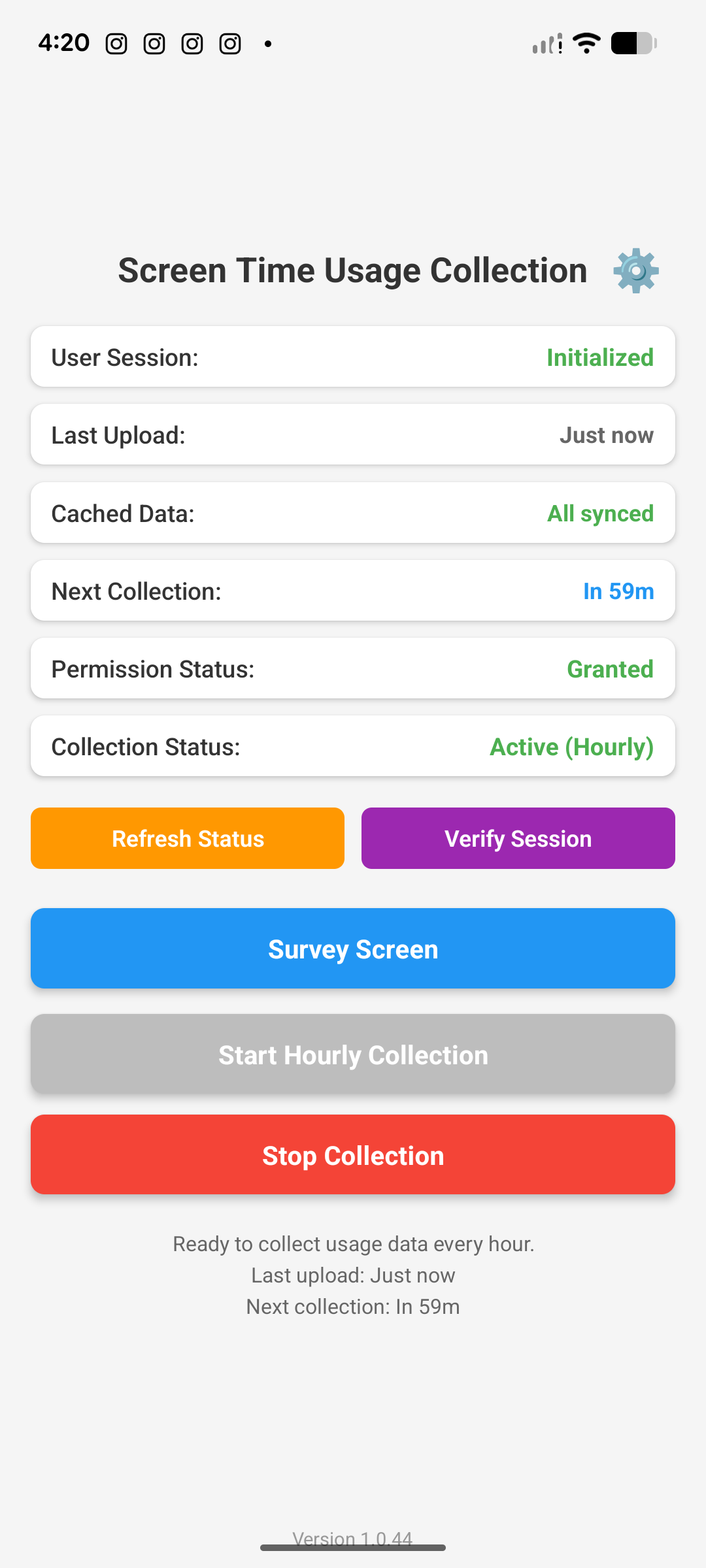}
    \small{(a)}
  \end{minipage}\hfill
  \begin{minipage}[t]{0.23\columnwidth}
    \centering
    \includegraphics[width=\linewidth]{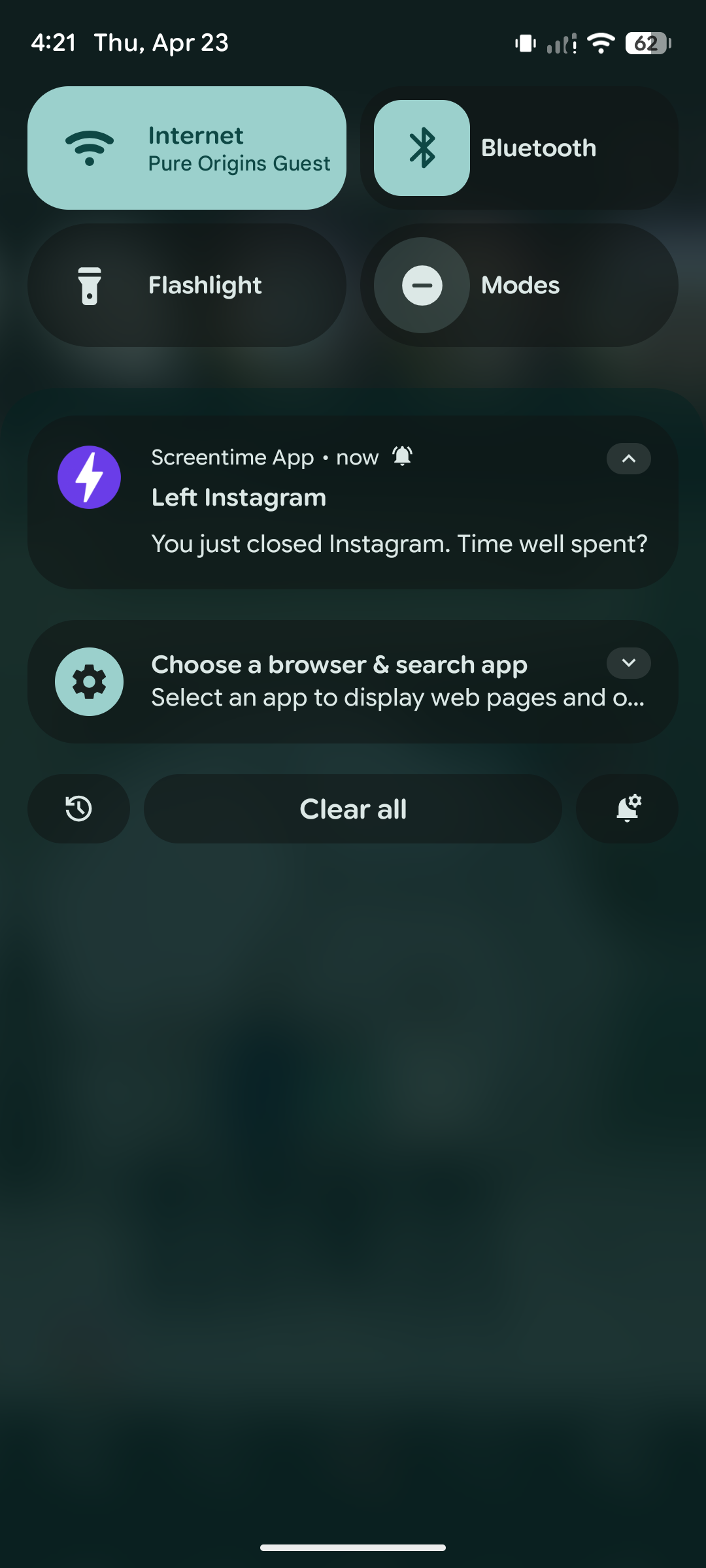}
    \small{(b)}
  \end{minipage}\hfill
  \begin{minipage}[t]{0.23\columnwidth}
    \centering
    \includegraphics[width=\linewidth]{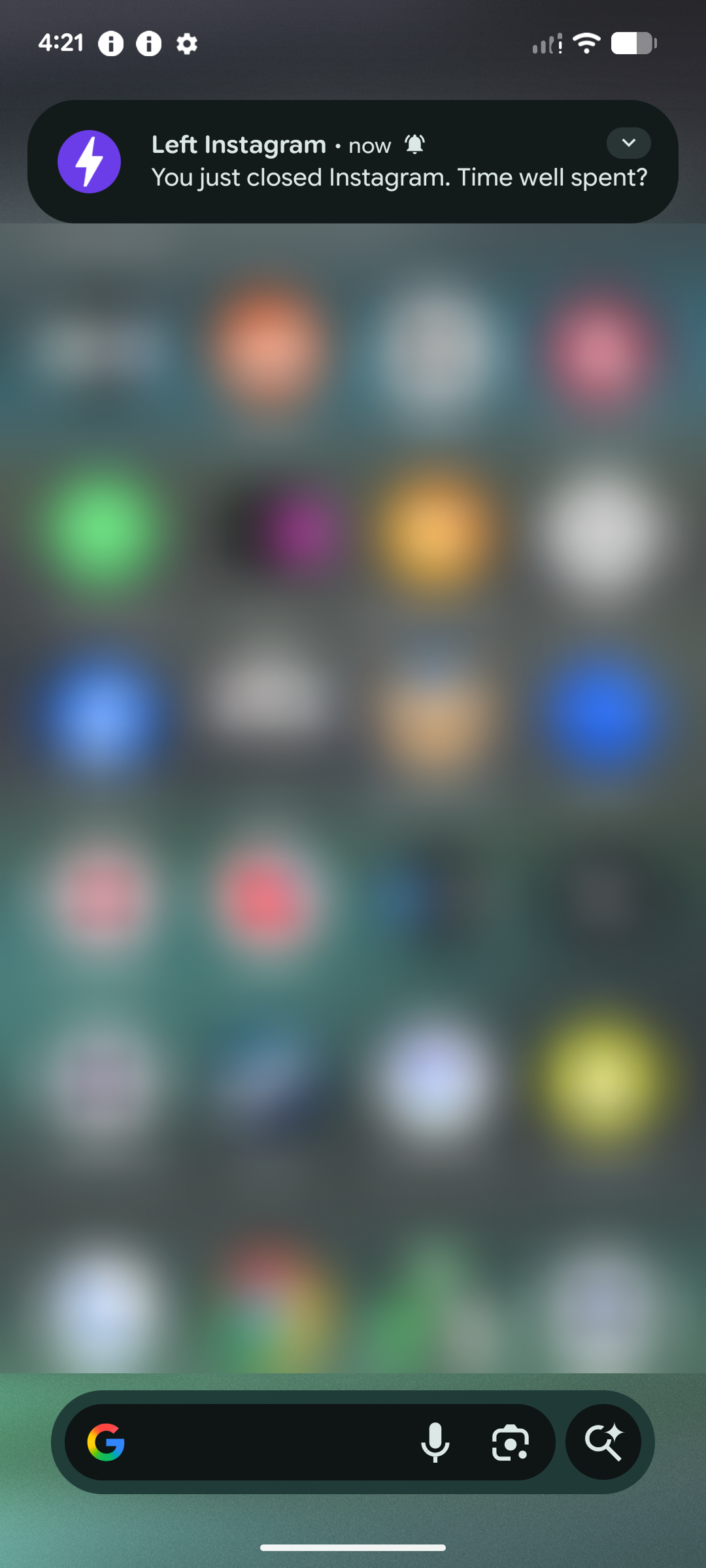}
    \small{(c)}
  \end{minipage}\hfill
  \begin{minipage}[t]{0.23\columnwidth}
    \centering
    \includegraphics[width=\linewidth]{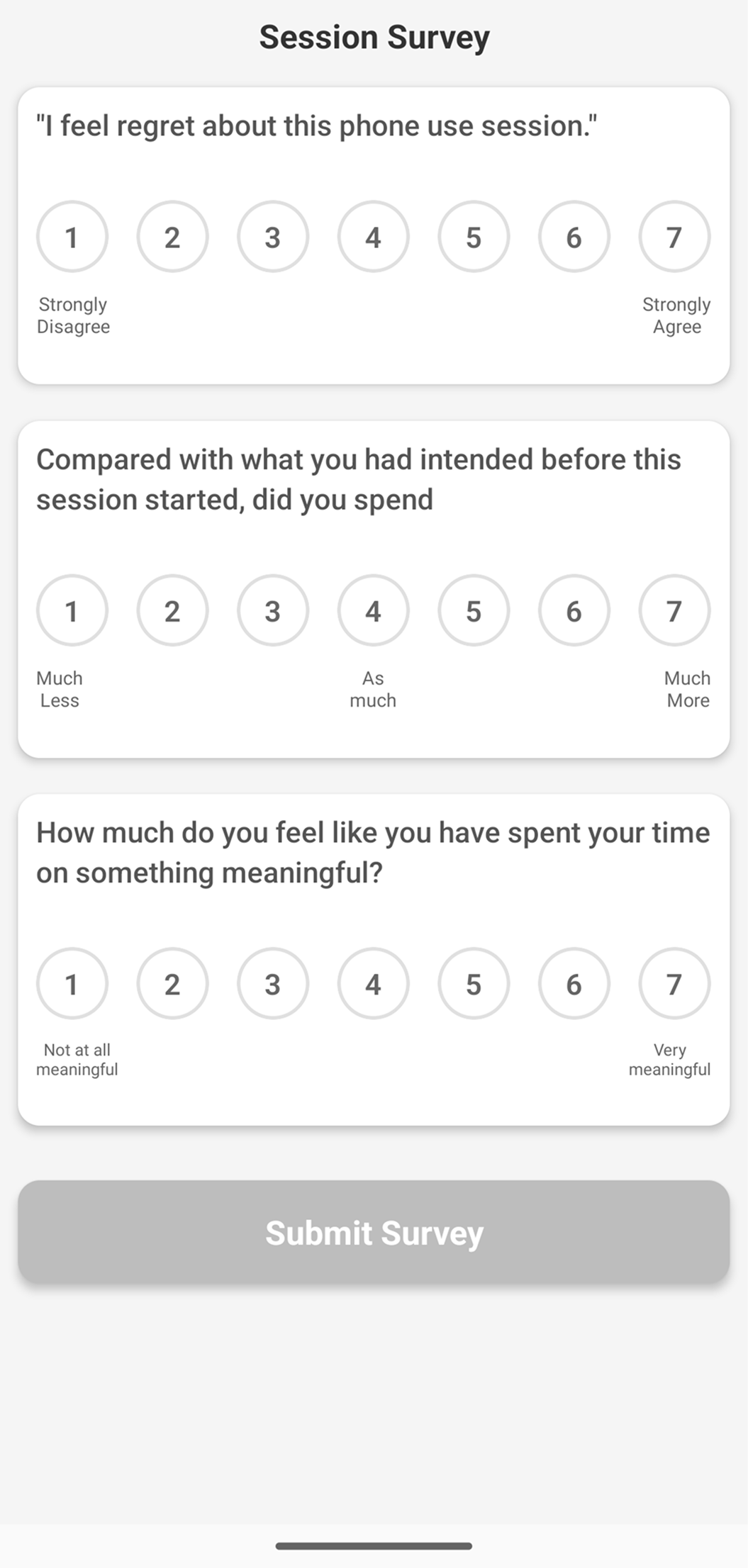}
    \small{(d)}
  \end{minipage}
  \caption{Screenshots of the custom Android observation application. (a) The main dashboard showing data collection status, sync state, and hourly collection schedule. (b) The notification panel showing the post-session prompt fired after the participant closed Instagram. (c) The notification banner appearing on the home screen (app icons blurred for participant privacy). (d) The post-session survey triggered by tapping the notification.}
  \label{fig:app-screens}
\end{figure}

A custom Android application ran in the background on participants' phones, syncing data to a database under an anonymous per-device identifier via two complementary channels. A \emph{batch} channel queried Android's \texttt{UsageStatsManager} API~\citep{android_usagestatsmanager} hourly via \texttt{WorkManager}~\citep{android_workmanager} to log app start/stop events, while a \emph{real-time} channel used an \texttt{AccessibilityService}~\citep{android_accessibilityservice} to detect exits from the 12 monitored applications (Appendix~\ref{sec:monitored-apps}) and fire a local notification deep-linking to the post-session survey, with transient overlays and system packages filtered and a per-app cooldown suppressing consecutive triggers. The service ran in the background and persisted logging across reboots, and implements local caching mechanisms if uploads failed due to internet connection failures. By construction, the app collected \textbf{no} in-app content: \texttt{UsageStatsManager} exposes \textbf{only} package-level metadata, and the \texttt{AccessibilityService} consumed \textbf{only} window-state transitions. Figure~\ref{fig:app-screens} shows screenshots of the application.

\subsection{Measures and Logged Data}\

\subsubsection{Baseline Measures}
\label{sec:baseline}

At the beginning of the study, participants completed a baseline questionnaire battery designed to capture mental health and social well-being prior to the in-the-wild phase. The survey included:
\begin{itemize}
    \item the Bergen Social Media Addiction Scale (BSMAS) to screen for problematic social media use \citep{andreassen_relationship_2016},
    \item the Patient Health Questionnaire (PHQ-9) to assess depressive symptoms \citep{kroenke_phq-9_2001},
    \item the Generalized Anxiety Disorder scale (GAD-7) to assess generalized anxiety \citep{spitzer_brief_2006},
    \item the UCLA Loneliness Scale to assess perceived social isolation \citep{russell_ucla_1996}, and
    \item the Brief Self-Control Scale (BSCS) to assess trait self-control \citep{tangney_high_2004}.
\end{itemize}

These validated measures provide a snapshot of participants' psychological functioning and social connectedness and were used to contextualize individual differences in patterns of social media use and associated emotional and physiological responses.

\subsubsection{Self-report Measures}

Across the study, we collected several types of self-report data:
\begin{itemize}
    \item \textbf{Post-session surveys}: regret\cite{guo_what_2025}, meaningfulness\cite{lukoff_what_2018}, and perceived time spent on the application for individual phone-use sessions. All were rated on a likert scale (1-7) and can be seen on Figure \ref{fig:app-screens}
    \item \textbf{End-of-day reflections}: overall daily regret, meaningfulness, and perceived intention-usage gap with respect to that day's phone use and single-item measures of perceived day-level stress, healthy eating, physical activity, restful sleep, and quality of social interactions. Full End-of-day reflections can be found in Appendix\ref{appendix:eod}
\end{itemize}
All in-the-wild items and most end-of-day items were assessed using Likert scales (1-7) to facilitate quantitative analysis; the EOD survey also included four open-ended questions (what typically started and ended social media sessions, cross-device usage, and usage contexts) analyzed qualitatively (see Section~\ref{sec:eod-qualitative}). We operationalized a session as the continuous interval between an app foreground event and the next app close or switch event.



\subsection{Empirical Approach}
\subsubsection{Qualitative Analysis}

\label{sec:eod-qualitative}
To complement our quantitative findings, we conducted semi-structured exit interviews with 16 participants who agreed to interviewing after completing the seven-day study. Interviews were conducted in person or via zoom. Each session began by presenting participants a personalized dashboard of their own usage statistics including regret ratings, session frequency and time-of-day patterns in order to ground the conversation in their lived experiences. We then asked participants to reflect on the sessions they found most and least regretful, their perceived triggers and stopping points, and their feelings toward existing screen time tools. Interviews were audio-recorded with participant consent and transcribed verbatim. Two researchers independently coded the transcripts using inductive thematic analysis following Braun and Clark \cite{braun_using_2006}\cite{braun_reflecting_2019}, with disagreements resolved through discussion. Full sample interview questions in Appendix \ref{appendix:interview}.

\subsubsection{Quantitative Analysis}
\label{sec:quantitative-analysis}


\paragraph{Data Preprocessing.}
We merged app-session metadata, smartwatch physiological recordings, in-phone contextual logs, experience-sampling responses, end-of-day surveys, and baseline questionnaires into a single session-level dataset. To safeguard outcome quality, we excluded all surveyed sessions for which the survey was completed more than 60 minutes after the session ended (42 sessions); a Spearman rank correlation between survey delay and regret was negligible ($\rho = -0.01$, $p = .733$), as was the correlation with quadratic regret ($\rho = -0.05$, $p = .073$), confirming no systematic delay-related drift toward the mean or extremes. We further restricted session data to the seven study days plus the on- and offboarding days, dropping 77 sessions whose offboarding had been rescheduled, and excluded three participants with fewer than 21 total sessions (38 sessions). The final dataset comprised 21 participants and 1{,}445 sessions. 

\paragraph{Feature construction.}
For each social media session, we extracted physiological features from the smartwatch data in two temporal windows: a \emph{pre-session} window (the measurement period immediately preceding app launch) and a \emph{during-session} window (measurements concurrent with app use). Within each window, we computed the mean and standard deviation of heart rate, filtered heart rate, skin temperature, and tri-axial accelerometer readings, as well as the fraction of time classified as active and the cumulative step count.

Contextual features captured in-phone behavioral patterns available before each session, including time of day, day of week, time since the last session, screen time in the preceding period, the number of distinct apps and app categories used, app-switching frequency (in 30- and 60-minute windows), and daily cumulative screen time. Rolling experience sampling aggregates were computed as the mean regret, meaningfulness, and time--intention gap over 30- and 60-minute windows preceding each session, as well as same-day and previous-day means and standard deviations. Baseline trait measures (age, social media addiction, PHQ-9, GAD-7, UCLA Loneliness, and Brief Self-Control Scale) were time-invariant and repeated for every session of a given participant.


\paragraph{Outcome operationalization.}
We analyzed session-level regret as both a continuous outcome (1--7 Likert scale) and a binary classification target. For the binary outcome, a session is positive if its regret score strictly exceeds that participant's own median (ties are negatives). This per-participant above-median rule anchors the positive class to each user's own distribution and isolates the construct an intervention would need to detect: when a user crosses into their personal above-typical-regret zone.

\paragraph{Quantitative analyses of associations with regret.}
To examine the determinants of self-reported regret (RQ1), we estimated Bayesian hierarchical linear regression models with session-level regret (1--7 scale) as the outcome and by-participant random intercepts. Effect sizes are reported as posterior means with 95\% highest density intervals (HDI). Full sampler configuration, prior specifications, and convergence diagnostics are reported in Appendix~\ref{sec:appendix-bayesian-spec}.

\paragraph{Feature selection.}
\label{sec:feature-selection-rq2}
For predicting regret (RQ2), we restricted our feature set to pre-session features only (35 variables), excluding during-session physiology, co-outcome ratings, and same-day aggregates that would be unavailable in a real-time intervention system. We partitioned the feature set into three a-priori subsets by data source, so that the ablation directly measures the contribution of each data channel. \textbf{Smartwatch} contains 19 pre-session smartwatch features (heart rate, filtered heart rate, temperature, tri-axial accelerometer, activity fraction, step counts). \textbf{Phone Context} contains 16 phone-derived pre-session features (temporal markers, session gaps, screen time, app-switch statistics). \textbf{Smartwatch+Phone Context} is the union of the two (35 features). None of the subsets include end-of-day survey ratings or baseline trait scales; predictions therefore depend only on passive pre-session sensing channels.
Within each pre-session feature subset, we applied recursive feature elimination with cross-validation (RFECV~\cite{sklearn_rfecv}) using a CatBoost estimator (3-fold CV, minimum 3 features, step size 1, \texttt{random\_state}=42), retaining the smallest subset whose CV score was statistically indistinguishable from the best. For within-person evaluation, RFECV was fit once on the pooled 70\% training set, yielding a single subset shared across individuals; for LOPO, it was refit on each fold's training participants so selection never saw the held-out individual. Retained sizes were 3 (Smartwatch), 8 (Phone Context), and 4 (Smartwatch+Phone Context); Table~\ref{tab:feature-importance} lists the features and their mean $|\text{SHAP}|$ contributions.

\begin{table*}[t]
\caption{Features retained by recursive feature elimination in each variant, with mean $|\text{SHAP}|$ measuring each feature's average absolute contribution to the model prediction in output space. All three variants use CatBoost (native) trained within-person on a 70/30 split. RFE retained 3 for Smartwatch, 8 for Phone Context, 4 for Smartwatch+Phone Context; blank rows indicate the variant retained fewer features.}
\label{tab:feature-importance}
\centering
\small
\begin{tabular}{rlr@{\hspace{1em}}rlr@{\hspace{1em}}rlr}
\toprule
\multicolumn{3}{c}{\textbf{Smartwatch}} & \multicolumn{3}{c}{\textbf{Phone Context}} & \multicolumn{3}{c}{\textbf{Smartwatch+Phone Context}} \\
\cmidrule(r){1-3}\cmidrule(lr){4-6}\cmidrule(l){7-9}
\# & Feature & $|\text{SHAP}|$ & \# & Feature & $|\text{SHAP}|$ & \# & Feature & $|\text{SHAP}|$ \\
\midrule
1 & Step count (prev day) & 0.366 & 1 & Screen time (prev day) & 0.251 & 1 & Screen time (prev day) & 0.327 \\
2 & Step count (day) & 0.281 & 2 & Screen time (day) & 0.220 & 2 & Step count (prev day) & 0.273 \\
3 & Acc-Z std (pre) & 0.183 & 3 & Study day & 0.187 & 3 & Day of week & 0.224 \\
 & &  & 4 & Hour of day & 0.180 & 4 & Screen time (day) & 0.153 \\
 & &  & 5 & Day of week & 0.131 &  & &  \\
 & &  & 6 & App switches (pre 60 min) & 0.130 &  & &  \\
 & &  & 7 & Mean gap (pre 60 min) & 0.113 &  & &  \\
 & &  & 8 & Time since last session & 0.096 &  & &  \\
\bottomrule
\end{tabular}
\end{table*}

\paragraph{Evaluation strategy.}
We evaluated models under two complementary protocols:
\begin{itemize}
    \item \textbf{Within-person evaluation} assessed how well regret can be predicted for a known individual. For RQ1, we used stratified 5-fold cross-validation within each participant (shuffled). For RQ2, we used a per-participant 70/30 random split, requiring a minimum of 13 sessions per participant to ensure adequate train and test sizes.
    \item \textbf{Between-person evaluation} assessed generalization to new, unseen individuals. We used Leave-One-Participant-Out (LOPO) cross-validation: in each fold, one participant's sessions were held out for testing while the model was trained on all remaining participants. This was repeated for all participants in the analysis (N=21 for context-only models, N=19 for physio-inclusive models), yielding per-participant performance estimates.
\end{itemize}

For classification, we report AUC-ROC against a naive baseline (each participant's mean regret).

\section{Results}
\label{sec:results}

\subsection{Descriptive Overview}
\label{sec:data-quality}

We collected 1,445 social media sessions with post-session experience sampling surveys from 21 participants over a 7-day field study. Session-level regret ratings (1--7 Likert scale) had a mean of 3.46 (SD~=~1.72), with the modal response at 4. The per-person above-median binarization yielded 495 regretful and 950 non-regretful sessions (approximately 34\% positive class). An intraclass correlation coefficient of ICC~=~.201 (95\% CI [.123,~.351]) indicated that approximately 20\% of regret variance was attributable to stable between-person differences, with 80\% reflecting within-person fluctuations across sessions, confirming that regret is predominantly a state-level phenomenon that varies across sessions rather than a fixed trait.

Per-participant session counts ranged from 22 to 245 (M~=~68.8), reflecting natural variation in social media engagement frequency (Figure~\ref{fig:data-overview}a). Figure~\ref{fig:regret-distribution} shows that participants exhibited markedly different regret distributions.

Through interview data from 16 participants, we identified four areas of exploration for RQ1: time blindness, intent drift, escape hatch, and why timers fail.

\subsection{Usage Patterns}
\label{sec:usage-patterns}
\begin{figure*}[h]
  \centering
  \includegraphics[width=\textwidth]{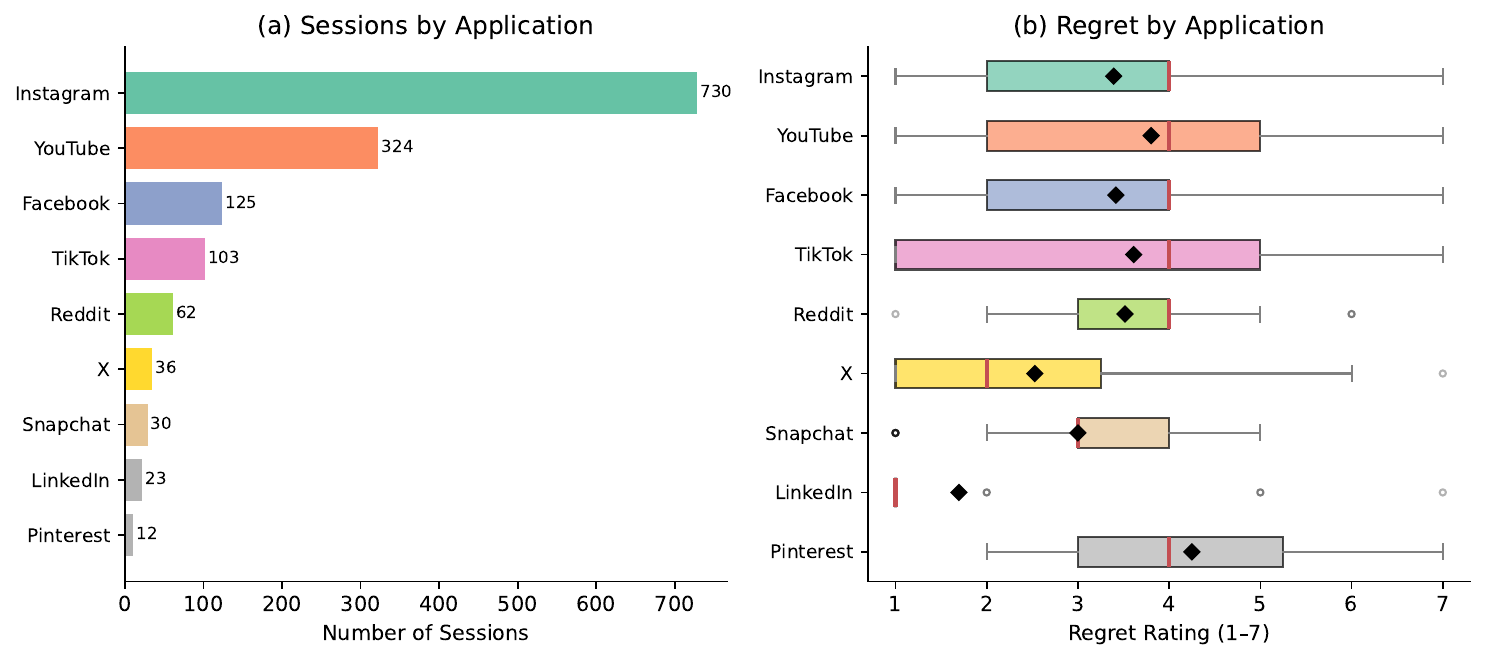}
\caption{Application usage patterns. (a) Number of sessions per application. Instagram, YouTube, and Facebook together accounted for 81.6\% of all sessions. (b) Regret distributions by application, with box plots showing medians (red lines) and means (black diamonds). TikTok and YouTube had the highest average regret.}
  \label{fig:app-usage}
  \Description{Left panel shows horizontal bar chart of sessions per app. Right panel shows box plots of regret by app.}
\end{figure*}

Instagram was the most frequently used application, accounting for 50.5\% of all sessions (730 of 1,445), followed by YouTube (22.4\%, 324 sessions) and Facebook (8.7\%, 125 sessions; Figure~\ref{fig:app-usage}a). The remaining apps (TikTok, Reddit, X, Snapchat, LinkedIn, and Pinterest) collectively accounted for 18.4\% of sessions. Regret varied by application (Figure~\ref{fig:app-usage}b): YouTube (\textit{M}~=~3.80, \textit{SD}~=~1.87) and TikTok (\textit{M}~=~3.61, \textit{SD}~=~2.08) elicited the highest mean regret, while LinkedIn (\textit{M}~=~1.70, \textit{SD}~=~1.64) and X (\textit{M}~=~2.53, \textit{SD}~=~1.70) had the lowest. This pattern is consistent with prior work suggesting that algorithmically driven, feed-based content consumption is more likely to produce regret than communication-oriented use~\cite{guo_what_2025}.

Social media use followed a bimodal temporal pattern (Figure~\ref{fig:app-usage}a), with peaks in the late evening and nighttime hours (9pm--2am) and a secondary peak in the late afternoon (3pm--7pm). Nighttime sessions (9pm--7am) accounted for 25\% of all sessions (360 of 1,445). Session durations were highly right-skewed (Figure~\ref{fig:app-usage}b): the median session lasted 1.4 minutes, while the mean was 6.1 minutes (\textit{SD}~=~13.4 min), indicating that most sessions were brief but a long tail of extended sessions pulled the average upward. The interquartile range spanned 19 seconds to 5.2 minutes, with the longest session reaching 149 minutes. Weekday sessions (70.7\%) outnumbered weekend sessions (29.3\%), roughly proportional to the 5:2 weekday-to-weekend ratio.

Participants rated most sessions as having low to moderate meaningfulness (\textit{M}~=~3.15, \textit{SD}~=~1.74; Figure~\ref{fig:outcome-distributions}a) and tended to perceive that they spent more time than intended (\textit{M~}=~4.19, \textit{SD}~=~1.37; Figure~\ref{fig:outcome-distributions}b), with 81.3\% of sessions rated at 4 or above on the time comparison scale. A pearson correlation between regret and meaningfulness showed a weak negative association (\textit{r}~=~-.18, \textit{p} < .001; Figure~\ref{fig:outcome-distributions}c, Table~\ref{tab:correlations}), suggesting that while meaningful sessions tend to be less regretful, the two constructs capture distinct aspects of the user experience. Importantly, the intention-usage gap (perceived mismatch between the intended time to spend and actual duration) showed a moderate correlation with regret (\textit{r} = .48, \textit{p} < .001) while app session duration itself was weakly associated (\textit{r} = .20, \textit{p} < .001).

\begin{table}
\centering
\small
\caption{Pairwise Pearson correlations between session-level variables.}
\label{tab:correlations}
\begin{threeparttable}
\begin{tabular}{lcccc}
\toprule
 & (1) & (2) & (3) & (4) \\
\midrule
(1) Regret            & ---           &                &              &  \\
(2) Meaningfulness    & $-0.18^{***}$ & ---            &              &  \\
(3) Intention-usage gap& $0.48^{***}$  & $-0.05^{*}$   & ---          &  \\
(4) Duration (min)    & $0.20^{***}$  & $-0.14^{***}$ & $0.27^{***}$ & --- \\
\bottomrule
\end{tabular}
\begin{tablenotes}
  \footnotesize
  \item \textit{Note.} $^{*}p < .05$; $^{***}p < .001$.
\end{tablenotes}
\end{threeparttable}
\end{table}



\subsection{RQ1: What factors explain regretful social media use}
\label{sec:rq1}

\subsubsection{Social Media Apps Let You Lose Sense of Time}

\begin{figure}[H]
    \centering
    \includegraphics[width=0.6\textwidth]{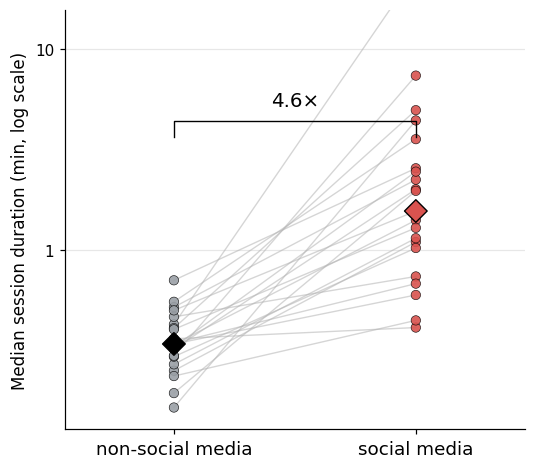}
    \caption{Median session duration for social media versus non-social media app sessions, computed per participant and aggregated across the sample. Social media sessions were substantially longer than sessions on other apps.}
    \label{fig:sm-vs-nonsm-duration}
\end{figure}

Given that our study centers on regret in social-media sessions specifically, we first examined whether these sessions are structurally distinct from other phone use. Through our analysis, we found a distinction between the length of a social-media session versus non social-media session. Across all 21 participants, social-media sessions were substantially longer than non-social-media sessions in every individual case Figure \ref{fig:sm-vs-nonsm-duration}, with a grand median roughly four times larger(Mann-Whitney $U = 1.23 \times 10^{8}$, $p < .001$).We find that the median value for non-social media sessions is 20.46s while the median value for a social media session is 94.02s. 
 
\subsubsection*{Regret is associated with an intention-usage gap and less with duration alone}

Although prior work often considers usage duration (e.g., screen time) as a proxy for problematic phone use when designing interventions \cite{hiniker2016mytime, ko2015nugu}, a more precise signal is the mismatch between explicit intentions and actual time spent, which we term the \textit{intention-usage gap}.
To disentangle whether duration itself or the intention-usage gap is more strongly associated with experiencing regret after a social media session, we ran three separate Bayesian hierarchical linear regression model variants (V1, V2, V3): V1 with session duration as a single predictor and V2 with subjective intention-usage gap as single predictor (derived from the EMA survey question "Compared with what you had intended before this session started, did you spend..." with (1) much less to (7) much more). Lastly, V3 included both predictors. All models had session-level regret as the dependent variable and added random intercepts per participant to account for the nested structure of our data (sessions nested within participants).

\begin{figure}[H]
    \centering
    \includegraphics[width=1\linewidth]{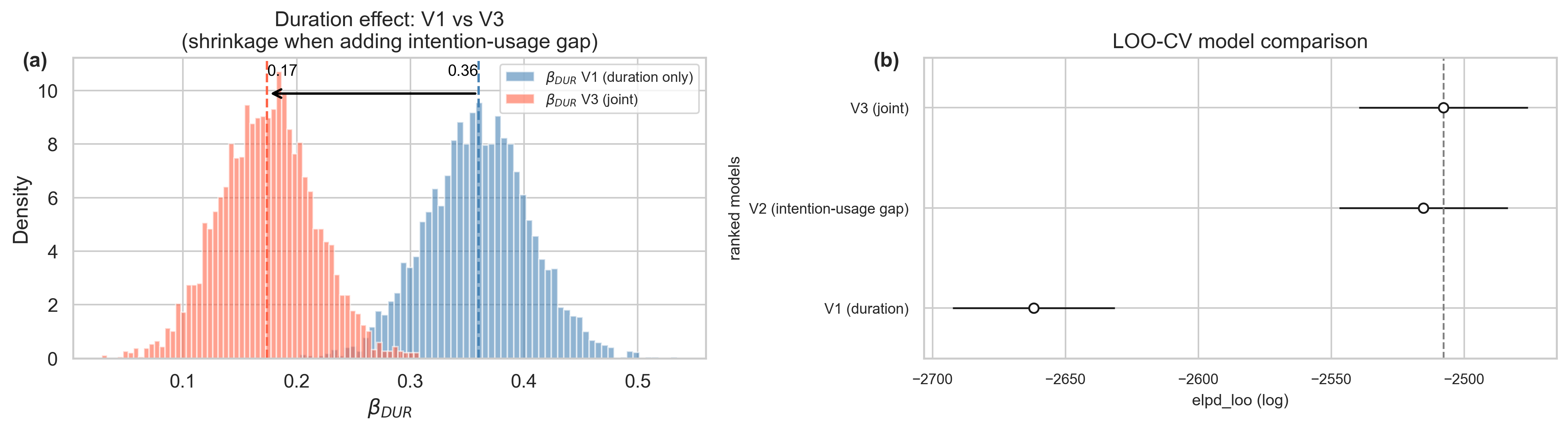}
    \caption{Perceived intention-usage gap explains regret better than session duration. (a) The effect of duration ($\beta_{\text{DUR}}$) reduces when taking perceived intention-usage gap into account. (b) Model comparison. Lower values (towards the right of the x-axis) indicate better model fit. Perceived intention-usage gap explains the data better than duration (V2 vs. V1). A model containing both (V3) does not explain the data much better than perceived intention-usage gap alone (V2).}
    \label{fig:duration_vs_overuse}
\end{figure}

As shown in Table~\ref{tab:duration-overuse}, perceived intention-usage gap is strongly associated with regret ($\beta_{\text{IUG}} = 0.76$, 95\% HDI = [0.686, 0.833]) while duration is only moderately associated ($\beta_{\text{DUR}} = 0.36$, 95\% HDI = [0.270, 0.44]). Importantly, when regressing regret onto both predictors inside the same model (V3), the estimate for the effect of duration on regret shrinks considerably to $\beta_{\text{DUR}} = 0.18$ (95\% HDI = [0.092, 0.260]) while intention-usage gap still remains strongly associated with regret ($\beta_{\text{IUG}} = 0.72$, 95\% HDI = [0.651, 0.796] (see Figure ~\ref{fig:duration_vs_overuse} A). We then compared the models directly regarding their relative fit to the data using leave-one-out cross-validation (LOO; see Figure \ref{fig:duration_vs_overuse} B). While V3 (ELPD\,$=-2508$) and V2 (ELPD\,$=-2515$) explain the data almost equally well, V1 performs worse (ELPD\,$=-2662$), suggesting that duration is not only a weaker predictor for regret compared to the intention-usage gap, but that large parts of the association between duration and regret is "explained away" when taking intention-usage gap into account. 
 
Together, these results suggest that regret does not necessarily increase as a function of spending longer on a social media app, but when usage time contrasts with more explicit intentions.

\begin{table}[htbp]
\centering
\caption{Bayesian hierarchical regression variants comparing duration and perceived intention-usage gap as predictors of session-level regret.}
\label{tab:duration-overuse}
\small
\begin{tabular}{l l c c c}
\toprule
\textbf{Variant} & \textbf{Predictor} & $\boldsymbol{\beta}$ & \textbf{95\% HDI} & \textbf{ELPD (LOO)} \\
\midrule
V1 & duration (DUR)  & 0.360 & $[0.272,\,0.443]$ & $-2661.935$ \\
V2 & intention-usage gap  (IUG)& 0.760 & $[0.688,\,0.830]$ & $-2515.316$ \\
\addlinespace
\multirow{2}{*}{V3} & duration (DUR)  & 0.174 & $[0.096,\,0.251]$ & \multirow{2}{*}{$-2507.814$}\\
   & intention-usage gap  (IUG)& 0.722 & $[0.651,\,0.796]$ & \\
\bottomrule
\end{tabular}
\end{table}
 
\subsubsection*{Salient Opportunity Costs Drive Regret}
 
Prior literature has shown that regret is driven by a post-event comparison between a reward obtained and the opportunity costs of salient options~\cite{zeelenberg_anticipated_1999}. Hence, regret should be higher whenever alternative higher-order goals (e.g., more valuable in the long run) are particularly salient, such as when being aware that one should sleep or work instead. We directly explored these hypotheses, again using Bayesian hierarchical linear regressions. 

\paragraph{Night vs.\ day.}
First, we tested whether regret was higher during night hours compared to during the day. For this, we regressed regret onto a binary variable encoding whether the session happened at night (between 12am and 9:59am) or during the day (between 10am and 11:59pm). To account for the confound of night sessions being longer than day sessions ($Median_{\text{night}} = 2.73$\,min vs. $Median_{\text{day}} = 1.30$\,min; $U = 157028.5$, $p < .001$), we also included session-level duration as a predictor. We included random intercepts and by-participant random slopes for the focal predictor (night vs.\ day), allowing each individual's effect to deviate from the population mean. 
 
\begin{figure}[h!]
     \centering
     \includegraphics[width=1\linewidth]{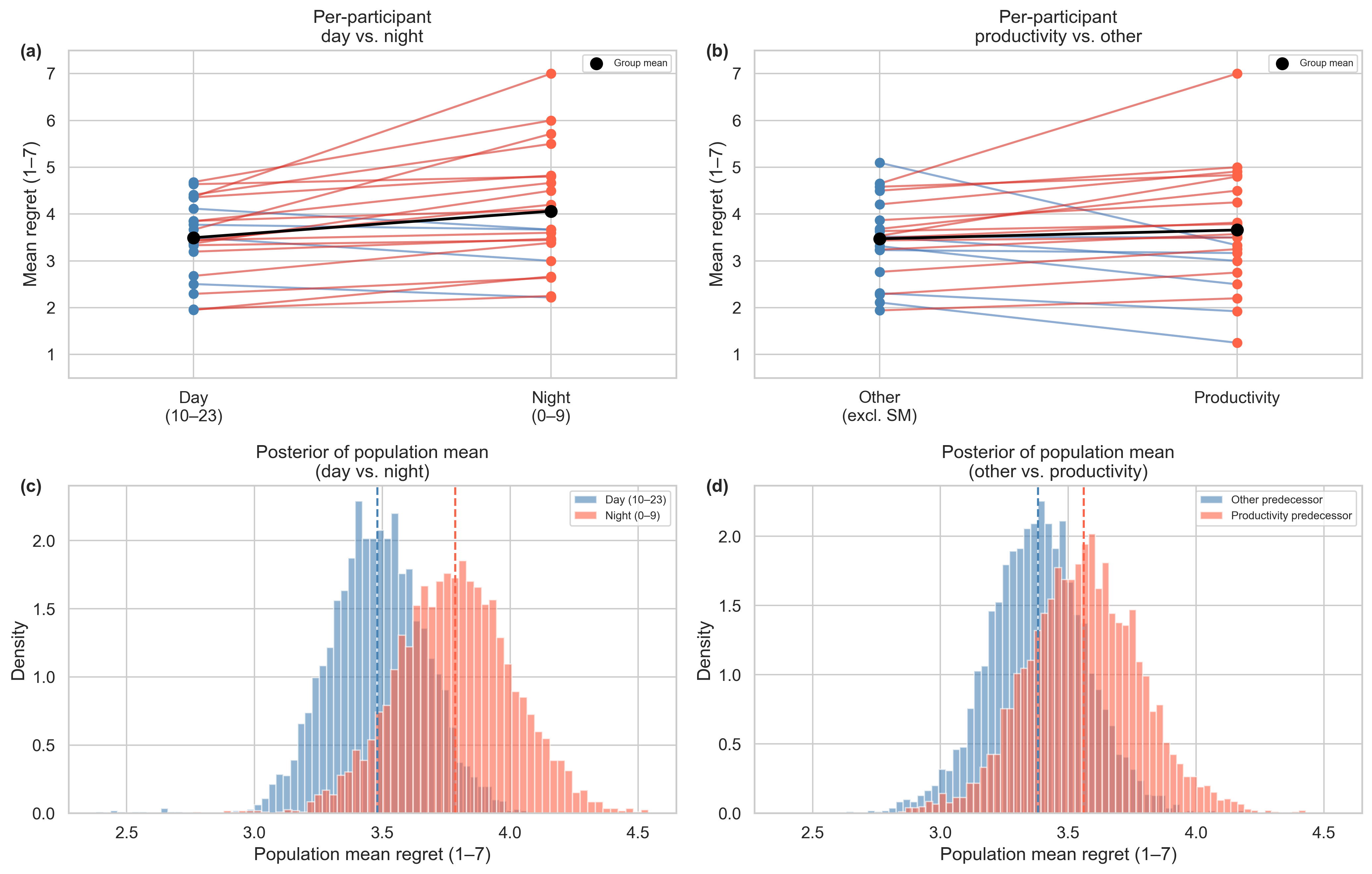}
     \caption{Regret is higher when opportunity costs are salient: Day vs. night (a, c) and productivity drift (b, d). (a) Difference in mean regret for day vs. night sessions per participant. (b) Difference in mean regret of a session when being preceded by a productivity app session vs. other app session. Blue and red dots indicate participant mean regret scores for Day / Other app category and Night / Productivity app category, respectively. Lines connect means of individual participants. (c) Difference in posterior population mean regret density for night vs. day sessions. (d) Difference in posterior population mean regret density for preceding sessions being productivity-related vs. other. Note that plots (c) and (d) show marginal effects when controlling for duration in the day vs. night comparison (c); and duration and day vs night for the app category comparison (d).}
     \label{fig:regret_night_productivity}
 \end{figure}

As shown in Figure  \ref{fig:regret_night_productivity} (a, c), regret was higher following social media sessions during night hours ($\beta = 0.304$, 95\% HDI\textit{ =} [0.029, 0.578]), suggesting that regret is higher when social media usage competes with sleep, even after accounting for the association between duration and regret ($\beta = 0.353$, 95\% HDI = [0.263, 0.44]).


\paragraph{Preceding app category.}
Second, we hypothesized that regret would be particularly high when the social media session interrupts productivity-related activities. For this, we categorized apps used right before a social media session into productivity-related (e.g., Obsidian, Outlook) vs.\ others and tested whether regret toward a social media app session was higher when directly preceded by usage of a productivity-related application. Note that we excluded social media apps from this analysis to exclude SM\,$\rightarrow$\,SM transitions (see \ref{tab:app_categories} for the detailed app categorization mapping).
We regressed regret onto the same predictor set as above (night vs.\ day, duration) but added app type (productivity\,=\,1, other\,=\,0) as a binary predictor and included random intercepts and random slopes for the app category predictor per participant as above.

As shown in Figure \ref{fig:regret_night_productivity} (b, d), the mean posterior estimate of the marginal effect of app type on regret shows a trend toward SM sessions preceded by productivity apps being regretted more ($\beta = 0.180$, 95\% HDI = [-0.094, 0.424]), although this effect should be treated with caution as the 95\% HDI is wide and still includes zero.

\subsection{RQ2: Predicting Regret from Pre-Session Features}
\label{sec:rq2-results}

We ask whether session-level regret can be predicted \emph{before} a session begins, using only features available prior to app launch. We defined three pre-session feature subsets a priori by data source, so that the ablation directly measures the contribution of each passive sensing channel: \textbf{Smartwatch} (19 features: heart rate, filtered heart rate, skin temperature, tri-axial accelerometer, activity fraction, step counts), \textbf{Phone Context} (16 features: temporal markers, session gaps, screen time, app-switch statistics), and \textbf{Smartwatch+Phone Context} (their union, 35 features). None of the subsets include self-report features, end-of-day surveys, or baseline trait scales. We trained CatBoost (native null handling) on each subset under both within-person and between-person protocols. Feature-level contributions for each variant are reported in Table~\ref{tab:feature-importance}.

Our goal is to flag when a session will be problematic for a user, meaning more regretful than usual for them, rather than to predict their 1-to-7 score. An absolute threshold would over-fire for heavy users and under-fire for light users whose regret only spikes against their own baseline. We therefore report per-participant AUC against each participant's own median, following the within-person template established for passive sensing of affect \cite{taylor2017personalized}.

\subsubsection{Classification Performance}

For within-person binary classification on the strongest variant, Smartwatch+Phone Context (per-participant above-median regret), all variations exceeded chance (.50), with the best configuration reaching AUC\,=\,.725. The curves cluster tightly, indicating that variations had a modest effect relative to the inherent difficulty of predicting regret from pre-session signals alone. The AUC values suggest that while pre-session features carry meaningful predictive signal, a substantial share of the variance in regret remains determined by in-session experiences not observable beforehand.


\paragraph{Global vs.\ per-participant AUC.}

Two AUC metrics appear in this paper and yield systematically different values: \emph{global AUC} pools all observations across participants before computing the ROC curve, while \emph{per-participant AUC} computes a separate ROC curve for each individual and then averages the resulting AUC values, weighting each participant equally. Because session counts are highly skewed (one participant contributed 17\% of all sessions), global AUC is dominated by high-volume individuals and consistently exceeds per-participant mean AUC across all variants. All primary results in Section~\ref{sec:rq2-ablation} use per-participant AUC, which treats each person equally and is the more conservative and interpretable metric for within-person prediction.

\subsubsection{Answering RQ2: Does Smartwatch Data Improve Regret Prediction?}
\label{sec:rq2-ablation}

To evaluate whether smartwatch-derived signals contribute to the prediction of self-reported scrolling regret, we compared three feature subsets: Smartwatch, Phone Context, and Smartwatch+Phone Context. Each subset was evaluated using CatBoost with native null handling under both within-person shuffled 70/30 splits and leave-one-participant-out (LOPO) cross-validation for binary classification (AUC). CatBoost was chosen due to it's native null handling instead of imputing which performs better with handling missing data.


\paragraph{Evaluation protocols for classification.} We report two complementary views of classification performance (Table~\ref{tab:rq2-methods-auc}). \textbf{Leave-one-participant-out (LOPO) AUC} is the primary deployability metric: it measures cold-start generalization to an unseen individual, the realistic scenario when a new user installs the app with no prior data. \textbf{Within-person AUC} (70/30 shuffled split) represents the personalized ceiling: performance achievable after collecting a participant's own data, corresponding to the scenario after an initial onboarding period. The gap between the two quantifies the cost of individual differences, and the count of participants whose individual within-person AUC reaches $.70$ (the Hosmer--Lemeshow ``acceptable discrimination'' threshold~\cite{hosmer2000applied}) provides a practical deployability bar. We focus on classification because the intervention target is binary detection of above-median-regret sessions.

\paragraph{Adding the smartwatch shifts who the model works for, not how well it works on average.}

On the full 21-participant sample, Phone Context reaches LOPO AUC $.536\scriptsize{\pm}.097$ and personalized-ceiling within-person AUC $.732\scriptsize{\pm}.017$ (mean $\pm$ SD across 50 random within-person seeds; Table~\ref{tab:rq2-methods-auc}). On the 19-participant smartwatch-eligible subset, Smartwatch+Phone Context reaches LOPO $.555\scriptsize{\pm}.122$ and within-person $.740\scriptsize{\pm}.020$. The cold-start lift over Phone Context is $+.019$ LOPO and the personalized-ceiling lift is $+.008$ within-person, both well inside one seed-level SD. By pooled AUC, adding the smartwatch is a wash.

Deployability tells a different story. Defining a participant as deployable when their mean within-person AUC across seeds is $\ge .70$, both variants qualify five participants on the shared 19-person subset, but the qualifying sets differ. P44 is deployable only under Phone Context (.760 with phone alone, .695 with the smartwatch added). P70 is deployable only under Smartwatch+Phone Context (.677 with phone alone, .791 with the smartwatch added). Adding physiology therefore shifts coverage rather than expanding it.

Two caveats. First, because the smartwatch-eligible subset is smaller and may differ compositionally from the full sample, the pooled-AUC comparison is not strictly apples-to-apples; the deployability comparison on the shared 19-person subset is more interpretable. Second, the top smartwatch predictors (skin temperature, accelerometer variability, filtered heart rate) are signals no phone-side channel can capture, and they are involuntary, continuous, and cheap to collect, but we have no independent autonomic measure to confirm whether they index autonomic state in this sample. The smartwatch's contribution here is best framed descriptively, as a coverage shift at the individual level rather than a pooled-AUC gain for the typical user.

\begin{table}[t]
\caption{Classification AUC across feature-set variants (above-median binarization). \textit{Top:} aggregate metrics. $n$: participants; WP AUC: within-person 70/30 shuffled split, reported as mean $\pm$ SD across 50 random seeds (the seed=42 single-shot underestimated all three variants and inflated the between-variant gap); LOPO AUC: leave-one-participant-out (mean $\pm$ SD across folds); $n_{\text{WP}}\ge.70$: participants whose \emph{mean} within-person AUC across seeds reaches the Hosmer--Lemeshow acceptable-discrimination threshold (denominators are 21 for Phone Context and 19 for smartwatch-based variants, which exclude 2 participants with 0\% smartwatch coverage). \textit{Bottom:} mean $\pm$ SD within-person AUC across seeds, for participants whose mean reaches $\ge.70$ under any variant. Bold AUC entries have mean $\ge.70$.}
\label{tab:rq2-methods-auc}
\centering
\small
\begin{tabular}{lccc}
\toprule
 & \textbf{Smartwatch} & \textbf{Phone Context} & \textbf{Smartwatch + Phone Context} \\
\midrule
\multicolumn{4}{l}{\textit{Aggregate}} \\
Participants ($n$) & 19 & 21 & 19 \\
LOPO AUC & \textbf{.565\scriptsize{$\pm$.116}} & .536\scriptsize{$\pm$.097} & .555\scriptsize{$\pm$.122} \\
WP AUC & .701\scriptsize{$\pm$.019} & .732\scriptsize{$\pm$.017} & \textbf{.740\scriptsize{$\pm$.020}} \\
$n_{\text{WP}}\ge.70$ & 5/19 & 5/21 & 5/19 \\
\midrule
\multicolumn{4}{l}{\textit{Participants near deployability threshold (within-person AUC, mean $\pm$ SD across seeds)}} \\
P72 & \textbf{.927\scriptsize{$\pm$.048}} & \textbf{.957\scriptsize{$\pm$.041}} & \textbf{.960\scriptsize{$\pm$.035}} \\
P66 & \textbf{.776\scriptsize{$\pm$.076}} & \textbf{.871\scriptsize{$\pm$.057}} & \textbf{.882\scriptsize{$\pm$.063}} \\
P70 & \textbf{.729\scriptsize{$\pm$.130}} & .677\scriptsize{$\pm$.129} & \textbf{.791\scriptsize{$\pm$.116}} \\
P50 & \textbf{.708\scriptsize{$\pm$.112}} & \textbf{.721\scriptsize{$\pm$.081}} & \textbf{.757\scriptsize{$\pm$.087}} \\
P71 & .655\scriptsize{$\pm$.160} & \textbf{.736\scriptsize{$\pm$.187}} & \textbf{.752\scriptsize{$\pm$.170}} \\
P44 & .600\scriptsize{$\pm$.155} & \textbf{.760\scriptsize{$\pm$.143}} & .695\scriptsize{$\pm$.126} \\
P54 & \textbf{.732\scriptsize{$\pm$.154}} & .605\scriptsize{$\pm$.213} & .635\scriptsize{$\pm$.247} \\
\bottomrule
\end{tabular}
\end{table}

\paragraph{Between-person evaluation (LOPO).}
Leave-one-participant-out cross-validation tested whether models generalize to \emph{unseen} participants, the critical test for a deployable system (Table~\ref{tab:rq2-methods-auc}). Smartwatch alone reached LOPO AUC\,=\,.565 (CatBoost native, $n=19$), Phone Context on all 21 participants reached .536, and Smartwatch+Phone Context on the 19-participant smartwatch-eligible subset reached .555. Smartwatch+Phone Context shows a small mean lift over Phone Context alone ($+.019$), but the SDs ($.097$--$.122$) are large relative to the between-variant gaps, so the three variants overlap substantially at the cold-start level.

\paragraph{Within-person evaluation (personalized ceiling).}
Within-person evaluation, with shuffled 70/30 splits repeated across 50 random seeds, gave mean AUCs of $.667$ on Smartwatch ($n=19$), $.732\!\pm\!.017$ on Phone Context ($n=21$), and $.740\!\pm\!.020$ on Smartwatch+Phone Context ($n=19$). All three subsets exceeded their LOPO counterparts by approximately $.10$--$.17$, the personalization premium from knowing an individual's history. The $+.008$ within-person lift from adding smartwatch features sits inside one seed-level SD, so the typical-user pooled benefit of the combined channel over phone context alone is small.

To interpret which features drive predictions, we computed SHAP values~\cite{lundberg_unified_2017} for CatBoost (native) in each variant (Table~\ref{tab:feature-importance}). Three patterns emerge. For \textbf{Phone Context}, the top features are day-level behavioral rhythm signals (previous-day screen time .251, current-day screen time .220, study day .187, hour of day .180) rather than immediate pre-session measurements, suggesting that regret is tied to habitual daily patterns more than to what happened in the moments before a session. For \textbf{Smartwatch}, recursive feature elimination retained only three features: previous-day step count (.366), current-day step count (.281), and pre-session Acc-Z standard deviation (.183); cumulative activity dominates over autonomic signals, and the absence of heart rate or skin temperature among the survivors helps explain why Smartwatch alone underperforms Phone Context. In the \textbf{Smartwatch+Phone Context} combined model, the top four features mix both channels (previous-day screen time .327, previous-day step count .273, day of week .224, current-day screen time .153), confirming that the smartwatch contribution is additive rather than dominant.

\paragraph{Who the model works for, and why}
Table~\ref{tab:rq2-methods-auc} breaks out deployability by participant after averaging within-person AUC across 50 seeds. Five participants reach mean AUC\,$\ge$\,.70 under Phone Context (P44, P50, P66, P71, P72), reflecting contextual regularity in their session timing, screen time, and app-switching. Under Smartwatch+Phone Context the deployable set shifts to \{P50, P66, P70, P71, P72\}: P44 drops below the threshold (.760\,$\to$\,.695) while P70 crosses it (.677\,$\to$\,.791), the largest smartwatch-driven gain in the table (+.114). Three participants (P50, P66, P72) are deployable under all variants, indicating that their regret tracks both contextual and autonomic regularity.

Per-participant AUCs should be read alongside test-set size: small-N participants like P54 (n$_{\text{test}}$\,=\,7) have multi-seed SDs of $.15$--$.25$, so any single-seed extreme is partly a grain artefact. P50 (n\,=\,74), P72 (41), and P66 (33) yield the most defensible estimates, and P70's +.114 gain on n$_{\text{test}}$\,=\,15 is the strongest ``smartwatch helps this person'' data point. Deployability thus depends on (a) within-person regret variability, (b) whether that variability is contextual or autonomic, and (c) whether the participant wore the watch.

Phone context provides a usable cold-start baseline (LOPO AUC\,=\,.536); the smartwatch adds a small cold-start lift ($+.019$, within LOPO SDs) and shifts which participants are covered post-personalization. A brief onboarding period closes the $\approx.10$--$.17$ gap between cold-start and personalized performance.


Phone-derived context provides a usable cold-start baseline (LOPO AUC\,=\,.536); physiology adds a small cold-start lift (LOPO $+.019$, within the LOPO SDs) and additionally expands and shifts which participants are covered post-personalization. A brief onboarding period closes the $\approx.10$--$.17$ gap between cold-start and personalized performance.

\section{Qualitative Findings: How Users Experience Regretful Sessions}
\label{sec:qualitative-findings}

Two researchers independently coded the transcripts using inductive thematic analysis following Braun and Clark \cite{braun2006using}\cite{braun_reflecting_2019}, with disagreements resolved through discussion. This process yielded four major themes, which we describe in the following subsections: Time Blindness, Intent Drift, and Why do timers fail. Quotebook with themes in codes in Appendix \ref{appendix:quotebook}.

\subsection{Time Blindness}

A pattern we observed was time blindness, a recurring collapse of temporal self-monitoring during scrolling. This theme emerged independently across participants and was among the most consistently reported experiences in the dataset.

Participants described two manifestations of this phenomenon. The first was a failure to act on existing awareness: participants recognized in the moment that they were exceeding their intentions but were unable to interrupt the session.
\begin{quote}
    \textit{I had that awareness, but I didn't do anything about it,}" ( P14)
\end{quote}
P4 reported scrolling despite an active deadline: \textit{"I have a deadline… and I just kept scrolling.}" P2 described the pattern as a daily recurrence, noting an explicit goal of not losing control of her time to technology while describing doing exactly that each morning.

The second manifestation was the retrospective discovery of the elapsed time: participants were unaware of the duration during the session and only registered the loss afterward. P13 described opening her phone to relax before sleep, only to find that "\textit{suddenly two hours has passed by}. P18 articulated the underlying mechanism: "\textit{You kind of lose your reference on what is happening around you... you are just stuck in an endless cycle, just watching stuff over and over again without being aware of what is happening around you.}"

\subsection{Intent Drift }
Participants find themselves drifting from their original intentions when they are on their phones. We refer to this pattern as \textit{intent drift}, which manifests at two levels: at the transition point between tasks and at the point of app entry. Participants described using social media in the context of other work and being pulled away from the task at hand(P9, P12). These accounts are consistent with our quantitative finding that productivity-adjacent context switches (sessions preceded by productivity app use) were associated with elevated mean regret score.

At the within-app level, participants described platform defaults as actively redirecting them away from their original intent upon entry. Participants had a pattern of entering an app with a specific intent and then getting sucked into recommended content (P9, P18). 
\begin{quote}
    \textit{I meant to look up something important, [but] the first thing the app opens to is reels and shorts… I'm not here for this, but I ended up getting caught in it." (P9)}
\end{quote}

Participants were not passive in these accounts,  they demonstrated clear awareness of the drift dynamic and articulated a desire for finer-grained intentional control. P4 described wanting to disable the Reels feature of Instagram entirely, framing it as damaging to attention. P17 expressed this most explicitly:
\begin{quote}
   \textit{ "I wish we could just take control of our algorithms… I would love to be able to just go on to my app and say, 'I want to see what my friends' updates are,' or 'I want to see only content that's creative right now.' But it seems to come to you in waves."}
\end{quote}
\subsection{Escape Hatch}
Participants most frequently attributed regret to sessions initiated not out of genuine interest but as a means of deferring an unwanted obligation, with social media serving as a low-effort escape from tasks that felt cognitively costly (P3, P4, P5, P9). Participants also used social media as an emotional escape (P8, P17)..
\begin{quote}
\textit{   " I feel regretful of social media usage when I know that I'm avoiding a task at work, that's usually the context in which I feel the most regret" (P5).}
\end{quote}
\begin{quote}
    \textit{"I use social media as an escape…. If I have problems and I’m escaping them, or if I’m using social media to pretend I have problems to escape. So it’s like a cycle.”}(P17)
\end{quote}
What made these sessions distinct was that participants were aware they were avoiding in the moment: regret was not a retrospective realization but a feeling present from the start, suggesting avoidance-driven use may be a more tractable target for just-in-time intervention than habitual or passive use.

\subsection{Why timers fail}
\begin{figure}[H]
     \centering
    \includegraphics[width=1\linewidth]{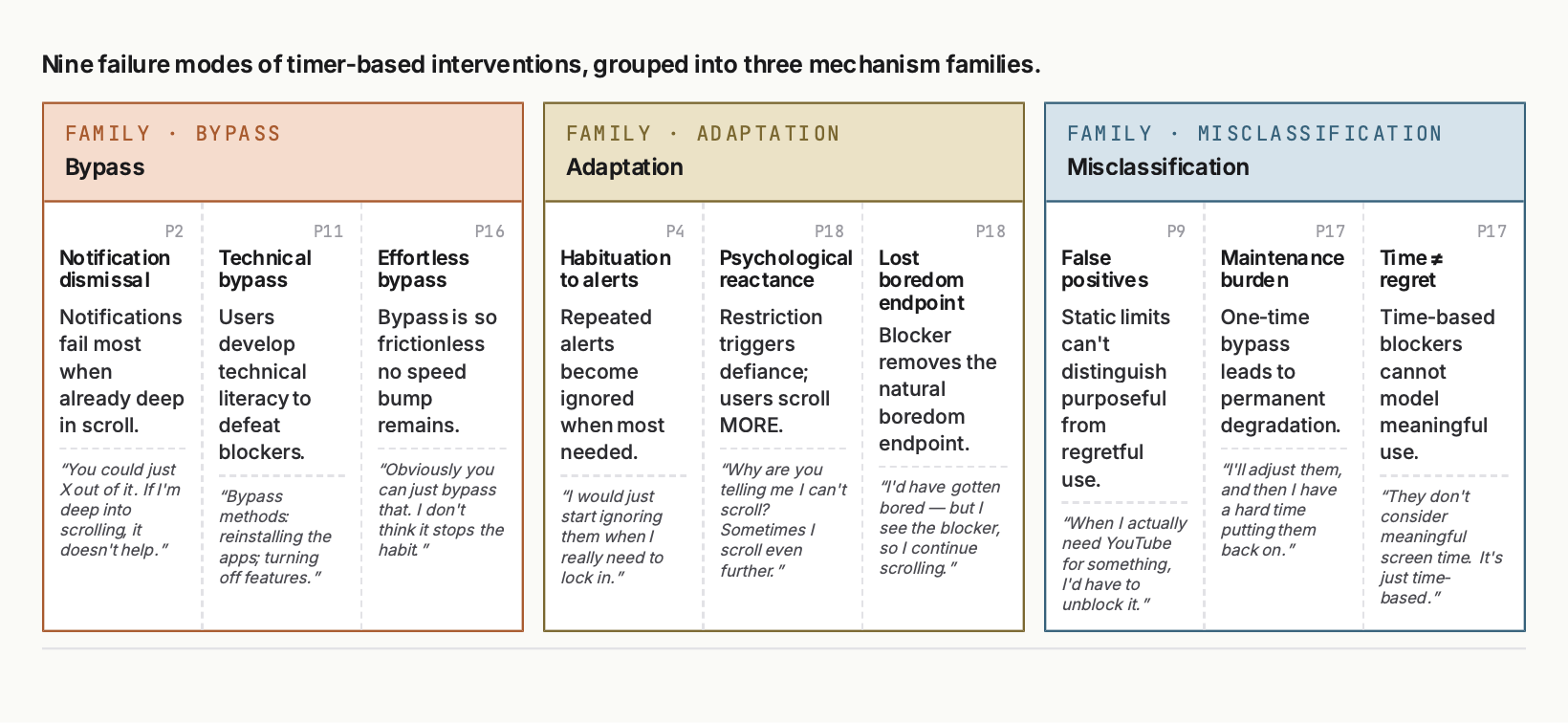}
     \caption{Across N=16 interviews, participants described nine ways timer-based interventions fail. They cluster into three families: bypass (the lock is circumvented), adaptation (users habituate to or defy the intervention), and misclassification (time-based rules cannot read intent). Quotes paraphrased for length; full text in body}
     \label{fig:timer-failures}
 \end{figure}
Participants were candid about why existing screen time interventions had failed them, identifying three recurring failure modes: ease of bypass, habituation and reactance, and misclassification of use.

\begin{quote}
  \textit{  "The blockers don't… they don't consider meaningful screen time or not. It's just time-based." }(P17).
\end{quote}

P17's critique cuts to the core limitation: current tools treat all screen time as equivalent, making no distinction between a session that felt wasteful and one that felt necessary or meaningful. This misclassification problem was compounded by the ease with which limits could be dismissed, with participants describing overriding timers without friction, and by a reactance effect in which the presence of a blocker paradoxically increased the salience of the blocked app. Together, these failure modes suggest that the problem is not one of exposure duration but of session quality, and that effective intervention requires the ability to distinguish regretful use from intentional use before it occurs.

%


\section{Discussion}
\label{sec:discussion}
Prior work on physiological prediction in social media contexts has been confined to the lab, using research-grade sensors and curated content\cite{gebhardt_detecting_2024}. Our 7-day in-the-wild study extends this line of work in two ways: participants engaged with their own feeds in everyday environments, and physiological signals came from a consumer-grade smartwatch (\$80 Bangle.js 2). Our outcome, session-level regret, is an evaluative judgment about how time was spent and is more directly actionable for just-in-time interventions than momentary affect. We organize the discussion around four points: why existing tools fail, the intention-usage gap as a better outcome than duration, opportunity-cost salience as its amplifier, and what our prediction results imply for a deployable sensing-based JITAI.


\subsection{Existing tools fail in predictable ways}
\label{sec:tools-fail}

Prior work has shown that people tend to bypass their interventions but little work asks participants why and when \cite{roffarello_achieving_2023, lyngs_hack_2020}. Early evidence has also shown that satisfaction with one's phone use is not a simple function of duration \cite{hiniker2016mytime} yet duration remains the dominant metric used by mainstream interventions\citep{roffarello_achieving_2023}. Our exit interviews surface a user-side taxonomy of failure modes that helps explain when and why these tools break down. We see that users exit the tools control through habituating repeated alerts, uninstalling the limiting apps, and bypassing screen-time limits. These findings relate to prior work that show that individuals tend to habituate their own interventions \cite{lyngs_hack_2020}.Participants reported that interruptions often arrived midsession instead at natural endpoints that may have otherwise been natural disengagement due to competing tasks or boredom. This pattern was also reinforced by our quantitative finding that the intention-usage gap and not session duration was the dominant predictor of session-level regret. Together, these findings motivate a shift away from static, duration-based restrictions toward interventions that are sensitive to the user’s pre-session intentions and momentary state.

\subsection{It is not time, it is losing control of time}

Participants in our sample were less concerned with time itself than with losing it. The intention-usage gap predicted regret roughly twice as strongly as duration alone ($\beta = 0.76$ vs.\ $\beta = 0.36$), and once the gap was in the model, duration's independent effect shrank to $\beta = 0.18$. The qualitative theme of time blindness tells the same story from a different angle: participants described both knowing in the moment that they were overrunning and being unable to stop, and registering only afterward how much time had actually passed. The objective logs reinforce the framing as well; across every participant, social-media sessions were roughly four times longer than non-social-media sessions, consistent with prior work on short-form-video time distortion \citep{yang2024time, jiang2025losing}. Together these strands extend the line of evidence from \citet{lukoff_what_2018}, \citet{cho2021reflect}, and \citet{guo_what_2025}: it is the alignment between intended and actual use, not the duration itself, that determines whether a session is regretted.

\subsection{Regret is highest when opportunity costs are salient}

Regret in our data was highest when scrolling competed with something the participant valued doing instead. Three converging strands support this. The qualitative interviews surfaced an ``escape hatch'' theme in which participants described regretting social-media use most when they had turned to it to avoid an unwanted task or feeling, consistent with prior experience-sampling work on procrastinatory media use \citep{Hofmann2012, Reinecke2016}. The quantitative analyses showed elevated regret for sessions immediately preceded by productivity-app use, and elevated regret for sessions occurring at night, when sleep was the plausible displaced alternative. All three patterns align with regret theory's claim that the intensity of regret tracks the perceived value of the alternative \citep{zeelenberg_anticipated_1999}: a session that displaces nothing in particular generates little regret, while a session that displaces work or sleep generates more. The felt cost of a session is therefore not fixed; it depends on what the session is competing with. The design implication is that interventions are likely to be most effective when they make the competing goal salient at session onset rather than enforce a duration cap after the fact.

\subsection{The strongest predictor of regret is the one passive sensing cannot observe}

Our results contain a tension that should be made explicit. RQ1 establishes that the intention-usage gap is by a wide margin the strongest predictor of regret. RQ2 deliberately excludes self-report features, because a deployable JITAI cannot ask each session whether it has met its intention before deciding whether to trigger. The variable that explains regret best is therefore precisely the variable that pre-session sensing cannot directly observe. Two things follow. First, the goal of pre-session sensing in this paradigm is not to recover the gap directly but to flag sessions in which the gap is likely to manifest, before any retrospective assessment is possible. Second, the within-person AUC of .740 (combined channel, 50-seed mean) is consistent with this framing: it is recovering a fraction of what an intention-aware system could in principle achieve, not approximating an upper bound. A natural next step is to test whether passive proxies (rolling history, contextual cues, behavioral rhythm) can stand in for the gap closely enough to support triggering decisions, and how much closer they bring the predictive ceiling to what an explicit intention measure would yield.

\subsection{What we can and cannot deploy}

The two evaluation metrics map to distinct phases of a deployed JITAI. \textit{LOPO AUC} mirrors the cold start: Phone Context alone is barely above chance, and smartwatch features add only a small lift within LOPO standard deviations. \textit{Within-person AUC} (70/30 shuffled, 50 seeds) mirrors the post-onboarding scenario and is substantially higher for both channels, supporting a one to two week passive-collection window before intervention triggers fire. Per-participant AUCs (Table~\ref{tab:rq2-methods-auc}) show this ceiling is unevenly distributed and tracks feature-side properties rather than data quality or regret variance: Smartwatch AUC tracks pre-session step count; Phone Context AUC tracks the within-person variance in hour of day and inversely the variance of inter-session gap; and Smartwatch+Phone Context produces the highest mean AUC, recovering participants who underperform under either single channel. Within-person regret SD ties only weakly to AUC, and missing-physio rate is uncorrelated with Smartwatch AUC, so participants score higher when their measured features align with their personal regret dimensions, not because they have more variance to predict or wore the watch more. A deployed system would default to the combined channel and route users matched by neither to a non-predictive baseline. Single-session prediction stays imperfect because much of what shapes regret happens during the session itself, which no pre-session sensor can observe. RQ1 found the intention-usage gap, not duration, to be the strongest predictor, and hard blocks and time limits habituate or provoke reactance (Section~\ref{sec:tools-fail}), so what fires should surface a brief reminder of the user's stated intention or a salient competing goal rather than restrict access, consistent with passive-sensing findings that personalized models outperform cross-subject ones for mood, stress, and mental health \cite{Jaques2017Predicting,Adler2020Predicting,taylor2017personalized}.

\section{Limitations}
\label{sec:limitations}
We modeled regret as a per-participant above-median binary, so predictions indicate only whether a session will be more or less regretful than typical for that person; a JITAI built on this signal would trigger on relative elevation, not absolute magnitude. The final sample comprises 21 participants (19 with usable smartwatch coverage), recruited primarily from a single university and skewed young (age 18--33, $M\!=\!21.8$), with no stratification by clinical severity or culture; generalization beyond this profile is unsupported. Participant 50 contributes roughly 17\% of all sessions, so we report per-participant and LOPO summaries alongside pooled metrics in Tables~\ref{tab:rq2-methods-auc} and~\ref{tab:rq2-within-detail}. Pre-session smartwatch features were missing for 24\% of sessions, reflecting the lower reliability of consumer wearables in the wild: participants forget the watch after charging, remove it for sleep or showers, and sensors drop samples under motion or loose contact. The seven-day window is short relative to habit change; longer deployments such as StudentLife \cite{nepal2024capturing} show what richer temporal coverage can recover, so the personalization premium we report ($\sim$.10--.17 LOPO-to-within-person gap) bounds neither what richer history nor a longer onboarding could provide. The Bangle.js~2 was chosen for openness and replicability, but its sensors are susceptible to motion and contact artifacts, and the Android-only observation app further narrows the deployable population. 

'

\section{Conclusion}

This paper asks whether and how a session of social-media use will be regretted can be anticipated before it begins, using passive signals from a phone and a low-cost consumer smartwatch. Three findings emerged. Regret tracks a person's intention-usage gap rather than session duration: the gap explains regret roughly twice as strongly as duration alone, and the qualitative pattern of time blindness, knowing in the moment that one is overrunning yet not stopping, converges on the same interpretation. Regret is amplified when the session displaces a valued alternative such as sleep or work, consistent with regret theory's claim that the cost of an action tracks the value of what it forecloses. Pre-session passive features carry sufficient predictive signal for the question we asked (within-person mean AUC of $.740$ with phone context plus smartwatch, against $.732$ for phone context alone), but the deployable population is small and heterogeneous: roughly a quarter of users in our sample reach the conventional acceptable-discrimination threshold, and adding the smartwatch shifts which users are covered rather than expanding overall coverage. The path to a deployable JITAI runs through intention-aware triggering rather than duration caps, and through staged personalization rather than one-size-fits-all population models.

\begin{acks}
We thank Md Nafis Shfat and Ivy Yip for their contributions to this research.
\end{acks}


\bibliographystyle{ACM-Reference-Format}
\bibliography{archive/references}


\begin{thebibliography}{108}


\ifx \showCODEN    \undefined \def \showCODEN     #1{\unskip}     \fi
\ifx \showISBNx    \undefined \def \showISBNx     #1{\unskip}     \fi
\ifx \showISBNxiii \undefined \def \showISBNxiii  #1{\unskip}     \fi
\ifx \showISSN     \undefined \def \showISSN      #1{\unskip}     \fi
\ifx \showLCCN     \undefined \def \showLCCN      #1{\unskip}     \fi
\ifx \shownote     \undefined \def \shownote      #1{#1}          \fi
\ifx \showarticletitle \undefined \def \showarticletitle #1{#1}   \fi
\ifx \showURL      \undefined \def \showURL       {\relax}        \fi
\providecommand\bibfield[2]{#2}
\providecommand\bibinfo[2]{#2}
\providecommand\natexlab[1]{#1}
\providecommand\showeprint[2][]{arXiv:#2}

\bibitem[Adler et~al\mbox{.}(2020)]%
        {Adler2020Predicting}
\bibfield{author}{\bibinfo{person}{Daniel~A. Adler}, \bibinfo{person}{Dror Ben-Zeev}, \bibinfo{person}{Vincent W.-S. Tseng}, \bibinfo{person}{John~M. Kane}, \bibinfo{person}{Rachel Brian}, \bibinfo{person}{Andrew~T. Campbell}, \bibinfo{person}{Marta Hauser}, \bibinfo{person}{Emily~A. Scherer}, {and} \bibinfo{person}{Tanzeem Choudhury}.} \bibinfo{year}{2020}\natexlab{}.
\newblock \showarticletitle{Predicting Early Warning Signs of Psychotic Relapse From Passive Sensing Data: An Approach Using Encoder-Decoder Neural Networks}.
\newblock \bibinfo{journal}{\emph{JMIR mHealth and uHealth}} \bibinfo{volume}{8}, \bibinfo{number}{8} (\bibinfo{year}{2020}), \bibinfo{pages}{e19962}.
\newblock
\href{https://doi.org/10.2196/19962}{doi:\nolinkurl{10.2196/19962}}


\bibitem[Andreassen et~al\mbox{.}(2016a)]%
        {Andreassen2016a}
\bibfield{author}{\bibinfo{person}{Cecilie~Schou Andreassen}, \bibinfo{person}{Jo{\"e}l Billieux}, \bibinfo{person}{Mark~D. Griffiths}, \bibinfo{person}{Daria~J. Kuss}, \bibinfo{person}{Zsolt Demetrovics}, \bibinfo{person}{Elvis Mazzoni}, {and} \bibinfo{person}{St{\aa}le Pallesen}.} \bibinfo{year}{2016}\natexlab{a}.
\newblock \showarticletitle{The Relationship Between Addictive Use of Social Media and Video Games and Symptoms of Psychiatric Disorders: A Large-Scale Cross-Sectional Study}.
\newblock \bibinfo{journal}{\emph{Psychology of Addictive Behaviors}} \bibinfo{volume}{30}, \bibinfo{number}{2} (\bibinfo{year}{2016}), \bibinfo{pages}{252--262}.
\newblock
\href{https://doi.org/10.1037/adb0000160}{doi:\nolinkurl{10.1037/adb0000160}}


\bibitem[Andreassen et~al\mbox{.}(2016b)]%
        {andreassen_relationship_2016}
\bibfield{author}{\bibinfo{person}{Cecilie~Schou Andreassen}, \bibinfo{person}{Joël Billieux}, \bibinfo{person}{Mark~D. Griffiths}, \bibinfo{person}{Daria~J. Kuss}, \bibinfo{person}{Zsolt Demetrovics}, \bibinfo{person}{Elvis Mazzoni}, {and} \bibinfo{person}{Ståle Pallesen}.} \bibinfo{year}{2016}\natexlab{b}.
\newblock \showarticletitle{The relationship between addictive use of social media and video games and symptoms of psychiatric disorders: {A} large-scale cross-sectional study.}
\newblock \bibinfo{journal}{\emph{Psychology of Addictive Behaviors}} \bibinfo{volume}{30}, \bibinfo{number}{2} (\bibinfo{date}{March} \bibinfo{year}{2016}), \bibinfo{pages}{252--262}.
\newblock
\showISSN{1939-1501, 0893-164X}
\href{https://doi.org/10.1037/adb0000160}{doi:\nolinkurl{10.1037/adb0000160}}


\bibitem[{Android Developers}(2024a)]%
        {android_accessibilityservice}
\bibfield{author}{\bibinfo{person}{{Android Developers}}.} \bibinfo{year}{2024}\natexlab{a}.
\newblock \bibinfo{title}{{AccessibilityService}}.
\newblock \bibinfo{howpublished}{\url{https://developer.android.com/reference/android/accessibilityservice/AccessibilityService}}.
\newblock
\newblock
\shownote{Accessed: 2024}.


\bibitem[{Android Developers}(2024b)]%
        {android_usagestatsmanager}
\bibfield{author}{\bibinfo{person}{{Android Developers}}.} \bibinfo{year}{2024}\natexlab{b}.
\newblock \bibinfo{title}{{UsageStatsManager}}.
\newblock \bibinfo{howpublished}{\url{https://developer.android.com/reference/android/app/usage/UsageStatsManager}}.
\newblock
\newblock
\shownote{Accessed: 2024}.


\bibitem[{Android Developers}(2024c)]%
        {android_workmanager}
\bibfield{author}{\bibinfo{person}{{Android Developers}}.} \bibinfo{year}{2024}\natexlab{c}.
\newblock \bibinfo{title}{{WorkManager}}.
\newblock \bibinfo{howpublished}{\url{https://developer.android.com/reference/androidx/work/WorkManager}}.
\newblock
\newblock
\shownote{Accessed: 2024}.


\bibitem[Bari et~al\mbox{.}(2020)]%
        {bari2020stressful}
\bibfield{author}{\bibinfo{person}{Rambabu Bari}, \bibinfo{person}{Rummana~J. Adams}, \bibinfo{person}{Md~Mahbubur Rahman}, \bibinfo{person}{Megan~Battles Lau}, \bibinfo{person}{Syed~Monowar Hossain}, \bibinfo{person}{William~T. Riley}, {et~al\mbox{.}}} \bibinfo{year}{2020}\natexlab{}.
\newblock \showarticletitle{Automated Detection of Stressful Conversations Using Wearable Physiological and Inertial Sensors}.
\newblock \bibinfo{journal}{\emph{Proceedings of the ACM on Interactive, Mobile, Wearable and Ubiquitous Technologies}} \bibinfo{volume}{4}, \bibinfo{number}{4} (\bibinfo{year}{2020}).
\newblock


\bibitem[Battalio et~al\mbox{.}(2021)]%
        {battalio2021sense2stop}
\bibfield{author}{\bibinfo{person}{Samuel Battalio} {et~al\mbox{.}}} \bibinfo{year}{2021}\natexlab{}.
\newblock \showarticletitle{{Sense2Stop}: A Micro-Randomized Trial Using Wearable Sensors to Optimize a Just-in-Time-Adaptive Stress Management Intervention for Smoking Relapse Prevention}.
\newblock \bibinfo{journal}{\emph{Contemporary Clinical Trials}}  \bibinfo{volume}{109} (\bibinfo{year}{2021}).
\newblock


\bibitem[Baumeister et~al\mbox{.}(2007)]%
        {Baumeister2007}
\bibfield{author}{\bibinfo{person}{Roy~F. Baumeister}, \bibinfo{person}{Kathleen~D. Vohs}, {and} \bibinfo{person}{Dianne~M. Tice}.} \bibinfo{year}{2007}\natexlab{}.
\newblock \showarticletitle{The Strength Model of Self-Control}.
\newblock \bibinfo{journal}{\emph{Current Directions in Psychological Science}} \bibinfo{volume}{16}, \bibinfo{number}{6} (\bibinfo{year}{2007}), \bibinfo{pages}{351--355}.
\newblock
\href{https://doi.org/10.1111/j.1467-8721.2007.00534.x}{doi:\nolinkurl{10.1111/j.1467-8721.2007.00534.x}}


\bibitem[Bayer et~al\mbox{.}(2022)]%
        {Bayer2022}
\bibfield{author}{\bibinfo{person}{Joseph~B. Bayer}, \bibinfo{person}{Ian~A. Anderson}, {and} \bibinfo{person}{Robert~S. Tokunaga}.} \bibinfo{year}{2022}\natexlab{}.
\newblock \showarticletitle{Building and Breaking Social Media Habits}.
\newblock \bibinfo{journal}{\emph{Current Opinion in Psychology}}  \bibinfo{volume}{45} (\bibinfo{year}{2022}), \bibinfo{pages}{101303}.
\newblock
\href{https://doi.org/10.1016/j.copsyc.2022.101303}{doi:\nolinkurl{10.1016/j.copsyc.2022.101303}}


\bibitem[Berkman et~al\mbox{.}(2017)]%
        {Berkman2017}
\bibfield{author}{\bibinfo{person}{Elliot~T. Berkman}, \bibinfo{person}{Cendri~A. Hutcherson}, \bibinfo{person}{Jordan~L. Livingston}, \bibinfo{person}{Lauren~E. Kahn}, {and} \bibinfo{person}{Michael Inzlicht}.} \bibinfo{year}{2017}\natexlab{}.
\newblock \showarticletitle{Self-Control as Value-Based Choice}.
\newblock \bibinfo{journal}{\emph{Current Directions in Psychological Science}} \bibinfo{volume}{26}, \bibinfo{number}{5} (\bibinfo{year}{2017}), \bibinfo{pages}{422--428}.
\newblock
\href{https://doi.org/10.1177/0963721417704394}{doi:\nolinkurl{10.1177/0963721417704394}}


\bibitem[Braun and Clarke(2006a)]%
        {braun_using_2006}
\bibfield{author}{\bibinfo{person}{Virginia Braun} {and} \bibinfo{person}{Victoria Clarke}.} \bibinfo{year}{2006}\natexlab{a}.
\newblock \showarticletitle{Using thematic analysis in psychology}.
\newblock \bibinfo{journal}{\emph{Qualitative Research in Psychology}} \bibinfo{volume}{3}, \bibinfo{number}{2} (\bibinfo{year}{2006}), \bibinfo{pages}{77--101}.
\newblock
\href{https://doi.org/10.1191/1478088706qp063oa}{doi:\nolinkurl{10.1191/1478088706qp063oa}}


\bibitem[Braun and Clarke(2006b)]%
        {braun2006using}
\bibfield{author}{\bibinfo{person}{Virginia Braun} {and} \bibinfo{person}{Victoria Clarke}.} \bibinfo{year}{2006}\natexlab{b}.
\newblock \showarticletitle{Using thematic analysis in psychology}.
\newblock \bibinfo{journal}{\emph{Qualitative Research in Psychology}} \bibinfo{volume}{3}, \bibinfo{number}{2} (\bibinfo{year}{2006}), \bibinfo{pages}{77--101}.
\newblock
\href{https://doi.org/10.1191/1478088706qp063oa}{doi:\nolinkurl{10.1191/1478088706qp063oa}}


\bibitem[Braun and Clarke(2019)]%
        {braun_reflecting_2019}
\bibfield{author}{\bibinfo{person}{Virginia Braun} {and} \bibinfo{person}{Victoria Clarke}.} \bibinfo{year}{2019}\natexlab{}.
\newblock \showarticletitle{Reflecting on reflexive thematic analysis}.
\newblock \bibinfo{journal}{\emph{Qualitative Research in Sport, Exercise and Health}} \bibinfo{volume}{11}, \bibinfo{number}{4} (\bibinfo{year}{2019}), \bibinfo{pages}{589--597}.
\newblock
\href{https://doi.org/10.1080/2159676X.2019.1628806}{doi:\nolinkurl{10.1080/2159676X.2019.1628806}}


\bibitem[Can et~al\mbox{.}(2019)]%
        {can2019continuous}
\bibfield{author}{\bibinfo{person}{Yekta~Said Can}, \bibinfo{person}{Niaz Chalabianloo}, \bibinfo{person}{Deniz Ekiz}, {and} \bibinfo{person}{Cem Ersoy}.} \bibinfo{year}{2019}\natexlab{}.
\newblock \showarticletitle{Continuous Stress Detection Using Wearable Sensors in Real Life: Algorithmic Programming Contest Case Study}.
\newblock \bibinfo{journal}{\emph{Sensors}} \bibinfo{volume}{19}, \bibinfo{number}{8} (\bibinfo{year}{2019}), \bibinfo{pages}{1849}.
\newblock
\href{https://doi.org/10.3390/s19081849}{doi:\nolinkurl{10.3390/s19081849}}


\bibitem[Cho et~al\mbox{.}(2021)]%
        {cho2021reflect}
\bibfield{author}{\bibinfo{person}{Hyunsung Cho}, \bibinfo{person}{Daeun Choi}, \bibinfo{person}{Donghwi Kim}, \bibinfo{person}{Wan~Ju Kang}, \bibinfo{person}{Eun~Kyoung Choe}, {and} \bibinfo{person}{Sung-Ju Lee}.} \bibinfo{year}{2021}\natexlab{}.
\newblock \showarticletitle{Reflect, Not Regret: Understanding Regretful Smartphone Use with App Feature-Level Analysis}.
\newblock \bibinfo{journal}{\emph{Proceedings of the ACM on Human-Computer Interaction}} \bibinfo{volume}{5}, \bibinfo{number}{CSCW2} (\bibinfo{year}{2021}), \bibinfo{pages}{1--28}.
\newblock
\href{https://doi.org/10.1145/3479600}{doi:\nolinkurl{10.1145/3479600}}


\bibitem[Connolly and Zeelenberg(2002)]%
        {connolly_regret_2002}
\bibfield{author}{\bibinfo{person}{Terry Connolly} {and} \bibinfo{person}{Marcel Zeelenberg}.} \bibinfo{year}{2002}\natexlab{}.
\newblock \showarticletitle{Regret in Decision Making}.
\newblock \bibinfo{journal}{\emph{Current Directions in Psychological Science}} \bibinfo{volume}{11}, \bibinfo{number}{6} (\bibinfo{year}{2002}), \bibinfo{pages}{212--216}.
\newblock
\href{https://doi.org/10.1111/1467-8721.00203}{doi:\nolinkurl{10.1111/1467-8721.00203}}


\bibitem[Cosme et~al\mbox{.}(2019)]%
        {Cosme2019}
\bibfield{author}{\bibinfo{person}{Danielle Cosme}, \bibinfo{person}{Rita~M. Ludwig}, {and} \bibinfo{person}{Elliot~T. Berkman}.} \bibinfo{year}{2019}\natexlab{}.
\newblock \showarticletitle{Comparing Two Neurocognitive Models of Self-Control During Dietary Decisions}.
\newblock \bibinfo{journal}{\emph{Social Cognitive and Affective Neuroscience}} \bibinfo{volume}{14}, \bibinfo{number}{9} (\bibinfo{year}{2019}), \bibinfo{pages}{957--966}.
\newblock
\href{https://doi.org/10.1093/scan/nsz068}{doi:\nolinkurl{10.1093/scan/nsz068}}


\bibitem[Dai et~al\mbox{.}(2021)]%
        {dai2021comparing}
\bibfield{author}{\bibinfo{person}{Ruixuan Dai}, \bibinfo{person}{Chenyang Lu}, \bibinfo{person}{Linda Yun}, \bibinfo{person}{Eric~J. Lenze}, \bibinfo{person}{Michael~S. Avidan}, {and} \bibinfo{person}{Thomas Kannampallil}.} \bibinfo{year}{2021}\natexlab{}.
\newblock \showarticletitle{Comparing Stress Prediction Models Using Smartwatch Physiological Signals and Participant Self-Reports}.
\newblock \bibinfo{journal}{\emph{Computer Methods and Programs in Biomedicine}}  \bibinfo{volume}{208} (\bibinfo{year}{2021}), \bibinfo{pages}{106207}.
\newblock
\href{https://doi.org/10.1016/j.cmpb.2021.106207}{doi:\nolinkurl{10.1016/j.cmpb.2021.106207}}


\bibitem[Duckworth et~al\mbox{.}(2016)]%
        {Duckworth2016}
\bibfield{author}{\bibinfo{person}{Angela~L. Duckworth}, \bibinfo{person}{Tamar~Szab{\'o} Gendler}, {and} \bibinfo{person}{James~J. Gross}.} \bibinfo{year}{2016}\natexlab{}.
\newblock \showarticletitle{Situational Strategies for Self-Control}.
\newblock \bibinfo{journal}{\emph{Perspectives on Psychological Science}} \bibinfo{volume}{11}, \bibinfo{number}{1} (\bibinfo{year}{2016}), \bibinfo{pages}{35--55}.
\newblock
\href{https://doi.org/10.1177/1745691615623247}{doi:\nolinkurl{10.1177/1745691615623247}}


\bibitem[Epstein et~al\mbox{.}(2016)]%
        {epstein2016reconsidering}
\bibfield{author}{\bibinfo{person}{Daniel~A. Epstein}, \bibinfo{person}{Jennifer~H. Kang}, \bibinfo{person}{Laura~R. Pina}, \bibinfo{person}{James Fogarty}, {and} \bibinfo{person}{Sean~A. Munson}.} \bibinfo{year}{2016}\natexlab{}.
\newblock \showarticletitle{Reconsidering the Device in the Drawer: Lapses as a Design Opportunity in Personal Informatics}. In \bibinfo{booktitle}{\emph{Proceedings of the 2016 ACM International Joint Conference on Pervasive and Ubiquitous Computing (UbiComp '16)}}. \bibinfo{publisher}{Association for Computing Machinery}, \bibinfo{address}{New York, NY, USA}, \bibinfo{pages}{829--840}.
\newblock
\href{https://doi.org/10.1145/2971648.2971656}{doi:\nolinkurl{10.1145/2971648.2971656}}


\bibitem[{Espruino}(2024)]%
        {banglejs2}
\bibfield{author}{\bibinfo{person}{{Espruino}}.} \bibinfo{year}{2024}\natexlab{}.
\newblock \bibinfo{title}{{Bangle.js 2}}.
\newblock \bibinfo{howpublished}{\url{https://www.espruino.com/Bangle.js2}}.
\newblock
\newblock
\shownote{Accessed: 2024}.


\bibitem[Flayelle et~al\mbox{.}(2023)]%
        {Flayelle2023}
\bibfield{author}{\bibinfo{person}{Ma{\`e}va Flayelle}, \bibinfo{person}{Damien Brevers}, \bibinfo{person}{Daniel~L. King}, \bibinfo{person}{Pierre Maurage}, \bibinfo{person}{Jos{\'e}~C. Perales}, {and} \bibinfo{person}{Jo{\"e}l Billieux}.} \bibinfo{year}{2023}\natexlab{}.
\newblock \showarticletitle{A Taxonomy of Technology Design Features That Promote Potentially Addictive Online Behaviours}.
\newblock \bibinfo{journal}{\emph{Nature Reviews Psychology}} \bibinfo{volume}{2}, \bibinfo{number}{3} (\bibinfo{year}{2023}), \bibinfo{pages}{136--150}.
\newblock
\href{https://doi.org/10.1038/s44159-023-00153-4}{doi:\nolinkurl{10.1038/s44159-023-00153-4}}


\bibitem[{Freeyourgadget}(2024)]%
        {gadgetbridge}
\bibfield{author}{\bibinfo{person}{{Freeyourgadget}}.} \bibinfo{year}{2024}\natexlab{}.
\newblock \bibinfo{title}{{Gadgetbridge}}.
\newblock \bibinfo{howpublished}{\url{https://codeberg.org/Freeyourgadget/Gadgetbridge/}}.
\newblock
\newblock
\shownote{Accessed: 2024}.


\bibitem[Garcia-Ceja et~al\mbox{.}(2018)]%
        {garciaceja2018engagement}
\bibfield{author}{\bibinfo{person}{Enrique Garcia-Ceja} {et~al\mbox{.}}} \bibinfo{year}{2018}\natexlab{}.
\newblock \showarticletitle{Unobtrusive Assessment of Students' Emotional Engagement during Lectures Using Electrodermal Activity Sensors}.
\newblock \bibinfo{journal}{\emph{Proceedings of the ACM on Interactive, Mobile, Wearable and Ubiquitous Technologies}} \bibinfo{volume}{2}, \bibinfo{number}{3} (\bibinfo{year}{2018}).
\newblock
\href{https://doi.org/10.1145/3264913}{doi:\nolinkurl{10.1145/3264913}}


\bibitem[Gebhardt et~al\mbox{.}(2024)]%
        {gebhardt_detecting_2024}
\bibfield{author}{\bibinfo{person}{Christoph Gebhardt}, \bibinfo{person}{Andreas Brombach}, \bibinfo{person}{Tiffany Luong}, \bibinfo{person}{Otmar Hilliges}, {and} \bibinfo{person}{Christian Holz}.} \bibinfo{year}{2024}\natexlab{}.
\newblock \showarticletitle{Detecting {Users}' {Emotional} {States} during {Passive} {Social} {Media} {Use}}.
\newblock \bibinfo{journal}{\emph{Proceedings of the ACM on Interactive, Mobile, Wearable and Ubiquitous Technologies}} \bibinfo{volume}{8}, \bibinfo{number}{2} (\bibinfo{date}{May} \bibinfo{year}{2024}), \bibinfo{pages}{1--30}.
\newblock
\showISSN{2474-9567}
\href{https://doi.org/10.1145/3659606}{doi:\nolinkurl{10.1145/3659606}}


\bibitem[Gilovich and Medvec(1995)]%
        {gilovich_experience_1995}
\bibfield{author}{\bibinfo{person}{Thomas Gilovich} {and} \bibinfo{person}{Victoria~Husted Medvec}.} \bibinfo{year}{1995}\natexlab{}.
\newblock \showarticletitle{The Experience of Regret: What, When, and Why}.
\newblock \bibinfo{journal}{\emph{Psychological Review}} \bibinfo{volume}{102}, \bibinfo{number}{2} (\bibinfo{year}{1995}), \bibinfo{pages}{379--395}.
\newblock
\href{https://doi.org/10.1037/0033-295X.102.2.379}{doi:\nolinkurl{10.1037/0033-295X.102.2.379}}


\bibitem[Grüning et~al\mbox{.}(2023)]%
        {gruning_directing_2023}
\bibfield{author}{\bibinfo{person}{David~J. Grüning}, \bibinfo{person}{Frederik Riedel}, {and} \bibinfo{person}{Philipp Lorenz-Spreen}.} \bibinfo{year}{2023}\natexlab{}.
\newblock \showarticletitle{Directing smartphone use through the self-nudge app one sec}.
\newblock \bibinfo{journal}{\emph{Proceedings of the National Academy of Sciences}} \bibinfo{volume}{120}, \bibinfo{number}{8} (\bibinfo{date}{Feb.} \bibinfo{year}{2023}), \bibinfo{pages}{e2213114120}.
\newblock
\showISSN{0027-8424, 1091-6490}
\href{https://doi.org/10.1073/pnas.2213114120}{doi:\nolinkurl{10.1073/pnas.2213114120}}


\bibitem[Guo et~al\mbox{.}(2025)]%
        {guo_what_2025}
\bibfield{author}{\bibinfo{person}{Longjie Guo}, \bibinfo{person}{Yue Fu}, \bibinfo{person}{Xiran Lin}, \bibinfo{person}{Xuhai Xu}, \bibinfo{person}{Yung-Ju Chang}, {and} \bibinfo{person}{Alexis Hiniker}.} \bibinfo{year}{2025}\natexlab{}.
\newblock \showarticletitle{What {Social} {Media} {Use} {Do} {People} {Regret}? {An} {Analysis} of {34K} {Smartphone} {Screenshots} with {Multimodal} {LLM}}. In \bibinfo{booktitle}{\emph{Proceedings of the 2025 {CHI} {Conference} on {Human} {Factors} in {Computing} {Systems}}}. \bibinfo{publisher}{ACM}, \bibinfo{address}{Yokohama Japan}, \bibinfo{pages}{1--23}.
\newblock
\showISBNx{979-8-4007-1394-1}
\href{https://doi.org/10.1145/3706598.3713724}{doi:\nolinkurl{10.1145/3706598.3713724}}


\bibitem[Harper and Southern(2019)]%
        {harper2019endtoend}
\bibfield{author}{\bibinfo{person}{Ross Harper} {and} \bibinfo{person}{Joshua Southern}.} \bibinfo{year}{2019}\natexlab{}.
\newblock \showarticletitle{End-To-End Prediction of Emotion from Heartbeat Data Collected by a Consumer Fitness Tracker}. In \bibinfo{booktitle}{\emph{Proceedings of the 8th International Conference on Affective Computing and Intelligent Interaction (ACII)}}. \bibinfo{pages}{1--7}.
\newblock
\href{https://doi.org/10.1109/acii.2019.8925520}{doi:\nolinkurl{10.1109/acii.2019.8925520}}


\bibitem[Hickey et~al\mbox{.}(2021)]%
        {hickey2021ppgstress}
\bibfield{author}{\bibinfo{person}{Brian~A. Hickey}, \bibinfo{person}{Taryn Chalmers}, \bibinfo{person}{Phillip Newton}, \bibinfo{person}{Chin-Teng Lin}, \bibinfo{person}{David Siber}, \bibinfo{person}{Han~Sean Leng}, {et~al\mbox{.}}} \bibinfo{year}{2021}\natexlab{}.
\newblock \showarticletitle{Smart Devices and Wearable Technologies to Detect and Monitor Mental Health Conditions and Stress: A Systematic Review}.
\newblock \bibinfo{journal}{\emph{Sensors}} \bibinfo{volume}{21}, \bibinfo{number}{10} (\bibinfo{year}{2021}), \bibinfo{pages}{3461}.
\newblock
\href{https://doi.org/10.3390/s21103461}{doi:\nolinkurl{10.3390/s21103461}}


\bibitem[Hiniker et~al\mbox{.}(2016)]%
        {hiniker2016mytime}
\bibfield{author}{\bibinfo{person}{Alexis Hiniker}, \bibinfo{person}{Sungsoo~(Ray) Hong}, \bibinfo{person}{Tadayoshi Kohno}, {and} \bibinfo{person}{Julie~A. Kientz}.} \bibinfo{year}{2016}\natexlab{}.
\newblock \showarticletitle{{MyTime}: Designing and Evaluating an Intervention for Smartphone Non-Use}. In \bibinfo{booktitle}{\emph{Proceedings of the 2016 CHI Conference on Human Factors in Computing Systems}} \emph{(\bibinfo{series}{CHI '16})}. \bibinfo{publisher}{ACM}, \bibinfo{address}{New York, NY, USA}, \bibinfo{pages}{4746--4757}.
\newblock
\href{https://doi.org/10.1145/2858036.2858403}{doi:\nolinkurl{10.1145/2858036.2858403}}


\bibitem[Hofmann et~al\mbox{.}(2012)]%
        {Hofmann2012}
\bibfield{author}{\bibinfo{person}{Wilhelm Hofmann}, \bibinfo{person}{Roy~F. Baumeister}, \bibinfo{person}{Georg F{\"o}rster}, {and} \bibinfo{person}{Kathleen~D. Vohs}.} \bibinfo{year}{2012}\natexlab{}.
\newblock \showarticletitle{Everyday Temptations: An Experience Sampling Study of Desire, Conflict, and Self-Control}.
\newblock \bibinfo{journal}{\emph{Journal of Personality and Social Psychology}} \bibinfo{volume}{102}, \bibinfo{number}{6} (\bibinfo{year}{2012}), \bibinfo{pages}{1318--1335}.
\newblock
\href{https://doi.org/10.1037/a0026545}{doi:\nolinkurl{10.1037/a0026545}}


\bibitem[Hofmann et~al\mbox{.}(2009)]%
        {Hofmann2009}
\bibfield{author}{\bibinfo{person}{Wilhelm Hofmann}, \bibinfo{person}{Malte Friese}, {and} \bibinfo{person}{Fritz Strack}.} \bibinfo{year}{2009}\natexlab{}.
\newblock \showarticletitle{Impulse and Self-Control From a Dual-Systems Perspective}.
\newblock \bibinfo{journal}{\emph{Perspectives on Psychological Science}} \bibinfo{volume}{4}, \bibinfo{number}{2} (\bibinfo{year}{2009}), \bibinfo{pages}{162--176}.
\newblock
\href{https://doi.org/10.1111/j.1745-6924.2009.01116.x}{doi:\nolinkurl{10.1111/j.1745-6924.2009.01116.x}}


\bibitem[Hosmer and Lemeshow(2000)]%
        {hosmer2000applied}
\bibfield{author}{\bibinfo{person}{David~W. Hosmer} {and} \bibinfo{person}{Stanley Lemeshow}.} \bibinfo{year}{2000}\natexlab{}.
\newblock \bibinfo{booktitle}{\emph{Applied Logistic Regression} (\bibinfo{edition}{2} ed.)}.
\newblock \bibinfo{publisher}{John Wiley \& Sons}, \bibinfo{address}{New York}.
\newblock
\href{https://doi.org/10.1002/0471722146}{doi:\nolinkurl{10.1002/0471722146}}


\bibitem[Hovsepian et~al\mbox{.}(2015)]%
        {hovsepian2015cstress}
\bibfield{author}{\bibinfo{person}{Karen Hovsepian}, \bibinfo{person}{Mustafa al Absi}, \bibinfo{person}{Emre Ertin}, \bibinfo{person}{Thomas Kamarck}, \bibinfo{person}{Motohiro Nakajima}, {and} \bibinfo{person}{Santosh Kumar}.} \bibinfo{year}{2015}\natexlab{}.
\newblock \showarticletitle{cStress: Towards a Gold Standard for Continuous Stress Assessment in the Field Using Body Sensor Networks}.
\newblock \bibinfo{journal}{\emph{Proceedings of the ACM International Joint Conference on Pervasive and Ubiquitous Computing}} (\bibinfo{year}{2015}), \bibinfo{pages}{493--504}.
\newblock
\href{https://doi.org/10.1145/2750858.2807526}{doi:\nolinkurl{10.1145/2750858.2807526}}


\bibitem[Huang(2022)]%
        {Huang2022}
\bibfield{author}{\bibinfo{person}{Chiungjung Huang}.} \bibinfo{year}{2022}\natexlab{}.
\newblock \showarticletitle{A Meta-Analysis of the Problematic Social Media Use and Mental Health}.
\newblock \bibinfo{journal}{\emph{International Journal of Social Psychiatry}} \bibinfo{volume}{68}, \bibinfo{number}{1} (\bibinfo{year}{2022}), \bibinfo{pages}{12--33}.
\newblock
\href{https://doi.org/10.1177/0020764020978434}{doi:\nolinkurl{10.1177/0020764020978434}}


\bibitem[Inzlicht et~al\mbox{.}(2021)]%
        {Inzlicht2021}
\bibfield{author}{\bibinfo{person}{Michael Inzlicht}, \bibinfo{person}{Kaitlyn~M. Werner}, \bibinfo{person}{Julia~L. Briskin}, {and} \bibinfo{person}{Brent~W. Roberts}.} \bibinfo{year}{2021}\natexlab{}.
\newblock \showarticletitle{Integrating Models of Self-Regulation}.
\newblock \bibinfo{journal}{\emph{Annual Review of Psychology}}  \bibinfo{volume}{72} (\bibinfo{year}{2021}), \bibinfo{pages}{319--345}.
\newblock
\href{https://doi.org/10.1146/annurev-psych-061020-105721}{doi:\nolinkurl{10.1146/annurev-psych-061020-105721}}


\bibitem[Jaques et~al\mbox{.}(2017)]%
        {Jaques2017Predicting}
\bibfield{author}{\bibinfo{person}{Natasha Jaques}, \bibinfo{person}{Ognjen Rudovic}, \bibinfo{person}{Sara Taylor}, \bibinfo{person}{Akane Sano}, {and} \bibinfo{person}{Rosalind Picard}.} \bibinfo{year}{2017}\natexlab{}.
\newblock \showarticletitle{Predicting Tomorrow's Mood, Health, and Stress Level Using Personalized Multitask Learning and Domain Adaptation}. In \bibinfo{booktitle}{\emph{Proceedings of {IJCAI} 2017 Workshop on Artificial Intelligence in Affective Computing}} \emph{(\bibinfo{series}{Proceedings of Machine Learning Research}, Vol.~\bibinfo{volume}{66})}. \bibinfo{publisher}{PMLR}, \bibinfo{pages}{17--33}.
\newblock


\bibitem[Jaques et~al\mbox{.}(2015)]%
        {jaques2015predicting}
\bibfield{author}{\bibinfo{person}{Natasha Jaques}, \bibinfo{person}{Sara Taylor}, \bibinfo{person}{Asaph Azaria}, \bibinfo{person}{Asma Ghandeharioun}, \bibinfo{person}{Akane Sano}, {and} \bibinfo{person}{Rosalind Picard}.} \bibinfo{year}{2015}\natexlab{}.
\newblock \showarticletitle{Predicting Students' Happiness from Physiology, Phone, Mobility, and Behavioral Data}. In \bibinfo{booktitle}{\emph{Proceedings of the 2015 International Conference on Affective Computing and Intelligent Interaction (ACII)}}. \bibinfo{pages}{222--228}.
\newblock
\href{https://doi.org/10.1109/ACII.2015.7344575}{doi:\nolinkurl{10.1109/ACII.2015.7344575}}


\bibitem[Jiang et~al\mbox{.}(2025)]%
        {jiang2025losing}
\bibfield{author}{\bibinfo{person}{Yaqi Jiang}, \bibinfo{person}{Zhihao Yan}, {and} \bibinfo{person}{Zeyang Yang}.} \bibinfo{year}{2025}\natexlab{}.
\newblock \showarticletitle{Losing Track of Time on {TikTok}? An Experimental Study of Short Video Users' Time Distortion}.
\newblock \bibinfo{journal}{\emph{Behavioral Sciences}} \bibinfo{volume}{15}, \bibinfo{number}{7} (\bibinfo{year}{2025}), \bibinfo{pages}{930}.
\newblock
\href{https://doi.org/10.3390/bs15070930}{doi:\nolinkurl{10.3390/bs15070930}}


\bibitem[Kapogianni et~al\mbox{.}(2025)]%
        {kapogianni2025using}
\bibfield{author}{\bibinfo{person}{Nikoletta-Anna Kapogianni}, \bibinfo{person}{Angeliki Sideraki}, {and} \bibinfo{person}{Christos-Nikolaos Anagnostopoulos}.} \bibinfo{year}{2025}\natexlab{}.
\newblock \showarticletitle{Using Smartwatches in Stress Management, Mental Health, and Well-Being: A Systematic Review}.
\newblock \bibinfo{journal}{\emph{Algorithms}} \bibinfo{volume}{18}, \bibinfo{number}{7} (\bibinfo{year}{2025}), \bibinfo{pages}{419}.
\newblock
\href{https://doi.org/10.3390/a18070419}{doi:\nolinkurl{10.3390/a18070419}}


\bibitem[Kim and Lee(2024)]%
        {kim_navigating_2024}
\bibfield{author}{\bibinfo{person}{Inyeop Kim} {and} \bibinfo{person}{Uichin Lee}.} \bibinfo{year}{2024}\natexlab{}.
\newblock \showarticletitle{Navigating {User}-{System} {Gaps}: {Understanding} {User}-{Interactions} in {User}-{Centric} {Context}-{Aware} {Systems} for {Digital} {Well}-being {Intervention}}. In \bibinfo{booktitle}{\emph{Proceedings of the {CHI} {Conference} on {Human} {Factors} in {Computing} {Systems}}}. \bibinfo{publisher}{ACM}, \bibinfo{address}{Honolulu HI USA}, \bibinfo{pages}{1--15}.
\newblock
\showISBNx{979-8-4007-0330-0}
\href{https://doi.org/10.1145/3613904.3641979}{doi:\nolinkurl{10.1145/3613904.3641979}}


\bibitem[Kim et~al\mbox{.}(2021)]%
        {kim_beneficial_2021}
\bibfield{author}{\bibinfo{person}{Minhyung Kim}, \bibinfo{person}{Inyeop Kim}, {and} \bibinfo{person}{Uichin Lee}.} \bibinfo{year}{2021}\natexlab{}.
\newblock \showarticletitle{Beneficial Neglect: Instant Message Notification Handling Behaviors and Academic Performance}.
\newblock \bibinfo{journal}{\emph{Proceedings of the ACM on Interactive, Mobile, Wearable and Ubiquitous Technologies}} \bibinfo{volume}{5}, \bibinfo{number}{1} (\bibinfo{year}{2021}), \bibinfo{pages}{1--26}.
\newblock
\href{https://doi.org/10.1145/3448089}{doi:\nolinkurl{10.1145/3448089}}


\bibitem[Ko et~al\mbox{.}(2015)]%
        {ko2015nugu}
\bibfield{author}{\bibinfo{person}{Minsam Ko}, \bibinfo{person}{Subin Yang}, \bibinfo{person}{Joonwon Lee}, \bibinfo{person}{Christian Heizmann}, \bibinfo{person}{Jinyoung Jeong}, \bibinfo{person}{Uichin Lee}, \bibinfo{person}{Daehee Shin}, \bibinfo{person}{Koji Yatani}, \bibinfo{person}{Junehwa Song}, {and} \bibinfo{person}{Kyong-Mee Chung}.} \bibinfo{year}{2015}\natexlab{}.
\newblock \showarticletitle{{NUGU}: A Group-based Intervention App for Improving Self-Regulation of Limiting Smartphone Use}. In \bibinfo{booktitle}{\emph{Proceedings of the 18th ACM Conference on Computer Supported Cooperative Work \& Social Computing}} \emph{(\bibinfo{series}{CSCW '15})}. \bibinfo{publisher}{ACM}, \bibinfo{address}{New York, NY, USA}, \bibinfo{pages}{1235--1245}.
\newblock
\href{https://doi.org/10.1145/2675133.2675244}{doi:\nolinkurl{10.1145/2675133.2675244}}


\bibitem[Kotabe and Hofmann(2015)]%
        {Kotabe2015}
\bibfield{author}{\bibinfo{person}{Hiroki~P. Kotabe} {and} \bibinfo{person}{Wilhelm Hofmann}.} \bibinfo{year}{2015}\natexlab{}.
\newblock \showarticletitle{On Integrating the Components of Self-Control}.
\newblock \bibinfo{journal}{\emph{Perspectives on Psychological Science}} \bibinfo{volume}{10}, \bibinfo{number}{5} (\bibinfo{year}{2015}), \bibinfo{pages}{618--638}.
\newblock
\href{https://doi.org/10.1177/1745691615593382}{doi:\nolinkurl{10.1177/1745691615593382}}


\bibitem[Kroenke et~al\mbox{.}(2001)]%
        {kroenke_phq-9_2001}
\bibfield{author}{\bibinfo{person}{Kurt Kroenke}, \bibinfo{person}{Robert~L. Spitzer}, {and} \bibinfo{person}{Janet B.~W. Williams}.} \bibinfo{year}{2001}\natexlab{}.
\newblock \showarticletitle{The {PHQ}-9: {Validity} of a brief depression severity measure}.
\newblock \bibinfo{journal}{\emph{Journal of General Internal Medicine}} \bibinfo{volume}{16}, \bibinfo{number}{9} (\bibinfo{date}{Sept.} \bibinfo{year}{2001}), \bibinfo{pages}{606--613}.
\newblock
\showISSN{0884-8734, 1525-1497}
\href{https://doi.org/10.1046/j.1525-1497.2001.016009606.x}{doi:\nolinkurl{10.1046/j.1525-1497.2001.016009606.x}}


\bibitem[Lee et~al\mbox{.}(2025)]%
        {lee_leveraging_2025}
\bibfield{author}{\bibinfo{person}{Hansoo Lee}, \bibinfo{person}{Taehyeon Park}, \bibinfo{person}{Youngji Koh}, \bibinfo{person}{Jae-Gil Lee}, {and} \bibinfo{person}{Uichin Lee}.} \bibinfo{year}{2025}\natexlab{}.
\newblock \showarticletitle{Leveraging Smartphone Human Interaction Routine Behavior Task Mining and Modeling for Daily Stress Monitoring}.
\newblock \bibinfo{journal}{\emph{Proceedings of the ACM on Interactive, Mobile, Wearable and Ubiquitous Technologies}} \bibinfo{volume}{9}, \bibinfo{number}{4} (\bibinfo{year}{2025}), \bibinfo{pages}{1--45}.
\newblock
\href{https://doi.org/10.1145/3770644}{doi:\nolinkurl{10.1145/3770644}}


\bibitem[Li and Washington(2024)]%
        {schmidt2024personalized}
\bibfield{author}{\bibinfo{person}{Joe Li} {and} \bibinfo{person}{Peter~Y. Washington}.} \bibinfo{year}{2024}\natexlab{}.
\newblock \showarticletitle{A Comparison of Personalized and Generalized Approaches to Emotion Recognition Using Consumer Wearable Devices: Machine Learning Study}.
\newblock \bibinfo{journal}{\emph{JMIR AI}}  \bibinfo{volume}{3} (\bibinfo{year}{2024}), \bibinfo{pages}{e52171}.
\newblock
\href{https://doi.org/10.2196/52171}{doi:\nolinkurl{10.2196/52171}}


\bibitem[Lin et~al\mbox{.}(2025)]%
        {lin2025beyond}
\bibfield{author}{\bibinfo{person}{Xuefei Lin} {et~al\mbox{.}}} \bibinfo{year}{2025}\natexlab{}.
\newblock \showarticletitle{Beyond Detection: Towards Actionable Sensing Research in Clinical Mental Healthcare}.
\newblock \bibinfo{journal}{\emph{Proceedings of the ACM on Interactive, Mobile, Wearable and Ubiquitous Technologies}} (\bibinfo{year}{2025}).
\newblock
\href{https://doi.org/10.1145/3699755}{doi:\nolinkurl{10.1145/3699755}}


\bibitem[Lindstr{\"o}m et~al\mbox{.}(2021)]%
        {Lindstrom2021}
\bibfield{author}{\bibinfo{person}{Bj{\"o}rn Lindstr{\"o}m}, \bibinfo{person}{Martin Bellander}, \bibinfo{person}{David~T. Schultner}, \bibinfo{person}{Allen Chang}, \bibinfo{person}{Philippe~N. Tobler}, {and} \bibinfo{person}{David~M. Amodio}.} \bibinfo{year}{2021}\natexlab{}.
\newblock \showarticletitle{A Computational Reward Learning Account of Social Media Engagement}.
\newblock \bibinfo{journal}{\emph{Nature Communications}} \bibinfo{volume}{12}, \bibinfo{number}{1} (\bibinfo{year}{2021}), \bibinfo{pages}{1311}.
\newblock
\href{https://doi.org/10.1038/s41467-020-19607-x}{doi:\nolinkurl{10.1038/s41467-020-19607-x}}


\bibitem[Lukoff et~al\mbox{.}(2023)]%
        {lukoff_switchtube_2023}
\bibfield{author}{\bibinfo{person}{Kai Lukoff}, \bibinfo{person}{Ulrik Lyngs}, \bibinfo{person}{Karina Shirokova}, \bibinfo{person}{Raveena Rao}, \bibinfo{person}{Larry Tian}, \bibinfo{person}{Himanshu Zade}, \bibinfo{person}{Sean~A. Munson}, {and} \bibinfo{person}{Alexis Hiniker}.} \bibinfo{year}{2023}\natexlab{}.
\newblock \showarticletitle{{SwitchTube}: A Proof-of-Concept System Introducing ``Adaptable Commitment Interfaces'' as a Tool for Digital Wellbeing}. In \bibinfo{booktitle}{\emph{Proceedings of the 2023 CHI Conference on Human Factors in Computing Systems}}.
\newblock
\href{https://doi.org/10.1145/3544548.3580703}{doi:\nolinkurl{10.1145/3544548.3580703}}


\bibitem[Lukoff et~al\mbox{.}(2018)]%
        {lukoff_what_2018}
\bibfield{author}{\bibinfo{person}{Kai Lukoff}, \bibinfo{person}{Cissy Yu}, \bibinfo{person}{Julie Kientz}, {and} \bibinfo{person}{Alexis Hiniker}.} \bibinfo{year}{2018}\natexlab{}.
\newblock \showarticletitle{What {Makes} {Smartphone} {Use} {Meaningful} or {Meaningless}?}
\newblock \bibinfo{journal}{\emph{Proceedings of the ACM on Interactive, Mobile, Wearable and Ubiquitous Technologies}} \bibinfo{volume}{2}, \bibinfo{number}{1} (\bibinfo{date}{March} \bibinfo{year}{2018}), \bibinfo{pages}{1--26}.
\newblock
\showISSN{2474-9567}
\href{https://doi.org/10.1145/3191754}{doi:\nolinkurl{10.1145/3191754}}


\bibitem[Lundberg and Lee(2017)]%
        {lundberg_unified_2017}
\bibfield{author}{\bibinfo{person}{Scott~M. Lundberg} {and} \bibinfo{person}{Su-In Lee}.} \bibinfo{year}{2017}\natexlab{}.
\newblock \showarticletitle{A Unified Approach to Interpreting Model Predictions}. In \bibinfo{booktitle}{\emph{Advances in Neural Information Processing Systems}}, Vol.~\bibinfo{volume}{30}. \bibinfo{pages}{4765--4774}.
\newblock


\bibitem[Lyngs et~al\mbox{.}(2020)]%
        {lyngs_hack_2020}
\bibfield{author}{\bibinfo{person}{Ulrik Lyngs}, \bibinfo{person}{Kai Lukoff}, \bibinfo{person}{Petr Slovak}, \bibinfo{person}{William Seymour}, \bibinfo{person}{Helena Webb}, \bibinfo{person}{Marina Jirotka}, \bibinfo{person}{Jun Zhao}, \bibinfo{person}{Max Van~Kleek}, {and} \bibinfo{person}{Nigel Shadbolt}.} \bibinfo{year}{2020}\natexlab{}.
\newblock \showarticletitle{{`I Just Want to Hack Myself to Not Get Distracted'}: Evaluating Design Interventions for Self-Control on Facebook}. In \bibinfo{booktitle}{\emph{Proceedings of the 2020 CHI Conference on Human Factors in Computing Systems}} \emph{(\bibinfo{series}{CHI '20})}. \bibinfo{publisher}{Association for Computing Machinery}, \bibinfo{address}{New York, NY, USA}, \bibinfo{pages}{1--15}.
\newblock
\href{https://doi.org/10.1145/3313831.3376672}{doi:\nolinkurl{10.1145/3313831.3376672}}


\bibitem[Marino et~al\mbox{.}(2021)]%
        {Marino2021}
\bibfield{author}{\bibinfo{person}{Claudia Marino}, \bibinfo{person}{Natale Canale}, \bibinfo{person}{Fiordalisa Melodia}, \bibinfo{person}{Marcantonio~M. Spada}, {and} \bibinfo{person}{Alessio Vieno}.} \bibinfo{year}{2021}\natexlab{}.
\newblock \showarticletitle{The Overlap Between Problematic Smartphone Use and Problematic Social Media Use: A Systematic Review}.
\newblock \bibinfo{journal}{\emph{Current Addiction Reports}} \bibinfo{volume}{8}, \bibinfo{number}{4} (\bibinfo{year}{2021}), \bibinfo{pages}{469--480}.
\newblock
\href{https://doi.org/10.1007/s40429-021-00398-0}{doi:\nolinkurl{10.1007/s40429-021-00398-0}}


\bibitem[Meegahapola et~al\mbox{.}(2023)]%
        {meegahapola2023generalization}
\bibfield{author}{\bibinfo{person}{Lakmal Meegahapola}, \bibinfo{person}{William Droz}, \bibinfo{person}{Peter Kun}, \bibinfo{person}{Amalia {De Gotzen}}, \bibinfo{person}{Chaitanya Nutakki}, \bibinfo{person}{Shyam Diber}, \bibinfo{person}{George Rubin}, {et~al\mbox{.}}} \bibinfo{year}{2023}\natexlab{}.
\newblock \showarticletitle{Generalization and Personalization of Mobile Sensing-Based Mood Inference Models: An Analysis of College Students in Eight Countries}.
\newblock \bibinfo{journal}{\emph{Proceedings of the ACM on Interactive, Mobile, Wearable and Ubiquitous Technologies}} \bibinfo{volume}{6}, \bibinfo{number}{4} (\bibinfo{year}{2023}).
\newblock
\href{https://doi.org/10.1145/3569483}{doi:\nolinkurl{10.1145/3569483}}


\bibitem[Mehrotra et~al\mbox{.}(2016)]%
        {mehrotra2016myphone}
\bibfield{author}{\bibinfo{person}{Abhinav Mehrotra}, \bibinfo{person}{Veljko Pejovi{\'c}}, \bibinfo{person}{Jo Vermeulen}, \bibinfo{person}{Robert Hendley}, {and} \bibinfo{person}{Mirco Musolesi}.} \bibinfo{year}{2016}\natexlab{}.
\newblock \showarticletitle{My Phone and Me: Understanding People's Receptivity to Mobile Notifications}. In \bibinfo{booktitle}{\emph{Proceedings of the 2016 CHI Conference on Human Factors in Computing Systems (CHI '16)}}. \bibinfo{publisher}{Association for Computing Machinery}, \bibinfo{address}{New York, NY, USA}, \bibinfo{pages}{1021--1032}.
\newblock
\href{https://doi.org/10.1145/2858036.2858566}{doi:\nolinkurl{10.1145/2858036.2858566}}


\bibitem[Mishra et~al\mbox{.}(2021)]%
        {mishra2021receptivity}
\bibfield{author}{\bibinfo{person}{Varun Mishra} {et~al\mbox{.}}} \bibinfo{year}{2021}\natexlab{}.
\newblock \showarticletitle{Detecting Receptivity for mHealth Interventions in the Natural Environment}.
\newblock \bibinfo{journal}{\emph{Proceedings of the ACM on Interactive, Mobile, Wearable and Ubiquitous Technologies}} \bibinfo{volume}{5}, \bibinfo{number}{2} (\bibinfo{year}{2021}).
\newblock


\bibitem[Montag et~al\mbox{.}(2021)]%
        {Montag2021}
\bibfield{author}{\bibinfo{person}{Christian Montag}, \bibinfo{person}{Elisa Wegmann}, \bibinfo{person}{Rayna Sariyska}, \bibinfo{person}{Zsolt Demetrovics}, {and} \bibinfo{person}{Matthias Brand}.} \bibinfo{year}{2021}\natexlab{}.
\newblock \showarticletitle{How to Overcome Taxonomical Problems in the Study of Internet Use Disorders and What to Do with ``Smartphone Addiction''?}
\newblock \bibinfo{journal}{\emph{Journal of Behavioral Addictions}} \bibinfo{volume}{9}, \bibinfo{number}{4} (\bibinfo{year}{2021}), \bibinfo{pages}{908--914}.
\newblock
\href{https://doi.org/10.1556/2006.8.2019.59}{doi:\nolinkurl{10.1556/2006.8.2019.59}}


\bibitem[Morshed et~al\mbox{.}(2019)]%
        {morshed2019mood}
\bibfield{author}{\bibinfo{person}{Mehrab~Bin Morshed}, \bibinfo{person}{Koustuv Saha}, \bibinfo{person}{Richard Li}, \bibinfo{person}{Sidney~K. D'Mello}, \bibinfo{person}{Munmun De~Choudhury}, \bibinfo{person}{Gregory~D. Abowd}, {and} \bibinfo{person}{Thomas Pl{\"o}tz}.} \bibinfo{year}{2019}\natexlab{}.
\newblock \showarticletitle{Prediction of Mood Instability with Passive Sensing}.
\newblock \bibinfo{journal}{\emph{Proceedings of the ACM on Interactive, Mobile, Wearable and Ubiquitous Technologies}} \bibinfo{volume}{3}, \bibinfo{number}{3} (\bibinfo{year}{2019}).
\newblock
\href{https://doi.org/10.1145/3351233}{doi:\nolinkurl{10.1145/3351233}}


\bibitem[Nahum-Shani and Murphy(2025)]%
        {nahum-shani_just--time_2025}
\bibfield{author}{\bibinfo{person}{Inbal Nahum-Shani} {and} \bibinfo{person}{Susan~A. Murphy}.} \bibinfo{year}{2025}\natexlab{}.
\newblock \showarticletitle{Just-in-{Time} {Adaptive} {Interventions}: {Where} {Are} {We} {Now} and {What} {Is} {Next}?}
\newblock \bibinfo{journal}{\emph{Annual Review of Psychology}} (\bibinfo{date}{Sept.} \bibinfo{year}{2025}).
\newblock
\showISSN{0066-4308, 1545-2085}
\href{https://doi.org/10.1146/annurev-psych-121024-044244}{doi:\nolinkurl{10.1146/annurev-psych-121024-044244}}


\bibitem[Nepal et~al\mbox{.}(2024)]%
        {nepal2024capturing}
\bibfield{author}{\bibinfo{person}{Subigya Nepal}, \bibinfo{person}{Weichen Wang}, \bibinfo{person}{Vlado Vojdanovski}, \bibinfo{person}{Jeremy~F. Huckins}, \bibinfo{person}{Alex daSilva}, \bibinfo{person}{Meghan Meyer}, {and} \bibinfo{person}{Andrew Campbell}.} \bibinfo{year}{2024}\natexlab{}.
\newblock \showarticletitle{Capturing the College Experience: A Four-Year Mobile Sensing Study of Mental Health, Resilience and Behavior of College Students during the Pandemic}.
\newblock \bibinfo{journal}{\emph{Proceedings of the ACM on Interactive, Mobile, Wearable and Ubiquitous Technologies}} \bibinfo{volume}{8}, \bibinfo{number}{1} (\bibinfo{year}{2024}), \bibinfo{pages}{1--37}.
\newblock
\href{https://doi.org/10.1145/3643501}{doi:\nolinkurl{10.1145/3643501}}


\bibitem[Orzikulova et~al\mbox{.}(2024)]%
        {orzikulova_time2stop_2024}
\bibfield{author}{\bibinfo{person}{Adiba Orzikulova}, \bibinfo{person}{Han Xiao}, \bibinfo{person}{Zhipeng Li}, \bibinfo{person}{Yukang Yan}, \bibinfo{person}{Yuntao Wang}, \bibinfo{person}{Yuanchun Shi}, \bibinfo{person}{Marzyeh Ghassemi}, \bibinfo{person}{Sung-Ju Lee}, \bibinfo{person}{Anind~K. Dey}, {and} \bibinfo{person}{Xuhai~"Orson" Xu}.} \bibinfo{year}{2024}\natexlab{}.
\newblock \showarticletitle{{Time2Stop}: {Adaptive} and {Explainable} {Human}-{AI} {Loop} for {Smartphone} {Overuse} {Intervention}}. In \bibinfo{booktitle}{\emph{Proceedings of the {CHI} {Conference} on {Human} {Factors} in {Computing} {Systems}}}. \bibinfo{pages}{1--20}.
\newblock
\href{https://doi.org/10.1145/3613904.3642747}{doi:\nolinkurl{10.1145/3613904.3642747}}
\newblock
\shownote{arXiv:2403.05584 [cs]}.


\bibitem[Oulasvirta et~al\mbox{.}(2012)]%
        {Oulasvirta2012}
\bibfield{author}{\bibinfo{person}{Antti Oulasvirta}, \bibinfo{person}{Tye Rattenbury}, \bibinfo{person}{Lingyi Ma}, {and} \bibinfo{person}{Eeva Raita}.} \bibinfo{year}{2012}\natexlab{}.
\newblock \showarticletitle{Habits Make Smartphone Use More Pervasive}.
\newblock \bibinfo{journal}{\emph{Personal and Ubiquitous Computing}} \bibinfo{volume}{16}, \bibinfo{number}{1} (\bibinfo{year}{2012}), \bibinfo{pages}{105--114}.
\newblock
\href{https://doi.org/10.1007/s00779-011-0412-2}{doi:\nolinkurl{10.1007/s00779-011-0412-2}}


\bibitem[Pakhomov et~al\mbox{.}(2020)]%
        {pakhomov2020consumer}
\bibfield{author}{\bibinfo{person}{Serguei V.~S. Pakhomov}, \bibinfo{person}{Paul~D. Thuras}, \bibinfo{person}{Raymond Finzel}, \bibinfo{person}{Jerika Eppel}, {and} \bibinfo{person}{Michael Kotlyar}.} \bibinfo{year}{2020}\natexlab{}.
\newblock \showarticletitle{Using consumer-wearable technology for remote assessment of physiological response to stress in the naturalistic environment}.
\newblock \bibinfo{journal}{\emph{PLOS ONE}} \bibinfo{volume}{15}, \bibinfo{number}{3} (\bibinfo{year}{2020}), \bibinfo{pages}{e0229942}.
\newblock
\href{https://doi.org/10.1371/journal.pone.0229942}{doi:\nolinkurl{10.1371/journal.pone.0229942}}


\bibitem[Park et~al\mbox{.}(2020)]%
        {park2020wellbeat}
\bibfield{author}{\bibinfo{person}{Sungkyu Park}, \bibinfo{person}{Marios Constantinides}, \bibinfo{person}{Luca~Maria Aiello}, \bibinfo{person}{Daniele Quercia}, {and} \bibinfo{person}{Paul van Gent}.} \bibinfo{year}{2020}\natexlab{}.
\newblock \showarticletitle{{WellBeat}: A Framework for Tracking Daily Well-Being Using Smartwatches}.
\newblock \bibinfo{journal}{\emph{IEEE Internet Computing}} \bibinfo{volume}{24}, \bibinfo{number}{5} (\bibinfo{year}{2020}), \bibinfo{pages}{10--17}.
\newblock
\href{https://doi.org/10.1109/mic.2020.3017867}{doi:\nolinkurl{10.1109/mic.2020.3017867}}


\bibitem[Pejovi{\'c} and Musolesi(2014)]%
        {pejovic2014interruptme}
\bibfield{author}{\bibinfo{person}{Veljko Pejovi{\'c}} {and} \bibinfo{person}{Mirco Musolesi}.} \bibinfo{year}{2014}\natexlab{}.
\newblock \showarticletitle{{InterruptMe}: Designing Intelligent Prompting Mechanisms for Pervasive Applications}. In \bibinfo{booktitle}{\emph{Proceedings of the 2014 ACM International Joint Conference on Pervasive and Ubiquitous Computing}}. \bibinfo{pages}{897--908}.
\newblock
\href{https://doi.org/10.1145/2632048.2632062}{doi:\nolinkurl{10.1145/2632048.2632062}}


\bibitem[Picard(1997)]%
        {picard1997affective}
\bibfield{author}{\bibinfo{person}{Rosalind~W. Picard}.} \bibinfo{year}{1997}\natexlab{}.
\newblock \bibinfo{booktitle}{\emph{Affective Computing}}.
\newblock \bibinfo{publisher}{MIT Press}, \bibinfo{address}{Cambridge, MA}.
\newblock


\bibitem[Picard et~al\mbox{.}(2001)]%
        {picard2001toward}
\bibfield{author}{\bibinfo{person}{Rosalind~W. Picard}, \bibinfo{person}{Elias Vyzas}, {and} \bibinfo{person}{Jennifer Healey}.} \bibinfo{year}{2001}\natexlab{}.
\newblock \showarticletitle{Toward Machine Emotional Intelligence: Analysis of Affective Physiological State}.
\newblock \bibinfo{journal}{\emph{IEEE Transactions on Pattern Analysis and Machine Intelligence}} \bibinfo{volume}{23}, \bibinfo{number}{10} (\bibinfo{year}{2001}), \bibinfo{pages}{1175--1191}.
\newblock
\href{https://doi.org/10.1109/34.954607}{doi:\nolinkurl{10.1109/34.954607}}


\bibitem[Quiroz et~al\mbox{.}(2018)]%
        {quiroz2018emotion}
\bibfield{author}{\bibinfo{person}{Juan~Carlos Quiroz}, \bibinfo{person}{Elena Geangu}, {and} \bibinfo{person}{Min~Hooi Yong}.} \bibinfo{year}{2018}\natexlab{}.
\newblock \showarticletitle{Emotion recognition using smart watch sensor data: Mixed-design study}.
\newblock \bibinfo{journal}{\emph{JMIR Mental Health}} \bibinfo{volume}{5}, \bibinfo{number}{3} (\bibinfo{year}{2018}), \bibinfo{pages}{e10153}.
\newblock
\href{https://doi.org/10.2196/10153}{doi:\nolinkurl{10.2196/10153}}


\bibitem[Rashid et~al\mbox{.}(2024)]%
        {rashid2024reproducible}
\bibfield{author}{\bibinfo{person}{Md~Mafijul~Islam Rashid} {et~al\mbox{.}}} \bibinfo{year}{2024}\natexlab{}.
\newblock \showarticletitle{A Reproducible Stress Prediction Pipeline with Mobile Sensor Data}.
\newblock \bibinfo{journal}{\emph{Proceedings of the ACM on Interactive, Mobile, Wearable and Ubiquitous Technologies}} \bibinfo{volume}{8}, \bibinfo{number}{3} (\bibinfo{year}{2024}).
\newblock
\href{https://doi.org/10.1145/3678578}{doi:\nolinkurl{10.1145/3678578}}


\bibitem[Reinecke and Hofmann(2016)]%
        {Reinecke2016}
\bibfield{author}{\bibinfo{person}{Leonard Reinecke} {and} \bibinfo{person}{Wilhelm Hofmann}.} \bibinfo{year}{2016}\natexlab{}.
\newblock \showarticletitle{Slacking Off or Winding Down? An Experience Sampling Study on the Drivers and Consequences of Media Use for Recovery versus Procrastination}.
\newblock \bibinfo{journal}{\emph{Human Communication Research}} \bibinfo{volume}{42}, \bibinfo{number}{3} (\bibinfo{year}{2016}), \bibinfo{pages}{441--461}.
\newblock
\href{https://doi.org/10.1111/hcre.12082}{doi:\nolinkurl{10.1111/hcre.12082}}


\bibitem[Rixen et~al\mbox{.}(2023)]%
        {rixen2023loop}
\bibfield{author}{\bibinfo{person}{Jan~Ole Rixen}, \bibinfo{person}{Luca-Maxim Meinhardt}, \bibinfo{person}{Michael Gl{\"o}ckler}, \bibinfo{person}{Marius-Lukas Ziegenbein}, \bibinfo{person}{Anna Schlothauer}, \bibinfo{person}{Mark Colley}, \bibinfo{person}{Enrico Rukzio}, {and} \bibinfo{person}{Jan Gugenheimer}.} \bibinfo{year}{2023}\natexlab{}.
\newblock \showarticletitle{The Loop and Reasons to Break It: Investigating Infinite Scrolling Behaviour in Social Media Applications and Reasons to Stop}.
\newblock \bibinfo{journal}{\emph{Proceedings of the ACM on Human-Computer Interaction}} \bibinfo{volume}{7}, \bibinfo{number}{MHCI}, Article \bibinfo{articleno}{228} (\bibinfo{date}{Sept.} \bibinfo{year}{2023}), \bibinfo{numpages}{22}~pages.
\newblock
\href{https://doi.org/10.1145/3604275}{doi:\nolinkurl{10.1145/3604275}}


\bibitem[Robinson and Berridge(2025)]%
        {Robinson2025}
\bibfield{author}{\bibinfo{person}{Terry~E. Robinson} {and} \bibinfo{person}{Kent~C. Berridge}.} \bibinfo{year}{2025}\natexlab{}.
\newblock \showarticletitle{The Incentive-Sensitization Theory of Addiction 30 Years On}.
\newblock \bibinfo{journal}{\emph{Annual Review of Psychology}}  \bibinfo{volume}{76} (\bibinfo{year}{2025}), \bibinfo{pages}{29--58}.
\newblock
\href{https://doi.org/10.1146/annurev-psych-011624-024031}{doi:\nolinkurl{10.1146/annurev-psych-011624-024031}}


\bibitem[Roffarello and De~Russis(2023)]%
        {roffarello_achieving_2023}
\bibfield{author}{\bibinfo{person}{Alberto~Monge Roffarello} {and} \bibinfo{person}{Luigi De~Russis}.} \bibinfo{year}{2023}\natexlab{}.
\newblock \showarticletitle{Achieving {Digital} {Wellbeing} {Through} {Digital} {Self}-control {Tools}: {A} {Systematic} {Review} and {Meta}-analysis}.
\newblock \bibinfo{journal}{\emph{ACM Transactions on Computer-Human Interaction}} \bibinfo{volume}{30}, \bibinfo{number}{4} (\bibinfo{date}{Aug.} \bibinfo{year}{2023}), \bibinfo{pages}{1--66}.
\newblock
\showISSN{1073-0516, 1557-7325}
\href{https://doi.org/10.1145/3571810}{doi:\nolinkurl{10.1145/3571810}}


\bibitem[Ruensuk et~al\mbox{.}(2020)]%
        {ruensuk_how_2020}
\bibfield{author}{\bibinfo{person}{Mintra Ruensuk}, \bibinfo{person}{Eunyong Cheon}, \bibinfo{person}{Hwajung Hong}, {and} \bibinfo{person}{Ian Oakley}.} \bibinfo{year}{2020}\natexlab{}.
\newblock \showarticletitle{How {Do} {You} {Feel} {Online}: {Exploiting} {Smartphone} {Sensors} to {Detect} {Transitory} {Emotions} during {Social} {Media} {Use}}.
\newblock \bibinfo{journal}{\emph{Proceedings of the ACM on Interactive, Mobile, Wearable and Ubiquitous Technologies}} \bibinfo{volume}{4}, \bibinfo{number}{4} (\bibinfo{date}{Dec.} \bibinfo{year}{2020}), \bibinfo{pages}{1--32}.
\newblock
\showISSN{2474-9567}
\href{https://doi.org/10.1145/3432223}{doi:\nolinkurl{10.1145/3432223}}


\bibitem[Russell(1996)]%
        {russell_ucla_1996}
\bibfield{author}{\bibinfo{person}{Daniel~W. Russell}.} \bibinfo{year}{1996}\natexlab{}.
\newblock \showarticletitle{{UCLA} {Loneliness} {Scale} ({Version} 3): {Reliability}, {Validity}, and {Factor} {Structure}}.
\newblock \bibinfo{journal}{\emph{Journal of Personality Assessment}} \bibinfo{volume}{66}, \bibinfo{number}{1} (\bibinfo{date}{Feb.} \bibinfo{year}{1996}), \bibinfo{pages}{20--40}.
\newblock
\showISSN{0022-3891, 1532-7752}
\href{https://doi.org/10.1207/s15327752jpa6601_2}{doi:\nolinkurl{10.1207/s15327752jpa6601_2}}


\bibitem[Saganowski et~al\mbox{.}(2020)]%
        {saganowski2020consumer}
\bibfield{author}{\bibinfo{person}{Stanislaw Saganowski}, \bibinfo{person}{Przemyslaw Kazienko}, \bibinfo{person}{Maciej Dziezyc}, \bibinfo{person}{Patrycja Jakimow}, \bibinfo{person}{Joanna Komoszynska}, \bibinfo{person}{Weronika Michalska}, \bibinfo{person}{Anna Dutkowiak}, \bibinfo{person}{Adam Polak}, \bibinfo{person}{Adam Dziadek}, {and} \bibinfo{person}{Michal Ujma}.} \bibinfo{year}{2020}\natexlab{}.
\newblock \showarticletitle{Consumer Wearables and Affective Computing for Wellbeing Support}. In \bibinfo{booktitle}{\emph{Proceedings of the 17th EAI International Conference on Mobile and Ubiquitous Systems (MobiQuitous)}}. \bibinfo{pages}{482--487}.
\newblock
\href{https://doi.org/10.1145/3448891.3450332}{doi:\nolinkurl{10.1145/3448891.3450332}}


\bibitem[Saganowski et~al\mbox{.}(2023)]%
        {saganowski2022emotion}
\bibfield{author}{\bibinfo{person}{Stanislaw Saganowski}, \bibinfo{person}{Bartosz Perz}, \bibinfo{person}{Adam~G. Polak}, {and} \bibinfo{person}{Przemyslaw Kazienko}.} \bibinfo{year}{2023}\natexlab{}.
\newblock \showarticletitle{Emotion Recognition for Everyday Life Using Physiological Signals From Wearables: A Systematic Literature Review}.
\newblock \bibinfo{journal}{\emph{IEEE Transactions on Affective Computing}} \bibinfo{volume}{14}, \bibinfo{number}{3} (\bibinfo{year}{2023}), \bibinfo{pages}{1876--1897}.
\newblock
\href{https://doi.org/10.1109/TAFFC.2022.3176135}{doi:\nolinkurl{10.1109/TAFFC.2022.3176135}}


\bibitem[Sano et~al\mbox{.}(2018)]%
        {sano2018identifying}
\bibfield{author}{\bibinfo{person}{Akane Sano}, \bibinfo{person}{Sara Taylor}, \bibinfo{person}{Andrew~W. McHill}, \bibinfo{person}{Andrew J.~K. Phillips}, \bibinfo{person}{Laura~K. Barger}, \bibinfo{person}{Elizabeth Klerman}, {and} \bibinfo{person}{Rosalind Picard}.} \bibinfo{year}{2018}\natexlab{}.
\newblock \showarticletitle{Identifying Objective Physiological Markers and Modifiable Behaviors for Self-Reported Stress and Mental Health Status Using Wearable Sensors and Mobile Phones: Observational Study}.
\newblock \bibinfo{journal}{\emph{Journal of Medical Internet Research}} \bibinfo{volume}{20}, \bibinfo{number}{6} (\bibinfo{year}{2018}), \bibinfo{pages}{e210}.
\newblock
\href{https://doi.org/10.2196/jmir.9410}{doi:\nolinkurl{10.2196/jmir.9410}}


\bibitem[Sarsenbayeva et~al\mbox{.}(2020)]%
        {sarsenbayeva2020smartphone}
\bibfield{author}{\bibinfo{person}{Zhanna Sarsenbayeva}, \bibinfo{person}{Gabriele Marini}, \bibinfo{person}{Niels van Berkel}, \bibinfo{person}{Chu Luo}, \bibinfo{person}{Weiwei Jiang}, \bibinfo{person}{Kangning Yang}, \bibinfo{person}{Greg Wadley}, \bibinfo{person}{Tilman Dingler}, \bibinfo{person}{Vassilis Kostakos}, {and} \bibinfo{person}{Jorge Goncalves}.} \bibinfo{year}{2020}\natexlab{}.
\newblock \showarticletitle{Does smartphone use drive our emotions or vice versa? {A} causal analysis}. In \bibinfo{booktitle}{\emph{Proceedings of the 2020 {CHI} Conference on Human Factors in Computing Systems}}. \bibinfo{publisher}{ACM}, \bibinfo{pages}{1--15}.
\newblock
\href{https://doi.org/10.1145/3313831.3376163}{doi:\nolinkurl{10.1145/3313831.3376163}}


\bibitem[Saylam and {\.I}ncel(2023)]%
        {saylam2023quantifying}
\bibfield{author}{\bibinfo{person}{Berrenur Saylam} {and} \bibinfo{person}{{\"O}zlem~Durmaz {\.I}ncel}.} \bibinfo{year}{2023}\natexlab{}.
\newblock \showarticletitle{Quantifying Digital Biomarkers for Well-Being: Stress, Anxiety, Positive and Negative Affect via Wearable Devices and Their Time-Based Predictions}.
\newblock \bibinfo{journal}{\emph{Sensors}} \bibinfo{volume}{23}, \bibinfo{number}{21} (\bibinfo{year}{2023}), \bibinfo{pages}{8987}.
\newblock
\href{https://doi.org/10.3390/s23218987}{doi:\nolinkurl{10.3390/s23218987}}


\bibitem[Schmidt et~al\mbox{.}(2018)]%
        {schmidt2018wesad}
\bibfield{author}{\bibinfo{person}{Philip Schmidt}, \bibinfo{person}{Attila Reiss}, \bibinfo{person}{Robert Duerichen}, \bibinfo{person}{Claus Marberger}, {and} \bibinfo{person}{Kristof Van~Laerhoven}.} \bibinfo{year}{2018}\natexlab{}.
\newblock \showarticletitle{Introducing {WESAD}, a Multimodal Dataset for Wearable Stress and Affect Detection}. In \bibinfo{booktitle}{\emph{Proceedings of the 20th ACM International Conference on Multimodal Interaction (ICMI '18)}}. \bibinfo{publisher}{Association for Computing Machinery}, \bibinfo{pages}{400--408}.
\newblock
\href{https://doi.org/10.1145/3242969.3242985}{doi:\nolinkurl{10.1145/3242969.3242985}}


\bibitem[Scholz et~al\mbox{.}(2022)]%
        {Scholz2022}
\bibfield{author}{\bibinfo{person}{Christin Scholz}, \bibinfo{person}{Hang-Yee Chan}, \bibinfo{person}{Russell~A. Poldrack}, \bibinfo{person}{Denise T.~D. de Ridder}, \bibinfo{person}{Ale Smidts}, {and} \bibinfo{person}{Laura~Nynke van~der Laan}.} \bibinfo{year}{2022}\natexlab{}.
\newblock \showarticletitle{Can We Have a Second Helping? A Preregistered Direct Replication Study on the Neurobiological Mechanisms Underlying Self-Control}.
\newblock \bibinfo{journal}{\emph{Human Brain Mapping}} \bibinfo{volume}{43}, \bibinfo{number}{16} (\bibinfo{year}{2022}), \bibinfo{pages}{4995--5016}.
\newblock
\href{https://doi.org/10.1002/hbm.26065}{doi:\nolinkurl{10.1002/hbm.26065}}


\bibitem[{scikit-learn developers}({[n.\,d.]})]%
        {sklearn_rfecv}
\bibfield{author}{\bibinfo{person}{{scikit-learn developers}}.} \bibinfo{year}{[n.\,d.]}\natexlab{}.
\newblock \bibinfo{title}{{RFECV}: Recursive Feature Elimination with Cross-Validation to Select Features}.
\newblock \bibinfo{howpublished}{\url{https://scikit-learn.org/stable/modules/generated/sklearn.feature_selection.RFECV.html}}.
\newblock
\newblock
\shownote{Accessed: 2026-04-28}.


\bibitem[Shah et~al\mbox{.}(2021)]%
        {shah2021personalized}
\bibfield{author}{\bibinfo{person}{Rutvik~V. Shah}, \bibinfo{person}{Gillian Grennan}, \bibinfo{person}{Mariam Zafar-Khan}, \bibinfo{person}{Fahad Alim}, \bibinfo{person}{Sujit Dey}, \bibinfo{person}{Dhakshin Ramanathan}, {and} \bibinfo{person}{Jyoti Mishra}.} \bibinfo{year}{2021}\natexlab{}.
\newblock \showarticletitle{Personalized machine learning of depressed mood using wearables}.
\newblock \bibinfo{journal}{\emph{Translational Psychiatry}} \bibinfo{volume}{11}, \bibinfo{number}{1} (\bibinfo{year}{2021}), \bibinfo{pages}{338}.
\newblock
\href{https://doi.org/10.1038/s41398-021-01445-0}{doi:\nolinkurl{10.1038/s41398-021-01445-0}}


\bibitem[Shahu et~al\mbox{.}(2024)]%
        {shahu2024beyond}
\bibfield{author}{\bibinfo{person}{Ambika Shahu}, \bibinfo{person}{Fabian Pechstein}, {and} \bibinfo{person}{Florian Michahelles}.} \bibinfo{year}{2024}\natexlab{}.
\newblock \showarticletitle{Beyond Screen Time: Exploring Smartwatch Interventions for Digital Well-Being}. In \bibinfo{booktitle}{\emph{Proceedings of Mensch und Computer 2024}}. \bibinfo{pages}{83--98}.
\newblock
\href{https://doi.org/10.1145/3670653.3670674}{doi:\nolinkurl{10.1145/3670653.3670674}}


\bibitem[Shannon et~al\mbox{.}(2022)]%
        {Shannon2022}
\bibfield{author}{\bibinfo{person}{Holly Shannon}, \bibinfo{person}{Katie Bush}, \bibinfo{person}{Paul~J. Villeneuve}, \bibinfo{person}{Kim G.~C. Hellemans}, {and} \bibinfo{person}{Synthia Guimond}.} \bibinfo{year}{2022}\natexlab{}.
\newblock \showarticletitle{Problematic Social Media Use in Adolescents and Young Adults: Systematic Review and Meta-Analysis}.
\newblock \bibinfo{journal}{\emph{JMIR Mental Health}} \bibinfo{volume}{9}, \bibinfo{number}{4} (\bibinfo{year}{2022}), \bibinfo{pages}{e33450}.
\newblock
\href{https://doi.org/10.2196/33450}{doi:\nolinkurl{10.2196/33450}}


\bibitem[Shannon et~al\mbox{.}(2025)]%
        {Shannon2025}
\bibfield{author}{\bibinfo{person}{Holly Shannon}, \bibinfo{person}{Matteo Montgomery}, \bibinfo{person}{Alison Funk}, \bibinfo{person}{Alireza Kamyabi}, \bibinfo{person}{Madison Hunt}, \bibinfo{person}{Ceinwen Pope}, \bibinfo{person}{Kim Hellemans}, {and} \bibinfo{person}{Synthia Guimond}.} \bibinfo{year}{2025}\natexlab{}.
\newblock \showarticletitle{Beyond Problematic Social Media Use and the Brain: A Public Health and Policy Perspective}.
\newblock \bibinfo{journal}{\emph{Annals of the New York Academy of Sciences}} \bibinfo{volume}{1550}, \bibinfo{number}{1} (\bibinfo{year}{2025}), \bibinfo{pages}{14--22}.
\newblock
\href{https://doi.org/10.1111/nyas.15409}{doi:\nolinkurl{10.1111/nyas.15409}}


\bibitem[Siirtola et~al\mbox{.}(2023)]%
        {siirtola2023wesad}
\bibfield{author}{\bibinfo{person}{Pekka Siirtola}, \bibinfo{person}{Satu Tamminen}, \bibinfo{person}{Gunjan Chandra}, \bibinfo{person}{Anusha Ihalapathirana}, {and} \bibinfo{person}{Juha Röning}.} \bibinfo{year}{2023}\natexlab{}.
\newblock \showarticletitle{Predicting Emotion with Biosignals: A Comparison of Classification and Regression Models for Estimating Valence and Arousal Level Using Wearable Sensors}.
\newblock \bibinfo{journal}{\emph{Sensors}} \bibinfo{volume}{23}, \bibinfo{number}{3} (\bibinfo{year}{2023}), \bibinfo{pages}{1598}.
\newblock
\href{https://doi.org/10.3390/s23031598}{doi:\nolinkurl{10.3390/s23031598}}


\bibitem[Spitzer et~al\mbox{.}(2006)]%
        {spitzer_brief_2006}
\bibfield{author}{\bibinfo{person}{Robert~L. Spitzer}, \bibinfo{person}{Kurt Kroenke}, \bibinfo{person}{Janet B.~W. Williams}, {and} \bibinfo{person}{Bernd Löwe}.} \bibinfo{year}{2006}\natexlab{}.
\newblock \showarticletitle{A {Brief} {Measure} for {Assessing} {Generalized} {Anxiety} {Disorder}: {The} {GAD}-7}.
\newblock \bibinfo{journal}{\emph{Archives of Internal Medicine}} \bibinfo{volume}{166}, \bibinfo{number}{10} (\bibinfo{date}{May} \bibinfo{year}{2006}), \bibinfo{pages}{1092}.
\newblock
\showISSN{0003-9926}
\href{https://doi.org/10.1001/archinte.166.10.1092}{doi:\nolinkurl{10.1001/archinte.166.10.1092}}


\bibitem[Sramek et~al\mbox{.}(2025)]%
        {sramek_beyond_2025}
\bibfield{author}{\bibinfo{person}{Zefan Sramek}, \bibinfo{person}{Sachinthya Lokuge}, \bibinfo{person}{Tia Sternat}, \bibinfo{person}{Martin~A. Katzman}, {and} \bibinfo{person}{Koji Yatani}.} \bibinfo{year}{2025}\natexlab{}.
\newblock \showarticletitle{Beyond the Feature Level: A Cluster Analysis of Feature-Level Social Media Behaviour Patterns, Maladaptive Use, and Psychological Well-Being}.
\newblock \bibinfo{journal}{\emph{Proceedings of the ACM on Interactive, Mobile, Wearable and Ubiquitous Technologies}} \bibinfo{volume}{9}, \bibinfo{number}{4} (\bibinfo{year}{2025}), \bibinfo{pages}{1--42}.
\newblock
\href{https://doi.org/10.1145/3770713}{doi:\nolinkurl{10.1145/3770713}}


\bibitem[Sultana et~al\mbox{.}(2020)]%
        {sultana2020using}
\bibfield{author}{\bibinfo{person}{Madeena Sultana}, \bibinfo{person}{Majed Al-Jefri}, {and} \bibinfo{person}{Joon Lee}.} \bibinfo{year}{2020}\natexlab{}.
\newblock \showarticletitle{Using Machine Learning and Smartphone and Smartwatch Data to Detect Emotional States and Transitions: Exploratory Study}.
\newblock \bibinfo{journal}{\emph{JMIR mHealth and uHealth}} \bibinfo{volume}{8}, \bibinfo{number}{9} (\bibinfo{year}{2020}), \bibinfo{pages}{e17818}.
\newblock
\href{https://doi.org/10.2196/17818}{doi:\nolinkurl{10.2196/17818}}


\bibitem[Tangney et~al\mbox{.}(2004)]%
        {tangney_high_2004}
\bibfield{author}{\bibinfo{person}{June~P. Tangney}, \bibinfo{person}{Roy~F. Baumeister}, {and} \bibinfo{person}{Angie~Luzio Boone}.} \bibinfo{year}{2004}\natexlab{}.
\newblock \showarticletitle{High self-control predicts good adjustment, less pathology, better grades, and interpersonal success}.
\newblock \bibinfo{journal}{\emph{Journal of Personality}} \bibinfo{volume}{72}, \bibinfo{number}{2} (\bibinfo{date}{April} \bibinfo{year}{2004}), \bibinfo{pages}{271--324}.
\newblock
\showISSN{0022-3506, 1467-6494}
\href{https://doi.org/10.1111/j.0022-3506.2004.00263.x}{doi:\nolinkurl{10.1111/j.0022-3506.2004.00263.x}}


\bibitem[Taylor et~al\mbox{.}(2020)]%
        {taylor2017personalized}
\bibfield{author}{\bibinfo{person}{Sara Taylor}, \bibinfo{person}{Natasha Jaques}, \bibinfo{person}{Ehimwenma Nosakhare}, \bibinfo{person}{Akane Sano}, {and} \bibinfo{person}{Rosalind Picard}.} \bibinfo{year}{2020}\natexlab{}.
\newblock \showarticletitle{Personalized Multitask Learning for Predicting Tomorrow's Mood, Stress, and Health}.
\newblock \bibinfo{journal}{\emph{IEEE Transactions on Affective Computing}} \bibinfo{volume}{11}, \bibinfo{number}{2} (\bibinfo{year}{2020}), \bibinfo{pages}{200--213}.
\newblock
\href{https://doi.org/10.1109/TAFFC.2017.2784832}{doi:\nolinkurl{10.1109/TAFFC.2017.2784832}}


\bibitem[Toshnazarov et~al\mbox{.}(2024)]%
        {toshnazarov2024sosw}
\bibfield{author}{\bibinfo{person}{Kobiljon Toshnazarov} {et~al\mbox{.}}} \bibinfo{year}{2024}\natexlab{}.
\newblock \showarticletitle{{SOSW}: Stress Sensing With Off-the-Shelf Smartwatches in the Wild}. In \bibinfo{booktitle}{\emph{Proceedings of the 2024 ACM International Joint Conference on Pervasive and Ubiquitous Computing}}.
\newblock


\bibitem[Troll et~al\mbox{.}(2021)]%
        {Troll2021}
\bibfield{author}{\bibinfo{person}{Eve~Sarah Troll}, \bibinfo{person}{Malte Friese}, {and} \bibinfo{person}{David~D. Loschelder}.} \bibinfo{year}{2021}\natexlab{}.
\newblock \showarticletitle{How Students' Self-Control and Smartphone-Use Explain Their Academic Performance}.
\newblock \bibinfo{journal}{\emph{Computers in Human Behavior}}  \bibinfo{volume}{117} (\bibinfo{year}{2021}), \bibinfo{pages}{106624}.
\newblock
\href{https://doi.org/10.1016/j.chb.2020.106624}{doi:\nolinkurl{10.1016/j.chb.2020.106624}}


\bibitem[Tutunji et~al\mbox{.}(2023)]%
        {tutunji2023detecting}
\bibfield{author}{\bibinfo{person}{Rayyan Tutunji}, \bibinfo{person}{Nikos Kogias}, \bibinfo{person}{Bob Kapteijns}, \bibinfo{person}{Martin Krentz}, \bibinfo{person}{Florian Krause}, \bibinfo{person}{Eliana Vassena}, {and} \bibinfo{person}{Erno~J. Hermans}.} \bibinfo{year}{2023}\natexlab{}.
\newblock \showarticletitle{Detecting prolonged stress in real life using wearable biosensors and ecological momentary assessments: Naturalistic experimental study}.
\newblock \bibinfo{journal}{\emph{Journal of Medical Internet Research}}  \bibinfo{volume}{25} (\bibinfo{year}{2023}), \bibinfo{pages}{e39995}.
\newblock
\href{https://doi.org/10.2196/39995}{doi:\nolinkurl{10.2196/39995}}


\bibitem[Umematsu et~al\mbox{.}(2020)]%
        {umematsu2020emotional}
\bibfield{author}{\bibinfo{person}{Terumi Umematsu}, \bibinfo{person}{Akane Sano}, \bibinfo{person}{Sara Taylor}, \bibinfo{person}{Masanori Tsujikawa}, {and} \bibinfo{person}{Rosalind~W. Picard}.} \bibinfo{year}{2020}\natexlab{}.
\newblock \showarticletitle{Forecasting Stress, Mood, and Health from Daytime Physiology in Office Workers and Students}. In \bibinfo{booktitle}{\emph{42nd Annual International Conference of the IEEE Engineering in Medicine and Biology Society (EMBC)}}. \bibinfo{publisher}{IEEE}, \bibinfo{pages}{5953--5957}.
\newblock
\href{https://doi.org/10.1109/EMBC44109.2020.9176706}{doi:\nolinkurl{10.1109/EMBC44109.2020.9176706}}


\bibitem[Wang et~al\mbox{.}(2014)]%
        {wang2014studentlife}
\bibfield{author}{\bibinfo{person}{Rui Wang}, \bibinfo{person}{Fanglin Chen}, \bibinfo{person}{Zhenyu Chen}, \bibinfo{person}{Tianxing Li}, \bibinfo{person}{Gabriella Harari}, \bibinfo{person}{Stefanie Tignor}, \bibinfo{person}{Xia Zhou}, \bibinfo{person}{Dror Ben-Zeev}, {and} \bibinfo{person}{Andrew~T. Campbell}.} \bibinfo{year}{2014}\natexlab{}.
\newblock \showarticletitle{{StudentLife}: Assessing Mental Health, Academic Performance and Behavioral Trends of College Students using Smartphones}.
\newblock  (\bibinfo{year}{2014}), \bibinfo{pages}{3--14}.
\newblock
\href{https://doi.org/10.1145/2632048.2632054}{doi:\nolinkurl{10.1145/2632048.2632054}}


\bibitem[Wang et~al\mbox{.}(2024)]%
        {wang2024college}
\bibfield{author}{\bibinfo{person}{Weichen Wang} {et~al\mbox{.}}} \bibinfo{year}{2024}\natexlab{}.
\newblock \showarticletitle{Capturing the College Experience: A Four-Year Mobile Sensing Study of Mental Health, Resilience and Behavior of College Students during the Pandemic}.
\newblock \bibinfo{journal}{\emph{Proceedings of the ACM on Interactive, Mobile, Wearable and Ubiquitous Technologies}} \bibinfo{volume}{8}, \bibinfo{number}{1} (\bibinfo{year}{2024}).
\newblock
\href{https://doi.org/10.1145/3643501}{doi:\nolinkurl{10.1145/3643501}}


\bibitem[Wood and R{\"u}nger(2016)]%
        {Wood2016}
\bibfield{author}{\bibinfo{person}{Wendy Wood} {and} \bibinfo{person}{Dennis R{\"u}nger}.} \bibinfo{year}{2016}\natexlab{}.
\newblock \showarticletitle{Psychology of Habit}.
\newblock \bibinfo{journal}{\emph{Annual Review of Psychology}} \bibinfo{volume}{67}, \bibinfo{number}{1} (\bibinfo{year}{2016}), \bibinfo{pages}{289--314}.
\newblock
\href{https://doi.org/10.1146/annurev-psych-122414-033417}{doi:\nolinkurl{10.1146/annurev-psych-122414-033417}}


\bibitem[Xu et~al\mbox{.}(2023)]%
        {xu2023globem}
\bibfield{author}{\bibinfo{person}{Xuhai Xu} {et~al\mbox{.}}} \bibinfo{year}{2023}\natexlab{}.
\newblock \showarticletitle{{GLOBEM}: Cross-Dataset Generalization of Longitudinal Human Behavior Modeling}.
\newblock \bibinfo{journal}{\emph{Proceedings of the ACM on Interactive, Mobile, Wearable and Ubiquitous Technologies}} \bibinfo{volume}{6}, \bibinfo{number}{4} (\bibinfo{year}{2023}).
\newblock


\bibitem[Yan et~al\mbox{.}(2020)]%
        {yan2020affect}
\bibfield{author}{\bibinfo{person}{Shen Yan}, \bibinfo{person}{Homa Hosseinmardi}, \bibinfo{person}{Hsien-Te Kao}, \bibinfo{person}{Shrikanth Narayanan}, \bibinfo{person}{Kristina Lerman}, {and} \bibinfo{person}{Emilio Ferrara}.} \bibinfo{year}{2020}\natexlab{}.
\newblock \showarticletitle{Affect Estimation with Wearable Sensors}.
\newblock \bibinfo{journal}{\emph{Journal of Healthcare Informatics Research}} \bibinfo{volume}{4}, \bibinfo{number}{3} (\bibinfo{year}{2020}), \bibinfo{pages}{261--294}.
\newblock
\href{https://doi.org/10.1007/s41666-019-00066-z}{doi:\nolinkurl{10.1007/s41666-019-00066-z}}


\bibitem[Yang et~al\mbox{.}(2024)]%
        {yang2024time}
\bibfield{author}{\bibinfo{person}{Yi Yang}, \bibinfo{person}{Ru-De Liu}, \bibinfo{person}{Yi Ding}, \bibinfo{person}{Jingmin Lin}, \bibinfo{person}{Zien Ding}, {and} \bibinfo{person}{Xiantong Yang}.} \bibinfo{year}{2024}\natexlab{}.
\newblock \showarticletitle{Time distortion for short-form video users}.
\newblock \bibinfo{journal}{\emph{Computers in Human Behavior}}  \bibinfo{volume}{151} (\bibinfo{year}{2024}), \bibinfo{pages}{108009}.
\newblock
\href{https://doi.org/10.1016/j.chb.2023.108009}{doi:\nolinkurl{10.1016/j.chb.2023.108009}}


\bibitem[Yu and Sano(2023)]%
        {yu2023semisupervised}
\bibfield{author}{\bibinfo{person}{Han Yu} {and} \bibinfo{person}{Akane Sano}.} \bibinfo{year}{2023}\natexlab{}.
\newblock \showarticletitle{Semi-Supervised Learning for Wearable-based Momentary Stress Detection in the Wild}.
\newblock \bibinfo{journal}{\emph{Proceedings of the ACM on Interactive, Mobile, Wearable and Ubiquitous Technologies}} \bibinfo{volume}{7}, \bibinfo{number}{2} (\bibinfo{year}{2023}).
\newblock
\href{https://doi.org/10.1145/3596246}{doi:\nolinkurl{10.1145/3596246}}


\bibitem[Zeelenberg(1999)]%
        {zeelenberg_anticipated_1999}
\bibfield{author}{\bibinfo{person}{Marcel Zeelenberg}.} \bibinfo{year}{1999}\natexlab{}.
\newblock \showarticletitle{Anticipated Regret, Expected Feedback and Behavioral Decision Making}.
\newblock \bibinfo{journal}{\emph{Journal of Behavioral Decision Making}} \bibinfo{volume}{12}, \bibinfo{number}{2} (\bibinfo{year}{1999}), \bibinfo{pages}{93--106}.
\newblock
\href{https://doi.org/10.1002/(SICI)1099-0771(199906)12:2<93::AID-BDM311>3.0.CO;2-S}{doi:\nolinkurl{10.1002/(SICI)1099-0771(199906)12:2<93::AID-BDM311>3.0.CO;2-S}}


\end{thebibliography}

\clearpage
\appendix
\title{Appendix}
\section{Participant Exclusion Flow}
\label{sec:exclusion-flow}

\begin{figure}[h]
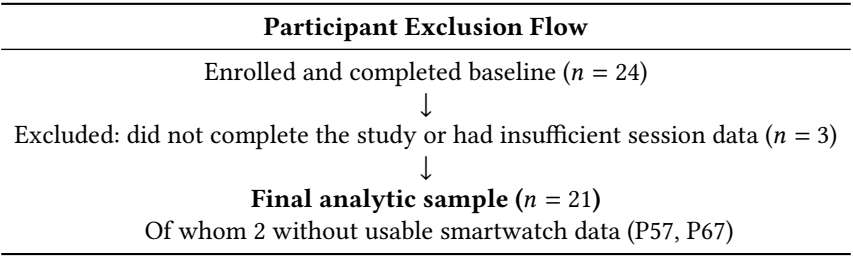

\centering
\begin{tabular}{c}
\toprule
\textbf{Participant Exclusion Flow} \\
\midrule
Enrolled and completed baseline ($n = 24$) \\
$\downarrow$ \\
Excluded: did not complete the study or had insufficient session data ($n = 3$) \\
$\downarrow$ \\
\textbf{Final analytic sample ($n = 21$)} \\
\quad Of whom 2 without usable smartwatch data (P57, P67) \\
\bottomrule
\end{tabular}
\caption{Participant exclusion flow from enrollment to the final analytic sample.}
\label{fig:exclusion-flow}
\end{figure}

\section{Participant Demographics}
\label{sec:demographics-table}

\begin{table}[h]
\caption{Participant demographics and baseline psychometric scores ($N = 21$). The sample comprised 11 female and 10 male participants, predominantly undergraduate students (age range 18--33, $Mdn = 20$).}
\label{tab:demographics}
\begin{tabular}{lcccc}
\toprule
\textbf{Measure} & \textbf{$M$} & \textbf{$SD$} & \textbf{Min} & \textbf{Max} \\
\midrule
Age (years)  & 21.8 & 4.5 & 18 & 33 \\
\midrule
BSMAS (Social Media Addiction) & 20.0 & 3.2 & 13 & 25 \\
PHQ-9 (Depression)             & 7.1  & 4.5 & 2  & 20 \\
GAD-7 (Anxiety)                & 5.3  & 5.0 & 0  & 18 \\
UCLA Loneliness                & 39.7 & 11.2 & 26 & 66 \\
BSCS (Self-Control)            & 34.5 & 6.2 & 22 & 46 \\
\bottomrule
\end{tabular}
\end{table}

\section{Monitored Applications}
\label{sec:monitored-apps}

We monitored 12 applications (Table~\ref{tab:monitored-apps}). We targeted
social media and short-form video apps, since feed-based scrolling on these
platforms is linked to regret and feelings of meaninglessness in earlier ESM
studies~\cite{guo_what_2025, lukoff_what_2018}. Short-form video apps like
TikTok and YouTube were of particular interest given their algorithmic feeds
designed for extended viewing sessions.

\begin{table}[h]
  \caption{Applications Monitored in the Study}
  \label{tab:monitored-apps}
  \begin{tabular}{lll}
    \toprule
    Application & Package Identifier & Category \\
    \midrule
    Instagram & com.instagram.android & Social Media \\
    Facebook & com.facebook.katana & Social Media \\
    Twitter/X & com.twitter.android & Social Media \\
    Snapchat & com.snapchat.android & Social Media \\
    TikTok & com.zhiliaoapp.musically & Short-form Video \\
    TikTok (Intl.) & com.ss.android.ugc.trill & Short-form Video \\
    Reddit & com.reddit.frontpage & Social Media \\
    LinkedIn & com.linkedin.android & Social Media \\
    Pinterest & com.pinterest & Social Media \\
    YouTube & com.google.android.youtube & Video Streaming \\
    Xiaohongshu & com.xingin.xhs & Social Media \\
    Truth Social & com.truthsocial.android.app & Social Media \\
    \bottomrule
  \end{tabular}
\end{table}

\section{Smartwatch Data Collection}
\label{sec:smartwatch-data}

The Bangle.js~2 smartwatch~\cite{banglejs2} collected the following physiological and behavioral signals throughout the 7-day study period:

\begin{table}[h]
  \caption{Signals Collected by the Bangle.js 2 Smartwatch}
  \label{tab:smartwatch-signals}
  \begin{tabular}{lll}
    \toprule
    Signal & Sensor & Sampling \\
    \midrule
    Heart Rate (BPM) & PPG & 0.5Hz (2s intervals) \\
    HRM Confidence & PPG & 0.5Hz (2s intervals) \\
    Accelerometer (X, Y, Z) & Accelerometer & 0.5Hz (2s intervals) \\
    Step Count & Accelerometer & 0.5Hz (2s intervals) \\
    Activity Level & Accelerometer & 0.5Hz (2s intervals) \\
    Skin Temperature & Temperature sensor & 0.5Hz (2s intervals) \\
    Barometric Pressure & Barometer & 0.5Hz (2s intervals) \\
    Altitude & Barometer (derived) & 0.5Hz (2s intervals) \\
    \bottomrule
  \end{tabular}
\end{table}

Data collection followed a duty cycle of one-minute measurement windows followed by four-minute breaks, resulting in twelve measurement periods per hour. All data were stored locally on the watch and retrieved during the offboarding session.

\section{Detailed Model Comparisons}
\label{sec:appendix-model-comparisons}

This appendix reports the full model-family and per-variant best-model breakdowns that underpin the main ablation (Section~\ref{sec:rq2-ablation}) and the per-protocol AUC summary (Table~\ref{tab:rq2-methods-auc})

\subsection{Per-Variant Best Model: Within-Person}

\begin{table}[t]
\caption{Within-person classification: best model per feature subset (shuffled 70/30 split). AUC is reported for the per-participant above-median binarization.}
\label{tab:rq2-within-detail}
\centering
\small
\begin{tabular}{llcc}
\toprule
\textbf{Subset} & \textbf{Best Model} & \textbf{AUC} & \textbf{$n$ participants} \\
\midrule
Smartwatch           & CatBoost (native)  & .667 & 19 \\
Phone Context          & CatBoost (native)  & .690 & 21 \\
Smartwatch+Phone Context   & CatBoost (native)  & .725 & 19 \\
\midrule
Na\"ive baseline & Person mean         & .500 & --- \\
\bottomrule
\end{tabular}
\end{table}

\subsection{Bayesian regression model specification}
\label{sec:appendix-bayesian-spec}

The Bayesian hierarchical regression models referenced in Section~\ref{sec:rq1} were implemented in PyMC and fitted using the No-U-Turn Sampler (NUTS) with 4 chains, 1{,}000 posterior draws per chain, and 2{,}000 warm-up steps. The target acceptance rate was set to 0.90.

All models included by-participant random intercepts to account for stable individual differences in baseline regret levels. Continuous predictors were $z$-scored within the analysis sample of each model to make regression coefficients directly comparable across predictors. Binary predictors were coded as 0/1 indicators and left unstandardized. We used weakly informative priors throughout: the grand intercept received $\mathcal{N}(3.5,\, 1.5)$, centered on the midpoint of the 1--7 regret scale; all fixed-effect slopes received $\mathcal{N}(0,\, 1)$, placing 95\% prior mass on effects smaller than $\pm$2 points per SD of the predictor; participant-level random intercepts were drawn from $\mathcal{N}(0,\, \sigma_u)$ with $\sigma_u \sim \text{HalfNormal}(1)$; residual SD received $\text{HalfNormal}(2)$. Models with by-participant random slopes used a non-centred parameterisation to avoid sampling pathologies near zero.

Convergence was assessed via the Gelman--Rubin statistic ($\hat{r}$) and bulk effective sample size (ESS). All models achieved zero divergent transitions. $\hat{r}$ reached a maximum of 1.01 for the grand intercept and was 1.00 for all focal parameters.

\section{Supplementary Figures}
\label{sec:supplementary-figures}

\begin{table}[ht]
\centering
\small
\caption{App category mapping used for analyses. Social media apps are those that triggered EMA surveys in the study; all other categories classify the
         \emph{preceding} app session. System apps (launcher, settings, etc.)
         were excluded from the preceding-app sequence entirely.}
\label{tab:app_categories}
\begin{threeparttable}
\begin{tabular}{lp{10cm}}
\toprule
\textbf{Category} & \textbf{Apps} \\
\midrule
Social media
  & Instagram, YouTube, Facebook, TikTok, Reddit, X, Snapchat, LinkedIn, Pinterest \\[4pt]
Productivity
  & Gmail, Outlook, Calendar, Calculator, Reminder, Samsung Notes, Drive, Docs,
    Sheets, Slides, Excel, Canvas, AnkiDroid, Duolingo, Notion, Obsidian,
    Keep Notes, Tasks, Adobe Scan, Camera, Gallery, Photos \\[4pt]
Communication
  & Telegram, WhatsApp, Messages, Discord, Messenger, Slack,
    Call, Phone, Signal, WeChat, LINE \\[4pt]
Information
  & Chrome, Google, Brave, Edge, Ecosia, Safari, Samsung Internet, Tor Browser,
    Maps, Weather, ChatGPT, Perplexity, Photomath \\[4pt]
Entertainment
  & Spotify, Netflix, YouTube Music, YT Music ReVanced, Samsung Music,
    MusicTube, SimpMusic, NewPipe, Twitch, Crunchyroll, Steam, WEBTOON,
    LiveScore, Formula 1, Pokémon GO, Pokémon Sleep, DreamyRoom, Retro Bowl,
    Offline Games, The Battle Cats, Block Blast!, Minesweeper,
    2 Player Games, Love\&deepspace, Heartopia \\
\bottomrule
\end{tabular}
\begin{tablenotes}
  \footnotesize
  \item \textit{Note.} For Finding 1.3, the preceding-app category
  is determined by the immediately preceding non-system app session
  (\texttt{app\_cat\_pre\_1}); SM-to-SM transitions ($n = 498$ sessions) are excluded,
  yielding $N = 947$ sessions in the H1.3 model.
\end{tablenotes}
\end{threeparttable}
\end{table}

\section{Open Source Materials}

To support replication and extension of this work, we release three artifacts as open source upon publication.

\textbf{Android Observation Application.}
The custom Android application used to instrument participants' phones will be released in full. The repository includes the background app-usage logging module (using \texttt{UsageStatsManager}~\cite{android_usagestatsmanager} for batch hourly queries via \texttt{WorkManager}~\cite{android_workmanager}), the real-time \texttt{AccessibilityService}~\cite{android_accessibilityservice} that detects exits from monitored apps and fires post-session survey notifications, the local caching and upload logic, and the configurable list of monitored applications. The code is available at [URL to be added upon acceptance].

\textbf{Bangle.js~2 Watch Application.}
The Espruino JavaScript application running on the Bangle.js~2 smartwatch~\cite{banglejs2} will also be released. It implements the 1-minute-on / 4-minute-off sampling protocol for heart rate (PPG), accelerometry, and skin temperature, stores timestamped readings to the watch's local flash, and exposes a Gadgetbridge-compatible~\cite{gadgetbridge} interface for clock synchronization with the paired phone. Because the Bangle.js~2 runs fully open-source firmware, the watch app can be sideloaded without any proprietary toolchain, lowering the barrier for researchers wishing to replicate the sensing pipeline on the same hardware. The code is available at [URL to be added upon acceptance].

\section{Quote Book}
\label{sec:appendix-quote-book}

\begingroup
\small
\setlength{\LTleft}{0pt}
\setlength{\LTright}{0pt}
\setlength{\tabcolsep}{4pt}
\renewcommand{\arraystretch}{1.08}

\begin{longtable}{@{}>{\RaggedRight\arraybackslash}p{0.24\textwidth}>{\RaggedRight\arraybackslash}p{0.21\textwidth}>{\RaggedRight\arraybackslash}p{0.51\textwidth}@{}}
\caption{Quote book for qualitative themes. Representative quotations are organized by theme, code, and illustrative quotation. Theme definitions are provided in the first column. Participant IDs are retained; names are removed.}
\label{tab:appendix-quote-book}\\
\toprule
\textbf{Theme and Definition} & \textbf{Code} & \textbf{Illustrative Quotations} \\
\midrule
\endfirsthead

\multicolumn{3}{@{}l}{\tablename\ \thetable{} -- Continued from previous page}\\
\toprule
\textbf{Theme and Definition} & \textbf{Code} & \textbf{Illustrative Quotations} \\
\midrule
\endhead

\midrule
\multicolumn{3}{r@{}}{Continued on next page}\\
\endfoot

\bottomrule
\endlastfoot
\label{appendix:quotebook}

\textbf{Time Blindness}\newline \emph{Definition:  Recurring collapse of temporal self-monitoring during scrolling.} & \textbf{Time distortion} & ``Something that would help with the time awareness. Because my sense of time goes completely blank when I'm doomscrolling. I think, oh, I've been here for two minutes, and I check it...'' P3 \\
\cmidrule(l){2-3}
 & \textbf{Prospective memory failure} & ``I forget I have to do something, and then I just scroll... And then after three hours of scrolling, I'm like, oh, that was so unproductive. Oh, wait, I had something to do.'' P6 \\
\cmidrule(l){2-3}
 & \textbf{Post-hoc time loss} & ``Time being lost and could have been spent some[thing] obviously better... waiting, I was like idling... before resting, it's kind of like clear [regret].'' P1 \\
\cmidrule(l){2-3}
 & \textbf{Missing time feedback} & ``The issue before was that I didn't have a meaningful way of measuring my time. I wouldn't be checking my screen time constantly, but getting the notifications, `oh, you have 10 minutes left'...'' P14 \\
\cmidrule(l){2-3}
 & \textbf{Routine time loss} & ``It happens a lot in bed... at home, I have some bad habits I have in place, such as a morning bed rot. So I'll wake up and the first thing I do is scroll for several hours and then get up.'' P2 \\
\midrule

\textbf{Intent Drift}\newline \emph{Definition: Drifting from an intended phone task into unintended social media use or recommended content.} & \textbf{Communication-to-scroll drift} & ``I started off with wanting to check one thing -- like a message from a friend on Instagram or on any social media -- and then I ended up... scrolling for a few hours, and then I feel like, oh no, I only meant to check that one thing.'' P3 \\
\cmidrule(l){2-3}
 & \textbf{Recommendation engine redirects} & ``There are some things that I use YouTube for, like tutorials... But then after that, then I start watching KSI and Mr. Beast.'' P9 \\
\cmidrule(l){2-3}
 & \textbf{Platform-part distinction} & ``Eventually I do want it back in my life, and so I would like to find a way of incorporating it in a healthy manner... There are parts of the platforms that I enjoy and there are parts that I don't.'' P2 \\
\cmidrule(l){2-3}
 & \textbf{Work-waiting drift} & ``If I'm waiting for something to finish computing in my work, I will open up Reddit or something else to read, to pass time. And then more time passes when I realize.'' P5 \\
\cmidrule(l){2-3}
 & \textbf{Quality loss after drift} & ``I wouldn't say I'm happy. I also wouldn't say I'm unhappy... it doesn't stop me from doing the things I have to do. However, I do find that it takes me longer to get things done because I oftentimes will get distracted on my phone or computer.'' P16 \\
\midrule

\textbf{Escape Hatch}\newline \emph{Definition: Using social media as a low-effort escape from tasks that felt cognitively costly.} & \textbf{Justified avoidance} & ``If I had to miss a commitment for something, and I had the possibility of making it up, I would end up procrastinating making up that commitment... instead of doing that, I would just end up doomscrolling or scrolling through social media, telling myself, you know, I deserve a break after the first fall semester.'' P7 \\
\cmidrule(l){2-3}
 & \textbf{Task pile-up avoidance} & ``I feel like I have a lot of things to do that I haven't yet done... I was just not studying last minute... I also have... emails, applications to fill and stuff, but I haven't gotten them done because, mostly because of YouTube.'' P9 \\
\cmidrule(l){2-3}
 & \textbf{Procrastination context} & ``Definitely when I'm back in my dorm, or at home, or at night time especially, or when I'm procrastinating doing work.'' P3 \\
\cmidrule(l){2-3}
 & \textbf{Mental fatigue avoidance} & ``Whenever there's a moment free, or [a] moment bored, or... even if I have something to do but I'm tapped down mentally, social media... there's always something new.'' P8 \\
\cmidrule(l){2-3}
 & \textbf{Break framing} & ``A lot of the time it's when I... let my guard down, or I'm just like, oh, it's okay if I take a break... and then I... can't stop. I can't get myself back together.'' P2 \\
\midrule

\textbf{Why Timers Fail}\newline \emph{Definition: Participants' explanations for why existing screen-time tools do not work for them.} & \textbf{Notification bypass} & ``YouTube has this thing where you can set a time thing where it'll give you notification every X minutes... For that one, you could just X out of it. So it's just like a notification.'' P2 \\
\cmidrule(l){2-3}
 & \textbf{Habituation to alerts} & ``They had the thing where every however many minutes it would have a pop-up... I would just start ignoring them when I really need to lock in. I will delete it, and I'll delete the app.'' P4 \\
\cmidrule(l){2-3}
 & \textbf{Technical bypass} & ``Bypass methods: reinstalling the apps; there's a menu where you can turn off certain features.'' P11 \\
\cmidrule(l){2-3}
 & \textbf{Habit below the timer threshold} & ``I have screen time limits on my phone, obviously you can just bypass that... the act of just picking up my phone as a habit... the screen time blocker [doesn't] stop that.'' P16 \\
\cmidrule(l){2-3}
 & \textbf{Reactance} & ``Sometimes I feel like [Instagram's built-in timer] has been more of a `I want to defy [it]. Like, why are you telling me I can't scroll? I want to scroll.' '' P18 \\

\end{longtable}
\endgroup

\section{End-of-Day Survey}
\label{appendix:eod}

Participants completed the following end-of-day survey each evening of the seven-day study period, delivered via Qualtrics. The survey took approximately 2--3 minutes to complete and combined open-ended reflection items with closed-ended Likert ratings covering phone use and lifestyle dimensions.

\paragraph{Open-ended reflection.}

\begin{enumerate}
    \item How did today's social media sessions usually start?
    \item What usually ended or interrupted your sessions today?
    \item Have you used any social media applications on devices other than your phone today? If yes, please specify (for example, your laptop).
    \item In what contexts did you use your phone most today? You can name multiple.
\end{enumerate}

\paragraph{Phone use ratings (7-point Likert).}
\begin{enumerate}
    \setcounter{enumi}{4}
    \item ``I feel regret about my overall phone use today.'' (1 = Strongly disagree, 7 = Strongly agree)
    \item Compared with what you had intended for the day, did you spend more or less time than intended? (1 = Much less than intended, 7 = Much more than intended)
    \item With respect to your phone usage today, how much do you feel like you have spent your time on something meaningful? (1 = Not at all meaningful, 7 = Very meaningful)
\end{enumerate}

\paragraph{Lifestyle ratings (7-point Likert).}
\begin{enumerate}
    \setcounter{enumi}{7}
    \item How healthy was your eating today? (1 = Very unhealthy, 7 = Very healthy)
    \item How physically active were you today? (1 = Not at all active, 7 = Very active)
    \item How restful was your sleep last night? (1 = Not at all restful, 7 = Very restful)
    \item How stressful was your day today? (1 = Not at all stressful, 7 = Very stressful)
    \item How satisfied were you with your social interactions today? (1 = Not at all connected, 7 = Very connected)
\end{enumerate}
\section{Sample Exit Interview Questions}
\label{appendix:interview}

The following questions formed the core of our semi-structured exit interviews, conducted in person at the end of the seven-day study period. Questions were used as anchors, with follow-up probes adapted to each participant's responses. All interviews were audio-recorded with consent and transcribed for thematic analysis.

\paragraph{Regretful and meaningful use.}
\begin{enumerate}
    \item Can you describe specific instances during the study week when you regretted a social media session, and what made those sessions feel regretful?
    \item Were there particular contexts, times of day, or emotional states in which regretful sessions tended to occur?
    \item Can you describe instances when your social media use felt meaningful, and what distinguished these from regretful sessions?
\end{enumerate}

\paragraph{Digital habits.}
\begin{enumerate}
    \setcounter{enumi}{3}
    \item How would you characterize your current relationship with social media? Are there aspects of your usage you would want to change?
\end{enumerate}

\paragraph{Prior interventions.}
\begin{enumerate}
    \setcounter{enumi}{4}
    \item Have you previously used any tools or strategies to manage your social media use? If so, how effective were they and why?
    \item Based on your experience this week, what kinds of interventions do you think would help align your social media use with your intentions?
\end{enumerate}
\clearpage

\section{Additional Figures }

\begin{figure*}[h]
  \centering
  \includegraphics[width=\textwidth]{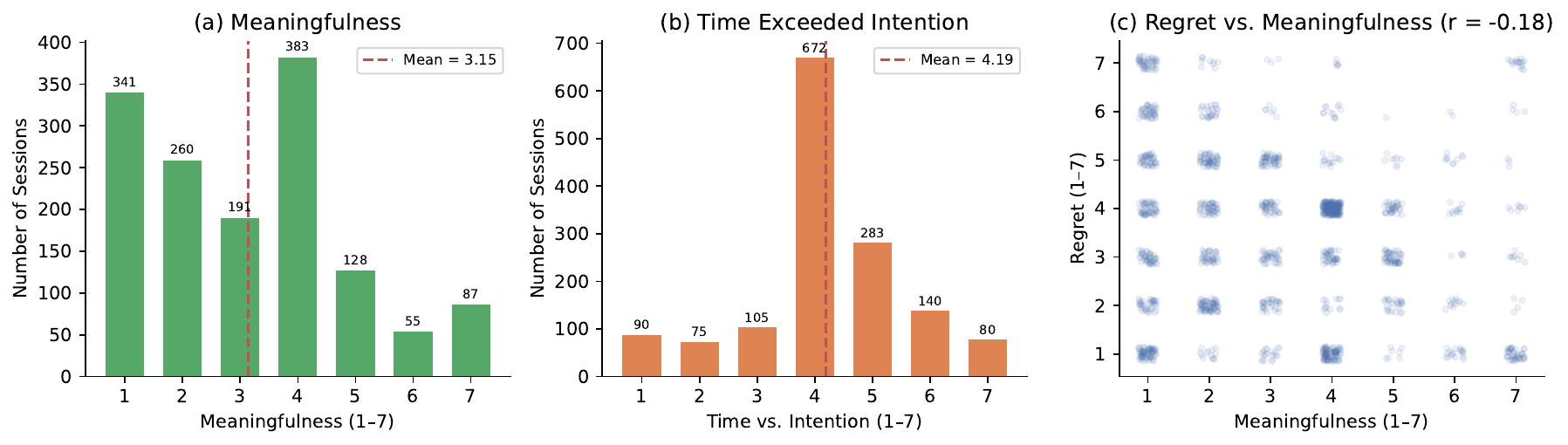}
   \caption{Co-outcome distributions. (a) Session meaningfulness ratings, with most sessions rated low to moderate. (b) Time exceeded intention ratings, skewed toward ``more than intended.'' (c) Scatter plot of regret versus meaningfulness with jitter, showing a modest negative correlation (r~=~-.18).}
  \label{fig:outcome-distributions}
  \Description{Three panels showing distributions of meaningfulness, time comparison, and a scatter plot of regret vs meaningfulness.}
\end{figure*}

\begin{figure*}[h]
  \centering
  \includegraphics[width=\textwidth]{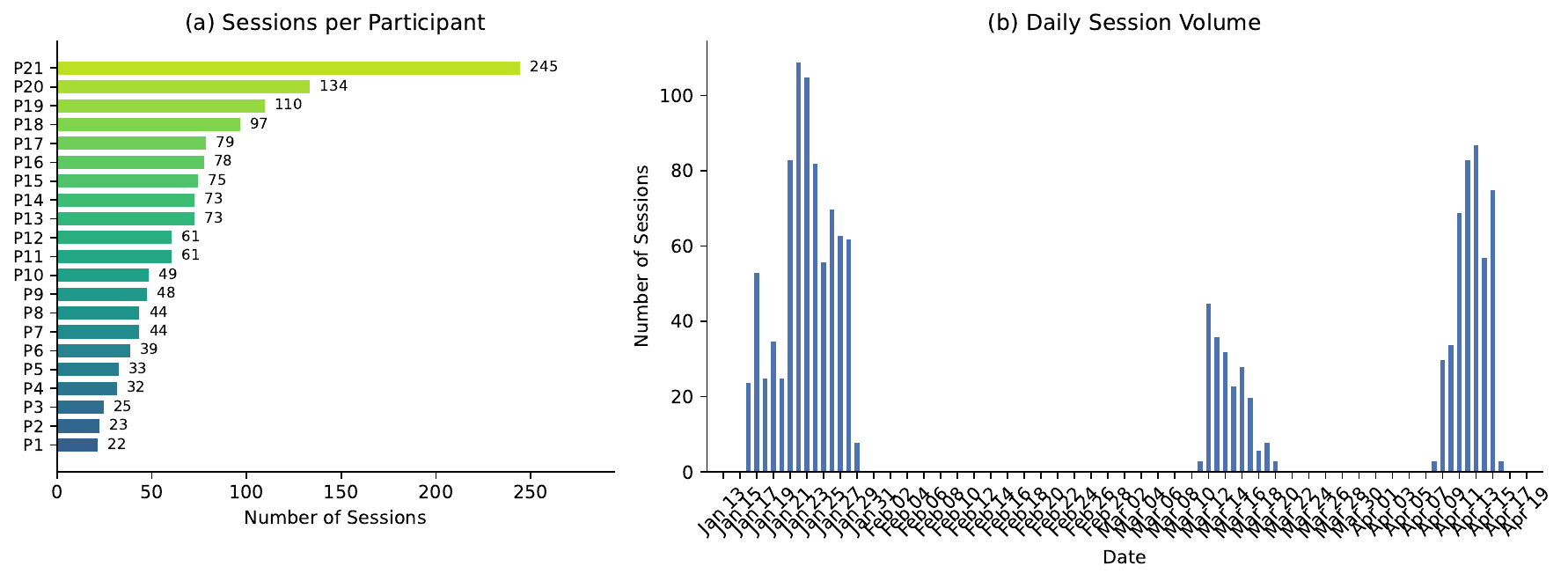}
  \caption{Data collection overview. (a) Number of logged social media sessions per participant, ranging from 22 to 245. (b) Daily session volume across the study period, showing a ramp-up as participants were onboarded in waves.}
  \label{fig:data-overview}
  \Description{Left panel shows a horizontal bar chart of sessions per participant. Right panel shows a bar chart of daily session counts over the study period.}
\end{figure*}

\begin{figure}[h]
  \centering
  \includegraphics[width=\columnwidth]{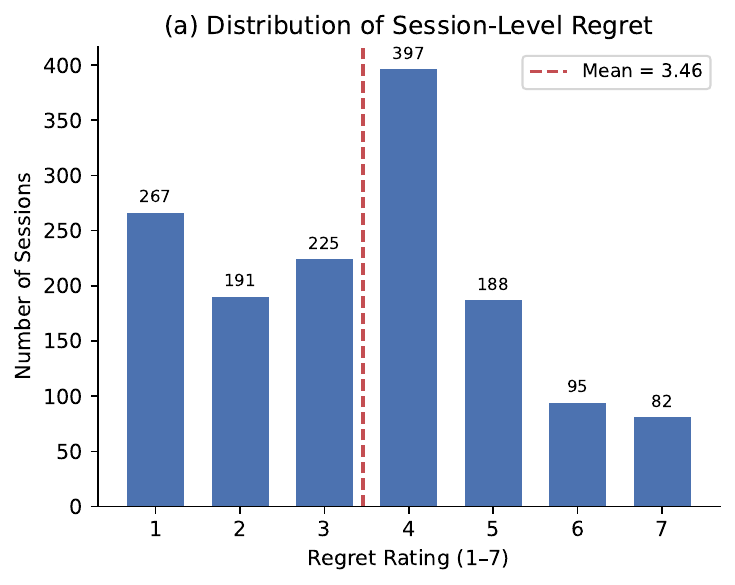}\\[4pt]
  \includegraphics[width=\columnwidth]{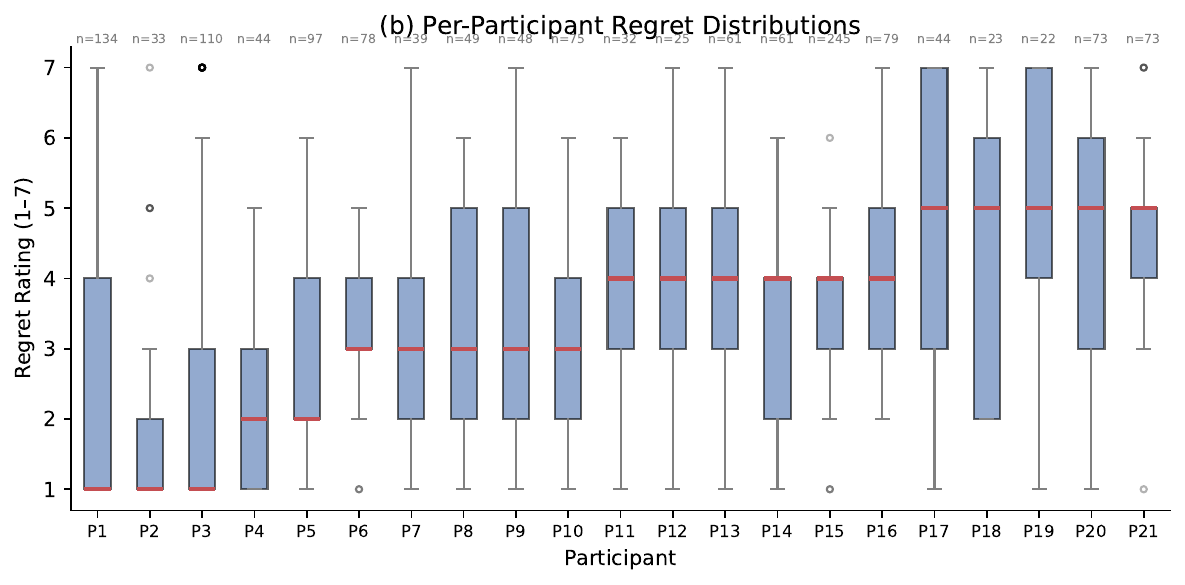}
  \caption{Distribution of session-level regret ratings. (a) Histogram of all 1{,}445 sessions showing a slight positive skew with a modal response of 4. The dashed line indicates the overall mean (3.46). (b) Per-participant box plots sorted by median regret, with session counts annotated above each box. Participants exhibited markedly different baseline regret levels and variability, motivating within-person modeling.}
  \label{fig:regret-distribution}
  \Description{Two stacked panels: (a) histogram of all sessions by regret rating 1--7, and (b) per-participant box plots for 21 participants sorted by median regret.}
\end{figure}

\begin{figure}[h]
  \centering
  \includegraphics[width=0.32\textwidth]{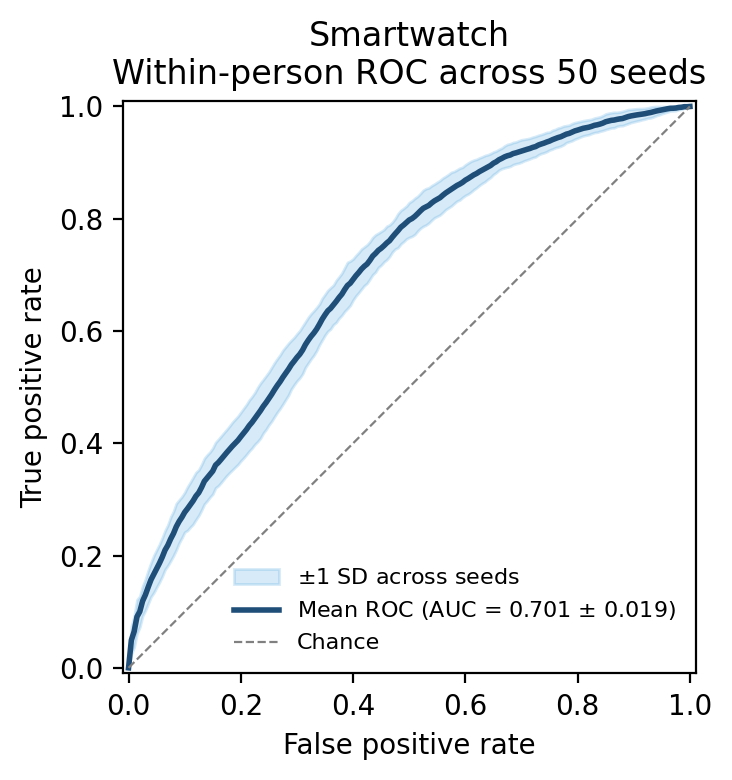}%
  \hfill%
  \includegraphics[width=0.32\textwidth]{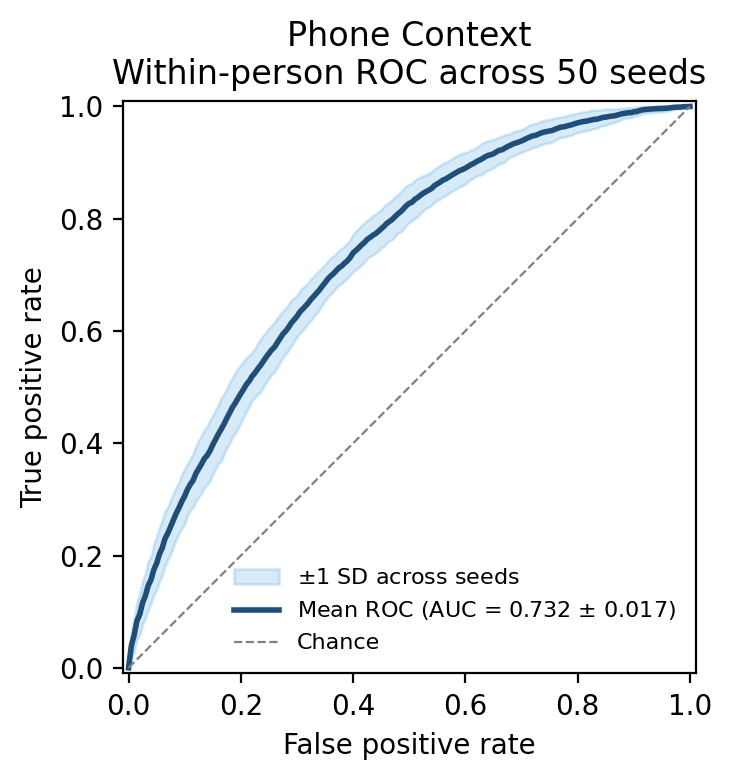}%
  \hfill%
  \includegraphics[width=0.32\textwidth]{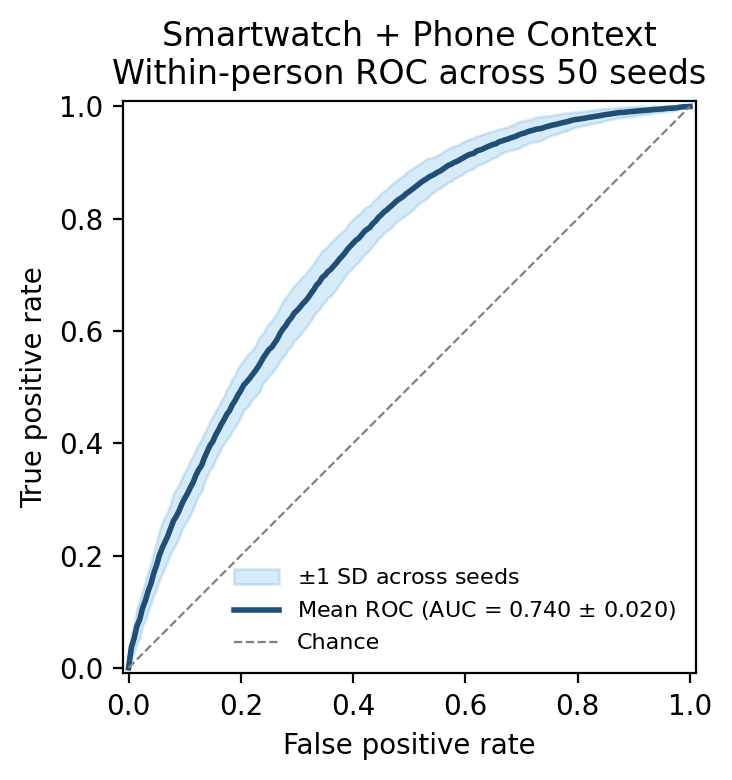}
  \caption{Within-person ROC curves across the three feature subsets: Smartwatch ($n=19$, best AUC\,=\,.667), Phone Context ($n=21$, best AUC\,=\,.690), and Smartwatch+Phone Context ($n=19$, best AUC\,=\,.725). On the smartwatch-eligible subset, Smartwatch+Phone Context yields the highest pooled-model AUC. Smartwatch-based panels use the 19 participants with any smartwatch coverage.}
  \label{fig:rq2-within-roc}
  \Description{Three ROC curve panels comparing model performance across Smartwatch, Phone Context, and Smartwatch+Phone Context feature subsets. Smartwatch+Phone Context shows the highest curves; Smartwatch alone shows the lowest.}
\end{figure}

\end{document}